\documentclass{article}
\usepackage{geometry}
\usepackage{enumitem}
\usepackage{hyperref}
\usepackage{graphicx}
\usepackage{url}
\usepackage{circus}

\begin{document}

\title{A Formal Specification of Operating System based on ARINC 653}
\author{Ziyan Wang$^1$  and Yan Zhang$^2$ \thanks{This work received financial support of the Youth Program of National Natural Science Foundation of China (No 61802191). Corresponding authors: 821651646@qq.com and zhangyan@njfu.edu.cn.} \\  
	$^{1,2}$ College of Information Science and Technology \\ Nanjing Forestry University, P. R. China  
}	

\date{\today}

\maketitle	

\begin{abstract}
	In this paper, by using the formal language \emph{Circus}, we give a formal specification of an operating system based on ARINC 653 standard.Our specification includes interrupt handling, time and memory management, partition and process scheduling, system call response and related APEX services. Especially, the concurrent behaviours of partitions and processes are also specified.
\end{abstract}

\section{Introduction}
\subsection{ARINC 653}
\hspace{1.1em}
In avionic industry, a new architecture of embedded real-time operating system called Integrated Modular Avionics (IMA) has gradually become the mainstream. Compared with old avionic architectures, IMA enables a variety of applications to execute on the same hardware and significantly reduces the power requirements, weight and costs of maintenance.

The guidelines for the design of IMA architecture have been developed by Aeronautical Radio Incorporated (ARINC). It is known as ARINC 653 \cite{ARINC653} and first released in 1997. The document proposes the baselines of operating environment for application software used in IMA architecture by defining a general-purpose APEX (APplication/EXecutive) interface between the operating system and application software. It also ensures the portability that application software, which uses the APEX interface, can execute on any operating system based on ARINC 653 standard.   
  
\begin{figure}[ht]
	\centering
	\includegraphics[scale=0.9]{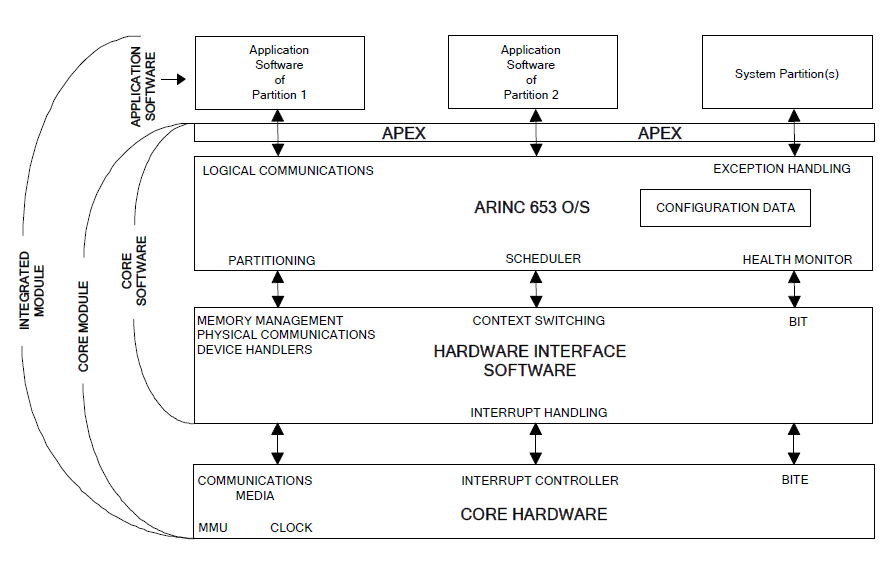}
	\caption{Architecture of ARINC 653 Module}
	\label{fig:architecture}
\end{figure}

The most important part of an IMA is ARINC 653 module. Figure 1 is the typical architecture of it. As shown in Figure 1, an integrated module contains a core module and application software running on it. The core module is composed of core hardware and core software.

An ARINC 653 module is divided into five layers: Core Hardware, Hardware Interface Software, ARINC 653 O/S, APEX and Application Software. The bottom layer Core Hardware is a layer of hardware units including CPU, MMU, clock, interrupt controller and etc. This layer provides hardware foundations for all the software of ARINC 653 architecture. The three layers in the middle make up the core software in an ARINC 653 module. The layer above Core Hardware is Hardware Interface Software, which provides functions to manipulate hardware units. ARINC 653 O/S layer is the most important part of the system because it makes decisions on scheduling and provides system calls to support the functions in APEX layer.There are also many functional modules in this layer, like Exception Handler and Health Monitor. APEX, which is similar to API in general-purpose operating system, provides an interface for application software.

The philosophy of ARINC 653 is partitioning. Applications resident in a module are partitioned with respect to space (spatial partitioning) and time (temporal partitioning). To support spatial partitioning, each partition is allocated with an amount of space in memory where the application of the partition resides. The space of each partition is specified during system configuration and initialization. To support temporal partitioning, the operating system maintains a major time frame of fixed duration, which is periodically repeated throughout the module's run time. Each partition is allocated with one or more time windows within a major time frame. The application of each partition is accessible to resources only during its own time window. 

In ARINC 653, a partition comprises one or more processes that operate alone or cooperate with each other to provide functions of the application. The processes in a partition which are scheduled according to their priority can be divided into two kinds: aperiodic and periodic. Each process is also associated with a deadline time before which the process should finish its execution if its deadline time is set to a finite value.

\subsection{Circus}
\hspace{1.1em}
Circus \cite{denotational,derivation,semantics} is a formal language that combines Z \cite{z}, CSP \cite{csp}, Morgan's specification statements, and Dijkstra's guarded commands. One main advantage of Circus is that the states and behaviours of a process can be captured in the same specification. It can be used for specification, programming, and verification by refinement. Its semantics is based on Hoare and He's Unifying Theories of Programming \cite{utp}.

A Circus program is composed of a sequence of Circus paragraphs and each paragraph can be a definition of a channel, a channelset or a Circus process.

Channels are used for communication between Circus processes. They are declared by their names and defined by the types of what they send or receive; if a declared channel does not deliver any value it only gives a signal to indicate that some synchronizing event happens. A channelset is a set of channels that are used to connect several Circus processes or several actions inside a Circus process. 

Circus process is the main part of the Circus language which can be defined explicitly or implicitly. The explicit definition of a Circus process is delimited by the key words \emph{begin} and \emph{end}, whose actions are enumerated between them. At the start of a Circus process, a schema is defined after the keyword \emph{state} if the process maintains a set of variables as its local state. At the end of a Circus process, a nameless action, which follows after a spot, defines the behaviours of the process. The implicit definitions of the Circus processes are compositions of defined processes with Circus operators. For example, given two Circus processes $ P_1 $ and $ P_2 $, combine them with a binary operator sequence, we get $ P_1  ; P_2 $ which means that it will execute the process $ P_2 $ after termination of execution of $ P_1 $.

For the whole definition of Circus, the readers could refer to \cite{denotational}. Here we only present some typical examples in our project to illustrate how to use Circus.

We can define types by declaration. For example,
\begin{zed}
[PSU] \\
[INTERRUPT]
\end{zed}
Here, PSU is short for Primary Storage Unit which is a 64-bit number. The type INTERRUPT denotes the set of all interrupts produced by hardware units which are electronic signals in core hardware layer.

Constants are defined by axiomatic definition. For example,
\begin{zed}
0_{PSU} : PSU \\
1_{PSU} : PSU \\
ClockTickInterval : PSU \\
CLOCK \_ TICK : INTERRUPT
\end{zed}
where $ 0_{PSU} $ and $ 1_{PSU} $ are the values zero and one respectively in a 64-bit system. ClockTickInterval is the time interval between two clock ticks and $ CLOCK \_ TICK $ is a kind of interrupt generated by the timer.

In Circus, we can use the reserved word \emph{channel} to define channels. For example,
\begin{zed}
\circchannel module \_ init,module \_ end \\
\circchannel rtclock \_ power \_ on,rtclock \_ power \_ off \\
\circchannel clock \_ pause \\
\circchannel hardware \_ interrupts : INTERRUPT
\end{zed}
$ module \_ init $ and $ module \_ end $ are two signals from the environment. They denote two life stages of ARINC 653 modules. Similarly, the two signals $ rtclock \_ power \_ on $ and $ rtclock \_ power \_ off $ denote two life stages of real-time clocks. $ clock \_ pause $ is a kind of signal generated by real-time clock and received by timer. The channel $ hardware \_ interrupts $ is used by various hardware units to send external interrupts to Interrupt Controller.

Next we give two Circus processes to model real-time clock and timer one by one.
\begin{zed}
	\circprocess RTClock \circdef \\
	\circbegin \\
	\circstate \ RTCLOCK \_ TYPE == [Value : PSU] \\
	\\
	\begin{array}{lll}
		Initialize & \circdef & Value := 0_{PSU} \\
		\\
	\end{array}
	\\
	\begin{array}{lll}
		RunClockPause & \circdef & \circmu X \circspot (\textbf{wait}(1) \circseq Value := Value + 1_{PSU} \circseq clock \_ pause) \circseq X \\
		\\
	\end{array}
	\\
	\begin{array}{lll}
		\circspot rtclock \_ power \_ on \circthen (Initialize \circseq RunClockPause) \Delta rtclock \_ power \_ off \circthen \Skip \\
		\\
	\end{array}
	\\		
	\circend
	\\
\end{zed}
In real-time clock, a local state is maintained. It is defined by a Z schema which includes a variable Value whose type is PSU. Then, two Circus actions, Initialize and RunClockPause, are defined.The former is simple that the local variable Value is set to $ 0_{PSU} $. The latter is a recursive composition that the sequent action
\begin{equation}
	\textbf{wait}(1) \circseq Value := Value + 1_{PSU} \circseq clock \_ pause
\end{equation}
will be repeated infinitely many times. (1) means that it signals a clock pause after waiting for a basic time unit which is a nanosecond according to ARINC 653. Then, the behaviour of the Circus process RTClock is defined at last. After receiving the power-on signal, it initializes its state and then generates clock pauses until the power runs out. $ \Delta $ is used to describe that $ Initialize \circseq RunClockPause $ is interrupted on occurrence of the signal $ rtclock \_ power \_ off $.

The model of timer is defined as follows:
\begin{schema*}{InitTimer}
	TIMER \_ TYPE' \\
	p? : PSU
	\where
	Value' = 0_{PSU} \\
	Period' = p?
\end{schema*}
\begin{zed}
	\circprocess Timer \circdef \\
	\circbegin \\
	\circstate \ TIMER \_ TYPE == [ Value,Period : PSU ] \\
	\\
	\begin{array}{lll}
		Initialize & \circdef & InitTimer[ClockTickInterval/p?] \\
		\\
	\end{array}
	\\
	\begin{array}{lll}
		ClockTick & \circdef & clock \_ pause \circthen Value := Value + 1_{PSU} \circseq \\
							 && \left(
								\begin{array}{ll}
									\circif & (Value \neq Period) \circthen \Skip \\
									\circelse & (Value = Period) \circthen Value := 0_{PSU} \circseq \\
											  & hardware \_ interrupts \ !CLOCK \_ TICK \circthen \Skip \\	
									\circfi
								\end{array}
								\right) \\
		\\
	\end{array}
	\\
	\begin{array}{lll}
		RunClockTick & \circdef & \circmu X \circspot ClockTick \circseq X \\
		\\
	\end{array}
	\\
	\begin{array}{lll}
		\circspot module \_ init \circthen (Initialize \circseq RunClockTick) \Delta module \_ end \circthen \Skip \\
		\\
	\end{array}	
	\\
	\circend
	\\
\end{zed}
The structure of Circus process Timer is similar to that of RTClock. We give an initialization by a Z operation, the schema InitTimer, to initialize the state of Timer. The action ClockTick expresses that the local variable Value will be updated by adding $ 1_{PSU} $ when a clock pause is received, then compare the value of two local variables, if Value is not equal to Period, terminate immediately, otherwise, reset Value to $ 0_{PSU} $ and send a signal $ CLOCK \_ TICK $ via Circus channel $ hardware \_ interrupts $.

Next by using a binary synchronization operator, we define a Circus process PeriodicTicker as follows.
\begin{zed}
	\circchannelset RTC \_ TIMER == \lchanset clock \_ pause \rchanset \\
	\circprocess PeriodicTicker \circdef RTClock \lpar RTC \_ TIMER \rpar Timer
\end{zed}
where two Circus processes RTClock and Timer communicate with each other by the channelset $ RTC \_ TIMER $ which contains only one channel $ clock \_ pause $.

We end this subsection by a complex example, the definition of ARINC 653 O/S Layer.
\begin{zed}
	\circprocess ARINC653 \_ OS \_ Layer \circdef OSKernelConfig : OS \_ KERNEL \_ CONFIG \_ TYPE \circspot \\
	\\
	\begin{array}{lll}
		CoreKernel(OSKernelConfig) \lpar OSLayerChannels \rpar (\Interleave partid : \dom (OSKernelConfig.Partitions) \circspot \\
		PartitionKernel(partid,OSKernelConfig.Partitions(partid))) \\
		\\
	\end{array}
\end{zed}
The process has an parameter OSKernelConfig which is the configuration of OS kernel. It can be used as a local variable. This Circus process is introduced by implicit definition of Circus processes CoreKernel and PartitionKernel with the binary synchronization operator $ \lpar \ \rpar $ and the general interleaving operator $ \Interleave $. Circus process CoreKernel and PartitionKernel are defined in Appendix D.3. The former has one argument OSKernelConfig and the latter has two: partid, which is a variable ranging over all identifiers of the partitions, and OSKernelConfig.Partitions(partid), which is the configuration of the partition indexed by partid (the structure of OSKernelConfig is in Appendix A.5). The specification of ARINC 653 OS layer reflects that several partition kernels interleave with each other according to time partitioning. They all synchronize with core kernel via the channelset OSLayerChannels in their own time windows.

\subsection{Related Work}
\hspace{1.1em}
There are plenty of researches on formal models of operating systems and ARINC 653 standard. 

Craig presents several formal models of classical system kernels, ranging from simple ones to complex ones, who also proves a lot of desired properties of the models \cite{formalmodels,formalrefinement}. These models of general-purpose kernels inspire us when we try to model a real-time kernel which meets ARINC 653 standard.Oliveira Gomes gives a formal model of an operating system with IMA architecture, which covers a small part of ARINC 653 services and kernel functions \cite{formalspecification}. In \cite{aadl} and \cite{twolevel}, the researchers focus on the partition scheduler and process scheduler of IMA-architecture system, the characteristics of temporal partitioning are captured while the spatial partitioning is omitted. In the series of Zhao's works \cite{zhaoaadl,zhaoeventb,zhaosk}, separation kernels as well as APEX interface have been modelled and verified. Moreover, three errors in ARINC 653 document have been detected. In our model, we also propose some solutions to avoid these errors.

In industry, there are several real-time operating systems based on ARINC 653 standard \cite{lynx,pikeos,vxworks653,vxworksmils}, such as VxWorks 653 and LynxOS-178. In VxWorks 653, the operating system is implemented as a Multi-Virtual Machine (MVM) framework. The core OS is the host OS and each partition OS is a guest OS. The advantage of this framework is the diversity of guest OS that each partition can install a different operating system from each other, such as real-time OS with APEX interface, general-purpose OS with POSIX interface. And what resides in a partition can even be some very old software that executes on bare machine. The disadvantage of this framework is the high cost of address translation and privilege instruction handling. A virtual machine manager for each partition is also necessary.

According to ARINC 653 document, we give a formal model of partition operating system where each partition OS is a real-time system with APEX interface.

\section{A Circus Model of OS based on ARINC 653}
\hspace{1.1em}
In this section, we give a brief illustration of the structure of our model in Appendix. Firstly, we list all titles of the sections and subsections of Appendix and their associated page numbers.

\begin{enumerate}[label={\Alph*}]
	\item Types and Constants
		  \begin{enumerate}[label={A.\arabic*}]
		  	\item Core Hardware Layer \pageref{section:types&constant_CoreHardware}
		  	\item Hardware Interface Software Layer \pageref{section:types&constant_HardwareInterfaceSoftware}
		  	\item ARINC 653 O/S Layer(Core Kernel) \pageref{section:types&constant_CoreOS}
		  	\item ARINC 653 O/S Layer(Partition Kernel) \pageref{section:types&constant_PartitionOS}
		  	\item ARINC 653 O/S Layer(Operating System Kernel) \pageref{section:types&constant_OS}
		  	\item Application/Executive(APEX) \pageref{section:types&constant_APEX}
		  	\item Core Module \pageref{section:types&constant_CoreModule}
		  \end{enumerate}
	\item Z Operations
		  \begin{enumerate}[label={B.\arabic*}]
		  	\item Core Hardware Layer \pageref{section:zoperation_CoreHardware}
		  	\item Hardware Interface Software Layer \pageref{section:zoperation_HardwareInterfaceSoftware}
		  	\item ARINC 653 O/S Layer \pageref{section:zoperation_OS}
		  \end{enumerate}
	\item Circus Channels
	\item Circus Processes
		  \begin{enumerate}[label={D.\arabic*}]
		  	\item Core Hardware Layer \pageref{section:circusprocess_CoreHardware}
		  	\item Hardware Interface Software Layer \pageref{section:circusprocess_HardwareInterfaceSoftware}
		  	\item ARINC 653 O/S Layer \pageref{section:circusprocess_OS}
		  	\item Application/Executive(APEX) \pageref{section:circusprocess_APEX}
		  	\item Core Module \pageref{section:circusprocess_CoreModule}
		  \end{enumerate}
\end{enumerate}

The Appendix is divided into four parts: Types and Constants (see Appendix A), Z Operations (see Appendix B), Circus Channels (see Appendix C) and Circus Processes (see Appendix D). In Appendix A, we define all types and constants of each layer as well as some related functions. They are simply declared, declared by abbreviation or axiom, or defined as schemas. In Appendix B, we define operations in Z schema to illustrate initializations and state transitions. All these schemas can be defined within the Circus processes where they belong, for simplicity, we put them in a separate section. In Appendix C, we define all channels. The channelsets are presented in Circus Processes where they are used. In Appendix D, for each layer, we define behaviours of all components in this layer, and then compose them to form its behaviours. Further, we connect the Circus processes of each layer with channelsets so that the behaviours of ARINC 653 Core Module can be specified.

In Core Hardware Layer, the model includes a generic CPU, a MMU, main memory, a real-time clock, a timer, a high-precision timer and an Interrupt Controller. We list their corresponding processes' definition associated with their page numbers as follows:

\begin{enumerate}[label={D.1.\arabic*}]
	\item Generic CPU \pageref{circusprocess_GenericCPU}
	\item Memory Management Unit (MMU) \pageref{circusprocess_MMU}
	\item Main Memory \pageref{circusprocess_MainMemory}
	\item Real-Time Clcok \pageref{circusprocess_RTClock}
	\item Timer \pageref{circusprocess_Timer}
	\item High-Precision Timer \pageref{circusprocess_HPTimer}
	\item Interrupt Controller \pageref{circusprocess_InterruptController}
\end{enumerate}
The generic CPU, which can be regarded as a single-core processor, is the centre of this layer because all the external events are reported to it as interrupts or exceptions. The MMU is used for address translation and the main memory is where the data and instructions are stored temporarily. The real-time clock is responsible for providing current time for operating system at its initialization phase. The timer is used to time after the initialization of the system and generate periodic clock ticks to update system time. The high-precision timer is used to time during the time window of a partition and generate aperiodic clock ticks to update the time counters of current partition. The interrupt controller has many duties including recording hardware interrupts, masking certain interrupt and reporting interrupts to CPU according to their priorities. Then the model of this layer, $ Core \_ Hardware \_ Layer $ (see Page \pageref{circusprocess_CoreHardwareLayer}), is constructed by connecting these hardware units together with the channelset in this layer so that they can synchronize with CPU directly or indirectly.

In Hardware Interface Software Layer, the component Circus processes consist of context switching, memory management, system clock, interrupt controlling and interrupt handler. We list their corresponding processes' definition associated with their page numbers as follows: 

\begin{enumerate}[label={D.2.\arabic*}]
	\item Context Switching \pageref{circusprocess_ContextSwitching}
	\item Memory Management \pageref{circusprocess_MemoryManagement}
	\item System Clock \pageref{circusprocess_SystemClock}
	\item Interrupt Controlling \pageref{circusprocess_InterruptControlling}
	\item Interrupt Handler \pageref{circusprocess_InterruptHandler}
\end{enumerate}
Context switching and interrupt controlling are both sets of instructions for specific functions. The former is used to switch contexts and the latter is used to mask and unmask, enable and disable interrupts. By contrast, memory management, system clock and interrupt handler are functional modules in this layer. Memory management is responsible for managing physical memory, creating virtual address space for processes (and loading pages of processes into main memory). System clock is responsible for providing system time to the whole system. Interrupt handler is the main part to deal with interrupts but it must support interrupt nesting in a real-time system so that all kinds of external interrupts can be handled in time. The model of this layer, $ Hardware \_ Interface \_ Software \_ Layer $ (see Page \pageref{circusprocess_HardwareInterfaceSoftwareLayer}), can be defined as context switching and interrupt controlling synchronize on the channelset of layer and interleave with memory management and system clock.

In ARINC 653 O/S Layer, there are two component Circus processes: CoreKernel (see Page \pageref{circusprocess_CoreKernel})and PartitionKernel (see Page \pageref{circusprocess_PartitionKernel}). The core kernel manages the kernel processes and all the partitions. The partitions are scheduled by core kernel based on round robin. Partition kernel manages the processes in its own partition and schedule them based on their priorities. The model of this layer, $ ARINC653 \_ OS \_ Layer $ (see Page \pageref{circusprocess_ARINC653OSLayer}), consists of a CoreKernel process and several PartitionKernel processes. Several partition kernels interleave with each other but they all synchronize with core kernel via the channelset in this layer.

In APEX Layer, the component Circus process is APEX Interface (see Page \pageref{circusprocess_APEXInterface}). It includes many services which we list below. The model of APEX Layer, $ APEX \_ Layer $ (see Page \pageref{circusprocess_APEXLayer}), can be seen as several APEX interface processes interleave with each other because the APEX interface is used by all the processes of all partitions of this ARINC 653 module.

\begin{enumerate}
	\item Partition Management Service
		  \begin{enumerate}
			  \item $ Get \_ Partition \_ Status $ \pageref{service_GetPartitionStatus}
			  \item $ Set \_ Partition \_ Mode $ \pageref{service_SetPartitionMode}
		  \end{enumerate}
	\item Process Management Sservice
		  \begin{enumerate}
		  	  \item $ Get \_ Process \_ Id $ \pageref{service_GetProcessId}
		  	  \item $ Get \_ Process \_ Status $ \pageref{service_GetProcessStatus}
		  	  \item $ Create \_ Process $ \pageref{service_CreateProcess}
		  	  \item $ Set \_ Priority $ \pageref{service_SetPriority}
		  	  \item $ Suspend \_ Self $ \pageref{service_SuspendSelf}
		  	  \item $ Suspend $ \pageref{service_Suspend}
		  	  \item $ Resume $ \pageref{service_Resume}
		  	  \item $ Stop \_ Self $ \pageref{service_StopSelf}
		  	  \item $ Stop $ \pageref{service_Stop}
		  	  \item $ Start $ \pageref{service_Start}
		  	  \item $ Delayed \_ Start $ \pageref{service_DelayedStart}
		  	  \item $ Lock \_ Preemption $ \pageref{service_LockPreemption}
		  	  \item $ Unlock \_ Preemption $ \pageref{service_UnlockPreemption}
		  	  \item $ Get \_ My \_ Id $ \pageref{service_GetMyId}
		  \end{enumerate}
	\item Time Management Service
		  \begin{enumerate}
		  	  \item $ Timed \_ Wait $ \pageref{service_TimedWait}
		  	  \item $ Periodic \_ Wait $ \pageref{service_PeriodicWait}
		  	  \item $ Get \_ Time $ \pageref{service_GetTime}
		  	  \item $ Replenish $ \pageref{service_Replenish}
		  \end{enumerate}
\end{enumerate}

\section{Conclusion and Future Work}
\hspace{1.1em}
In this paper, we present a formal model of a real-time operating system based on ARINC 653 standard. By defining the workflow of the core module of ARINC 653, our model specifies the mechanism how the kernel operates, including interrupt response, partitions scheduling, processes scheduling and system calls to support the related APEX functions.

Contrary to other formal models of operating system based on ARINC 653, the concurrent behaviours of partitions and processes are also specified in this model. In ARINC 653 O/S Layer of the model, several partition kernels interleave with each other. They run concurrently according to its time window in major time frame, it is the concurrent execution with a fixed order. In APEX Layer of the model, all the APEX interfaces, which is used by all the processes of all the partitions in this ARINC 653 Module, interleave. Since each APEX interface is specified by a unique process, the interleaving execution of APEX interfaces is the interleaving execution of processes. And the concurrent execution is not in a fixed order because the scheduling of process is based on its priority.

The following are some works we would do in the future.

Our work includes Partition Management, Process Management and Time Management of the document. A number of services, which are related to Interpartition/Intrapartition Communication as well as Health Monitor, are not included in this paper. We will add these services and their associated system calls in the future.

The second work is to prove the correctness of this model. There are two kinds of properties we should prove, basic ones of general operating systems and special ones of ARINC 653 standard. For the former, many essential properties are taken into consideration, for example, we can easily prove that `The kernel stack of different processes will not overlap both physically and virtually' in this model by contradiction. For the latter, our concentration will be focused on the rules of process scheduling as stated in the document.

As we know, in this paper, the hardware model we present is rather simple and generic. A lot of work is required when we try to build a model of a more complex system, where there will be a correspondingly complex hardware architecture. This area covers the hardware units, like DMA chips, disks and various sensors, as well as the processor architecture like Symmetrical Multi-Processing (SMP) and Asymmetric Multi-processing (AMP), Unified Memory Access (UMA) and Non-Unified Memory Access (NUMA) and so on. The software model can also be specified with more details, such as file system and exception handler, they both need support from corresponding hardware units.

 


\bibliographystyle{plain}
\bibliography{Reference}

\newpage

\appendix

\section{Types and Constants} \label{section:types&constant}
\subsection{Core Hardware Layer} \label{section:types&constant_CoreHardware}
\subsubsection{Basic Definition}
\paragraph{1.The Bit Length of the Hardware and the System, namely N, which should be 64 according to ARINC 653}
\begin{axdef}
	N : \nat_1
\end{axdef}

\paragraph{2.The type of the value stored in a Bit, which can be zero or one}
\begin{zed}
	\begin{array}{lll}
		BIT & ::= & 0 | 1
	\end{array}
\end{zed}

\paragraph{3.The type of the content stored in Primary Storage Unit (PSU), which is the set of N-Bit binary numbers}
\begin{zed}
	\begin{array}{lll}
		[PSU]
	\end{array}
\end{zed}

\paragraph{4.The value zero and one of PSU}
\begin{axdef}
	0_{PSU} : PSU \\
	1_{PSU} : PSU
\end{axdef}

\paragraph{5.The limit value of PSU, which appears to be 1 in each digit of the N-Bit binary number}
\begin{axdef}
	LIMIT_{PSU} : PSU
\end{axdef}

\paragraph{6.The type of System Address}
\begin{zed}
	\begin{array}{lll}
		SYSTEM \_ ADDRESS \_ TYPE & == & PSU
	\end{array}
\end{zed}

\paragraph{7.The type of the content stored in Memory Storage Unit (MSU)}
\begin{zed}
	\begin{array}{lll}
		MSU & == & PSU
	\end{array}
\end{zed}

\paragraph{8.The definition of function \textit{bintounsigned} and \textit{unsignedtobin}}
\begin{axdef}
	\mathit{bintounsigned} : PSU \fun \nat \\
	\mathit{unsignedtobin} : \nat \fun PSU
\end{axdef}

\paragraph{9.The definition of function \textit{bintosigned} and \textit{signedtobin}}
\begin{axdef}
	\mathit{bintosigned} : PSU \fun \num \\
	\mathit{signedtobin} : \num \fun PSU
\end{axdef}

\paragraph{10.The amount of interrupt kinds(which should be greater than or equal to 1 in this text)}
\begin{axdef}
	amount_i : \nat_1
\end{axdef}

\paragraph{11.The type of interrupt}
\begin{zed}
	\begin{array}{lll}
		INTERRUPT \_ TYPE & == & 0 \upto (amount_i - 1)
	\end{array}
\end{zed}

\paragraph{12.The Clock Tick}
\begin{axdef}
	CLOCK \_ TICK : INTERRUPT \_ TYPE
	\where
	CLOCK \_ TICK =  min(INTERRUPT \_ TYPE)
\end{axdef}

\paragraph{13.The amount of exception kinds(which should be greater than or equal to 2 in this text)}
\begin{zed}
	\begin{array}{lll}
		amount_e : \nat_1
	\end{array}
\end{zed}

\paragraph{14.The type of exception}
\begin{zed}
	\begin{array}{lll}
		EXCEPTION \_ TYPE & == & 0 \upto (amount_e - 1)
	\end{array}
\end{zed}

\paragraph{15.The Page Fault and Access Violation}
\begin{axdef}
	PAGE \_ FAULT : EXCEPTION \_ TYPE \\
	ACCESS \_ VIOLATION : EXCEPTION \_ TYPE
\end{axdef}

\subsubsection{Generic CPU}
\paragraph{1.The amount of General-Purpose Registers}
\begin{axdef}
	amount_{gpr} : \nat_1
\end{axdef}

\paragraph{2.The number of General-Purpose Registers}
\begin{zed}
	\begin{array}{lll}
		GPNumber & == & 0 \upto (amount_{gpr} - 1)
	\end{array}
\end{zed}

\paragraph{3.The type of General-Purpose Registers}
\begin{zed}
	\begin{array}{lll}
		GPREGISTER \_ TYPE & == & GPNumber \fun PSU
	\end{array}
\end{zed}

\paragraph{4.The type of Registers for Stack Segment}
\begin{schema*}{SSREGISTER \_ TYPE}
	SS : PSU \\
	SL : PSU \\
	BP : PSU \\
	SP : PSU
\end{schema*}

\paragraph{5.The type of Registers for Data Segment}
\begin{schema*}{DSREGISTER \_ TYPE}
	DS : PSU \\
	DL : PSU \\
	SI : PSU \\
	DI : PSU
\end{schema*}

\paragraph{6.The type of Registers for Code Segment}
\begin{schema*}{CSREGISTER \_ TYPE}
	CS : PSU \\
	CL : PSU \\
	IP : PSU 
\end{schema*}

\paragraph{7.The type of Registers for Program Status Word}
\begin{schema*}{PSWREGISTER \_ TYPE}
	CF : BIT \\
	PF : BIT \\
	AF : BIT \\
	ZF : BIT \\
	SF : BIT \\
	OF : BIT \\
	TF : BIT \\
	DF : BIT \\
	IF : BIT
\end{schema*}

\paragraph{8.The type of CPU Register}
\begin{schema*}{CPU \_ REGISTER \_ TYPE}
	GPRegister : GPREGISTER \_ TYPE \\
	SSRegister : SSREGISTER \_ TYPE \\ 
	DSRegister : DSREGISTER \_ TYPE \\
	CSRegister : CSREGISTER \_ TYPE \\
	PSWRegister : PSWREGISTER \_ TYPE
\end{schema*}

\subsubsection{Memory Management Unit (MMU)}
\paragraph{1.The Bit Length used for page, namely PAGE, and the Bit Length used for offset, namely OFFSET}
\begin{axdef}
	PAGE : \nat_1 \\
	OFFSET : \nat_1
	\where
	PAGE + OFFSET = N
\end{axdef}

\paragraph{2.The constant PAGE \_ SIZE}
\begin{axdef}
	PAGE \_ SIZE : \nat_1
	\where
	PAGE \_ SIZE = 2^{OFFSET}
\end{axdef}

\paragraph{3.The definition of function \textit{getpage} and \textit{getoffset}}
\begin{axdef}
	\mathit{getpage} : SYSTEM \_ ADDRESS \_ TYPE \fun SYSTEM \_ ADDRESS \_ TYPE \\
	\mathit{getoffset} : SYSTEM \_ ADDRESS \_ TYPE \fun SYSTEM \_ ADDRESS \_ TYPE
	\where
	\forall a : SYSTEM \_ ADDRESS \_ TYPE \spot \\
	\t1 \: \mathit{getpage}(a) = a - \mathit{getoffset}(a) \\
	\t1 \: \mathit{getoffset}(a) = a \mod \mathit{unsignedtobin}(PAGE \_ SIZE)
\end{axdef}

\paragraph{4.The type of Page}
\begin{zed}
	\begin{array}{lll}
		PAGE \_ TYPE & == & \{ a : SYSTEM \_ ADDRESS \_ TYPE | a \mod \mathit{unsignedtobin}(PAGE \_ SIZE) = 0_{PSU} \}
	\end{array}
\end{zed}

\paragraph{5.The definition of function \textit{pagespace}}
\begin{axdef}
	\mathit{pagespace} : PAGE \_ TYPE \fun \power SYSTEM \_ ADDRESS \_ TYPE
	\where
	\forall p : PAGE \_ TYPE \spot \\
	\t1 \: \mathit{pagespace}(p) = \{ a : SYSTEM \_ ADDRESS \_ TYPE | a \geq p \land a < p + \mathit{unsignedtobin}(PAGE \_ SIZE) \}
\end{axdef}

\paragraph{6.The definition of function \textit{nextpage}}
\begin{axdef}
	\mathit{nextpage} : SYSTEM \_ ADDRESS \_ TYPE \fun PAGE \_ TYPE
	\where
	\forall a : SYSTEM \_ ADDRESS \_ TYPE \spot \\
	\t1 \: \mathit{nextpage}(a) = \mathit{getpage}(a) + \mathit{unsignedtobin}(PAGE \_ SIZE)
\end{axdef}

\paragraph{7.The type of Page Table}
\begin{zed}
	\begin{array}{lll}
		PAGE \_ TABLE \_ TYPE & == & PAGE \_ TYPE \pinj PAGE \_ TYPE
	\end{array}
\end{zed}

\paragraph{8.The type of Piece which is the content of a Page}
\begin{zed}
	\begin{array}{lll}
		PIECE \_ TYPE & == & \{ \emptyset \} \cup \{ p : SYSTEM \_ ADDRESS \_ TYPE \ffun PSU | p \neq \emptyset \\
						   && \t1 \ \ \ \ \land min(\dom p) \in PAGE \_ TYPE \land \# p \leq PAGE \_ SIZE \}
	\end{array}
\end{zed}

\paragraph{9.The type of Memory Management Unit}
\begin{schema*}{MEMORY \_ MANAGEMENT \_ UNIT \_ TYPE}
	PageTableBuffer : PAGE \_ TABLE \_ TYPE
\end{schema*}

\subsubsection{Main Memory}
\paragraph{1.The type of Main Memory}
\begin{schema*}{MAIN \_ MEMORY \_ TYPE}
	Store : SYSTEM \_ ADDRESS \_ TYPE \pfun MSU
\end{schema*}

\subsubsection{Real-Time Clock}
\paragraph{1.The type of RTClock}
\begin{schema*}{RTCLOCK \_ TYPE}
	Value : PSU
\end{schema*}

\subsubsection{Timer}
\paragraph{1.The definition of Clock Tick Interval}
\begin{axdef}
	ClockTickInterval : PSU
\end{axdef}

\paragraph{2.The type of Timer}
\begin{schema*}{TIMER \_ TYPE}
	Value : PSU \\
	Period : PSU
\end{schema*}

\subsubsection{High-Precision Timer}
\paragraph{1.The type of HPTimer}
\begin{schema*}{HPTIMER \_ TYPE}
	Value : PSU \\
	Alarm : PSU
\end{schema*}

\subsubsection{Interrupt Controller}
\paragraph{1.The type of Interrupt Controller Registers}
\begin{zed}
	\begin{array}{lll}
		ICREGISTER \_ TYPE & == & INTERRUPT \_ TYPE \fun BIT
	\end{array}
\end{zed}

\paragraph{2.The type of Interrupt Controller}
\begin{schema*}{INTERRUPT \_ CONTROLLER \_ TYPE}
	IRRegister : ICREGISTER \_ TYPE \\
	IMRegister : ICREGISTER \_ TYPE \\
	ISRegister : ICREGISTER \_ TYPE
\end{schema*}

\subsection{Hardware Interface Software Layer} \label{section:types&constant_HardwareInterfaceSoftware}
\subsubsection{Basic Definition}
\paragraph{1.The definition of sequence of a generic type}
\begin{zed}
	\begin{array}{lll}
		seq[\textit{X}] & == & \{ s : \nat \ffun \textit{X} | \exists n : \nat \spot \dom s = 1 \upto n \}
	\end{array}
\end{zed}

\paragraph{2.The definition of non-empty sequence of a generic type}
\begin{zed}
	\begin{array}{lll}
		seq_1[\textit{X}] & == & \{ s : seq[\textit{X}] | s \neq \langle \rangle \}
	\end{array}
\end{zed}

\paragraph{3.The definition of function \textit{slide}}
\begin{axdef}
	\mathit{slide} : seq_1[\textit{X}] \cross \nat \fun (\nat_1 \ffun \textit{X})
	\where
	\forall s : seq_1[\textit{X}];n : \nat \spot \\
	\t1 \mathit{slide}(s,n) = \{ i : \nat | i \in \dom s \spot (i + n) \mapsto s(i) \}
\end{axdef}

\paragraph{4.The type of Unsigned Integer}
\begin{axdef}
	UNSIGNED \_ INTEGER \_ TYPE : \power \nat
	\where
	\forall n : \nat \spot  \\
	\t1 \: n \in UNSIGNED \_ INTEGER \_ TYPE \iff n \geq \mathit{bintounsigned}(0_{PSU}) \land n \leq \mathit{bintounsigned}(LIMIT_{PSU})
\end{axdef}

\paragraph{5.The type of Signed Integer}
\begin{axdef}
	SIGNED \_ INTEGER \_ TYPE : \power \num
	\where
	\forall z : \num \spot \\
	\t1 \: z \in SIGNED \_ INTEGER \_ TYPE \iff z \geq \mathit{bintosigned}(LIMIT_{PSU}) - 1 \land z \leq \negate \mathit{bintosigned}(LIMIT_{PSU})
\end{axdef}

\paragraph{6.The type of Option}
\begin{zed}
	\begin{array}{lll}
		OPTION \_ TYPE & ::= & PART | ALL
	\end{array}
\end{zed}

\paragraph{7.The type of Segment}
\begin{zed}
	\begin{array}{lll}
		SEGMENT \_ TYPE & == & \{ \emptyset \} \cup \{ seg : SYSTEM \_ ADDRESS \_ TYPE \ffun PSU | \exists end : \\
							 && \t1 \ \ \ \ SYSTEM \_ ADDRESS \_ TYPE \spot \dom seg = 0_{PSU} \upto end \}
	\end{array}
\end{zed}

\paragraph{8.The type of Executable File}
\begin{schema*}{EXECUTABLE \_ FILE \_ TYPE}
	Data : SEGMENT \_ TYPE \\
	Code : SEGMENT \_ TYPE
\end{schema*}

\paragraph{9.The definition of function \textit{pagecount}}
\begin{axdef}
	\mathit{pagecount} : UNSIGNED \_ INTEGER \_ TYPE \fun UNSIGNED \_ INTEGER \_ TYPE
	\where
	\forall n : UNSIGNED \_ INTEGER \_ TYPE \spot \\
	\t1 \: \mathit{pagecount}(n) = (n + PAGE \_ SIZE - 1) \div PAGE \_ SIZE
\end{axdef}

\paragraph{10.The definition of function \textit{sizecount}}
\begin{axdef}
	\mathit{sizecount} : UNSIGNED \_ INTEGER \_ TYPE \fun UNSIGNED \_ INTEGER \_ TYPE
	\where
	\forall n : UNSIGNED \_ INTEGER \_ TYPE \spot \\
	\t1 \: \mathit{sizecount}(n) = \mathit{pagecount}(n) * PAGE \_ SIZE
\end{axdef}

\paragraph{11.The definition of function \textit{paging}}
\begin{axdef}
	\mathit{paging} : SEGMENT \_ TYPE \fun seq[PIECE \_ TYPE]
	\where
	\forall seg : SEGMENT \_ TYPE \spot \\
	\t1 \ \ \; \mathit{paging}(seg) = \{i : \nat_1 | i \in 1 \upto \mathit{pagecount}(\# seg) \spot i \mapsto \{ a : SYSTEM \_ ADDRESS \_ TYPE | \\
	\t4 \ \ \ \ \ \: a \in \dom seg \land a \geq \mathit{unsignedtobin}((i - 1) * PAGE \_ SIZE) \land a < \mathit{unsignedtobin} \\
	\t4 \ \ \ \ \ \: (i * PAGE \_ SIZE) \} \dres seg \}
\end{axdef}

\paragraph{12.The definition of function \textit{randompage}}
\begin{axdef}
	\mathit{randompage} : \power PAGE \_ TYPE \fun PAGE \_ TYPE
	\where
	\mathit{randompage} = (\lambda ps : \power PAGE \_ TYPE | ps \neq \emptyset \spot (p : PAGE \_ TYPE | p \in ps))
\end{axdef}

\paragraph{13.The type of Kernel Identifier}
\begin{zed}
	\begin{array}{lll}
		KERNEL \_ ID \_ TYPE & == & UNSIGNED \_ INTEGER \_ TYPE
	\end{array}
\end{zed}

\paragraph{14.The constant Core Kernel}
\begin{axdef}
	CORE \_ KERNEL : KERNEL \_ ID \_ TYPE
	\where
	CORE \_ KERNEL = 0
\end{axdef}

\subsubsection{Context Switching}
\paragraph{1.The type of Context}
\begin{zed}
	\begin{array}{lll}
		CONTEXT \_ TYPE & == & CPU \_ REGISTER \_ TYPE \cross PAGE \_ TABLE \_ TYPE
	\end{array}
\end{zed}

\paragraph{2.The definition of function \textit{makecontext}}
\begin{axdef}
	\mathit{makecontext} : CPU \_ REGISTER \_ TYPE \cross PAGE \_ TABLE \_ TYPE \fun CONTEXT \_ TYPE
	\where
	\forall cpureg : CPU \_ REGISTER \_ TYPE;pt : PAGE \_ TABLE \_ TYPE \spot \\
	\t2 \ ~ \mathit{makecontext}(cpureg,pt) = (cpureg,pt)
\end{axdef}

\subsubsection{Memory Management}
\paragraph{1.The type of Memory Block}
\begin{schema*}{MEMORY \_ BLOCK \_ TYPE}
	Start : PAGE \_ TYPE \\
	Size : UNSIGNED \_ INTEGER \_ TYPE
	\where
	Size \mod PAGE \_ SIZE = 0
\end{schema*}

\paragraph{2.The constant NULL \_ BLOCK which denotes a Memory Block with no size}
\begin{axdef}
	NULL \_ BLOCK : MEMORY \_ BLOCK \_ TYPE
	\where
	NULL \_ BLOCK.Size = 0
\end{axdef}

\paragraph{3.The definition of function \textit{blockspace}}
\begin{axdef}
	\mathit{blockspace} : MEMORY \_ BLOCK \_ TYPE \fun \power SYSTEM \_ ADDRESS \_ TYPE
	\where
	\forall mb : MEMORY \_ BLOCK \_ TYPE \spot \\
	\t1 \ \ \ \mathit{blockspace}(mb) = \{ a : SYSTEM \_ ADDRESS \_ TYPE | a \geq mb.Start \\
	\t5 \ \ \ \ \: \land a < mb.Start + \mathit{unsignedtobin}(mb.Size) \}
\end{axdef}

\paragraph{4.The definition of function \textit{totalpage}}
\begin{axdef}
	\mathit{totalpage} : MEMORY \_ BLOCK \_ TYPE \fun \power PAGE \_ TYPE
	\where
	\forall mb : MEMORY \_ BLOCK \_ TYPE \spot \\
	\t1 \ \ \ \mathit{totalpage}(mb) = \{ p : PAGE \_ TYPE | p \in \mathit{blockspace}(mb) \}
\end{axdef}

\paragraph{5.The type of Physical Memory Block}
\begin{zed}
	\begin{array}{lll}
		PHYSICAL \_ MEMORY \_ BLOCK \_ TYPE & == & MEMORY \_ BLOCK \_ TYPE
	\end{array}
\end{zed}

\paragraph{6.The type of Virtual Memory Block}
\begin{schema*}{VIRTUAL \_ MEMORY \_ BLOCK \_ TYPE}
	Memory : MEMORY \_ BLOCK \_ TYPE \\
	Base : \power PAGE \_ TYPE \\
	PageTable : PAGE \_ TABLE \_ TYPE
	\where
	\dom PageTable \subseteq \mathit{totalpage}(Memory) \\
	\ran PageTable \subseteq Base
\end{schema*}

\paragraph{7.The constant NULL which is a reserved System Address in virtual address space for denoting empty}
\begin{axdef}
	NULL : SYSTEM \_ ADDRESS \_ TYPE
	\where
	NULL = 0_{PSU}
\end{axdef}

\paragraph{8.The constant PROCESS \_ VIRTUAL \_ ADDRESS \_ SPACE and KERNEL \_ VIRTUAL \_ ADDRESS \_ SPACE}
\begin{axdef}
	PROCESS \_ VIRTUAL \_ ADDRESS \_ SPACE : MEMORY \_ BLOCK \_ TYPE \\
	KERNEL \_ VIRTUAL \_ ADDRESS \_ SPACE : MEMORY \_ BLOCK \_ TYPE
	\where
	NULL \notin \mathit{blockspace}(PROCESS \_ VIRTUAL \_ ADDRESS \_ SPACE) \\
	NULL \notin \mathit{blockspace}(KERNEL \_ VIRTUAL \_ ADDRESS \_ SPACE) \\
	\mathit{blockspace}(PROCESS \_ VIRTUAL \_ ADDRESS \_ SPACE) \cap \\
	\mathit{blockspace}(KERNEL \_ VIRTUAL \_ ADDRESS \_ SPACE) = \emptyset
\end{axdef}

\paragraph{9.The constant KERNEL \_ STACK \_ SIZE which is the size of kernel stack of a process}
\begin{axdef}
	KERNEL \_ STACK \_ SIZE : UNSIGNED \_ INTEGER \_ TYPE
	\where
	KERNEL \_ STACK \_ SIZE \neq 0 \\
	KERNEL \_ STACK \_ SIZE \mod PAGE \_ SIZE = 0
\end{axdef}

\paragraph{10.The type of Process Virtual Memory}
\begin{schema*}{PROCESS \_ VIRTUAL \_ MEMORY \_ TYPE}
	VMBlock : VIRTUAL \_ MEMORY \_ BLOCK \_ TYPE \\
	Stack : MEMORY \_ BLOCK \_ TYPE \\
	Data : MEMORY \_ BLOCK \_ TYPE \\
	Code : MEMORY \_ BLOCK \_ TYPE
	\where
	\mathit{blockspace}(VMBlock.Memory) \subseteq \mathit{blockspace}(PROCESS \_ VIRTUAL \_ ADDRESS \_ SPACE) \\
	\mathit{blockspace}(VMBlock.Memory) = \mathit{blockspace}(Stack) \cup \mathit{blockspace}(Data) \cup \mathit{blockspace}(Code) \\
	\mathit{blockspace}(Stack) \cap \mathit{blockspace}(Data) = \emptyset \\
	\mathit{blockspace}(Stack) \cap \mathit{blockspace}(Code) = \emptyset \\
	\mathit{blockspace}(Data) \cap \mathit{blockspace}(Code) = \emptyset
\end{schema*}

\paragraph{11.The type of Area Memory Management}
\begin{schema*}{AREA \_ MEMORY \_ MANAGEMENT \_ TYPE}
	Memory : PHYSICAL \_ MEMORY \_ BLOCK \_ TYPE \\
	Allocated : \power PAGE \_ TYPE \\
	Free : \power PAGE \_ TYPE
	\where
	Allocated \cap Free = \emptyset \\
	Allocated \cup Free = \mathit{totalpage}(Memory)
\end{schema*}

\paragraph{12.The type of Memory Management State}
\begin{schema*}{MEMORY \_ MANAGEMENT \_ STATE \_ TYPE}
	AreaMemoryManagementRef : AREA \_ MEMORY \_ MANAGEMENT \_ TYPE
\end{schema*}

\subsubsection{System Clock}
\paragraph{1.The type of System Time}
\begin{zed}
	\begin{array}{lll}
		SYSTEM \_ TIME \_ TYPE & == & SIGNED \_ INTEGER \_ TYPE
	\end{array}
\end{zed}

\paragraph{2.The constant DEFAULT \_ TIME which is a default value for uncertain time}
\begin{axdef}
	DEFAULT \_ TIME : SYSTEM \_ TIME \_ TYPE
	\where
	DEFAULT \_ TIME = \negate 1
\end{axdef}

\paragraph{3.The type of System Clock State}
\begin{schema*}{SYSTEM \_ CLOCK \_ STATE \_ TYPE}
	BaseTime : SYSTEM \_ TIME \_ TYPE \\
	TickInterval : SYSTEM \_ TIME \_ TYPE \\
	TickCounter : UNSIGNED \_ INTEGER \_ TYPE \\
	Time : SYSTEM \_ TIME \_ TYPE \\
	InterruptTime : SYSTEM \_ TIME \_ TYPE
\end{schema*}

\subsubsection{Interrupt Handler}
\paragraph{1.The type of Interrupt Handler State}
\begin{schema*}{INTERRUPT \_ HANDLER \_ STATE \_ TYPE}
	KernelPageTableRef : PAGE \_ TABLE \_ TYPE \\
	InterruptStack : MEMORY \_ BLOCK \_ TYPE \\
	TempContext : seq[CONTEXT \_ TYPE]
\end{schema*}

\paragraph{2.The type of Interrupt Handler Configuration}
\begin{schema*}{INTERRUPT \_ HANDLER \_ CONFIG \_ TYPE}
	InterruptStackSize : UNSIGNED \_ INTEGER \_ TYPE
	\where
	InterruptStackSize \neq 0 \\
	InterruptStackSize \mod PAGE \_ SIZE = 0
\end{schema*}

\subsection{ARINC 653 O/S Layer(Core Kernel)} \label{section:types&constant_CoreOS}
\subsubsection{Basic Definition}
\paragraph{1.The definition of non-empty injective sequence of a generic type}
\begin{zed}
	\begin{array}{lll}
		iseq_1[\textit{X}] & == & \{ s : seq[\textit{X}] | s \neq \langle \rangle \land s \in \nat \inj \textit{X} \}
	\end{array}
\end{zed}

\paragraph{2.The type of Boolean}
\begin{zed}
	\begin{array}{lll}
		BOOLEAN & ::= & TRUE | FALSE
	\end{array}
\end{zed}

\paragraph{3.The type of String}
\begin{zed}
	\begin{array}{lll}
		[STRING]
	\end{array}
\end{zed}

\paragraph{4.The constant PARTITION \_ NUMBER \_ LIMIT which is the Limit Number of Partitions of the System}
\begin{axdef}
	PARTITION \_ NUMBER \_ LIMIT : UNSIGNED \_ INTEGER \_ TYPE
\end{axdef}

\subsubsection{Core Kernel}
\paragraph{1.The type of Kernel Process Management}
\begin{schema*}{KERNEL \_ PROCESS \_ MANAGEMENT \_ TYPE}
	ProcessTable : PROCESS \_ TABLE \_ TYPE \\
	CurrentProcess : PROCESS \_ ID \_ TYPE
	\where
	(\forall i,j : PROCESS \_ ID \_ TYPE | i,j \in \dom ProcessTable \spot \\
	\ i \neq j \implies \mathit{blockspace}(ProcessTable(i).KernelStack) \cap \mathit{blockspace}(ProcessTable(j).KernelStack) = \emptyset) \\
	(\forall i : PROCESS \_ ID \_ TYPE | i \in \dom ProcessTable \spot \\
	\ ProcessTable(i).VirtualMemory.VMBlock.Memory = NULL \_ BLOCK) \\
	(\forall i : PROCESS \_ ID \_ TYPE | i \in \dom ProcessTable \spot \\
	\ ProcessTable(i).VirtualMemory.VMBlock.Base = \emptyset) \\
	(\forall i : PROCESS \_ ID \_ TYPE | i \in \dom ProcessTable \spot \\
	\ ProcessTable(i).UserTempContext = \langle \ \rangle) \\
	CurrentProcess \in \{ NULL \_ PROCESS \_ ID \} \cup \dom ProcessTable \\
	(\forall i,j : PROCESS \_ ID \_ TYPE | i,j \in \dom ProcessTable \spot \\
	\ ProcessTable(i).ProcessState = RUNNING \land ProcessTable(j).ProcessState = RUNNING \implies i = j = \\
	\ CurrentProcess)
\end{schema*}

\paragraph{2.The type of Partition Table}
\begin{zed}
	\begin{array}{lll}
		PARTITION \_ TABLE \_ TYPE & == & PARTITION \_ ID \_ TYPE \pinj PARTITION \_ CONTROL \_ BLOCK \_ TYPE
	\end{array}
\end{zed}

\paragraph{3.The type of Partition Time Window}
\begin{schema*}{PARTITION \_ TIME \_ WINDOW \_ TYPE}
	PartitionId : PARTITION \_ ID \_ TYPE \\
	Periodicity : PARTITION \_ PERIODICITY \_ TYPE \\
	Offset : SYSTEM \_ TIME \_ TYPE \\
	PeriodicProcStart : BOOLEAN
\end{schema*}

\paragraph{4.The type of Partition Time Window Configuration}
\begin{schema*}{PARTITION \_ TIME \_ WINDOW \_ CONFIG \_ TYPE}
	Name : PARTITION \_ NAME \_ TYPE \\
	PeriodicProcStart : BOOLEAN
\end{schema*}

\paragraph{5.The type of MajorTimeFrame}
\begin{schema*}{MAJOR \_ TIME \_ FRAME \_ TYPE}	
	PartitionQueue : iseq_1[PARTITION \_ TIME \_ WINDOW \_ TYPE] \\
	Length : SYSTEM \_ TIME \_ TYPE
\end{schema*}

\paragraph{6.The type of Partition Management}
\begin{schema*}{PARTITION \_ MANAGEMENT \_ TYPE}
	PartitionTable : PARTITION \_ TABLE \_ TYPE \\
	MajorTimeFrame : MAJOR \_ TIME \_ FRAME \_ TYPE \\
	CurrentPartition : PARTITION \_ ID \_ TYPE
	\where
	(\forall i,j : PARTITION \_ ID \_ TYPE | i,j \in \dom PartitionTable \spot \\
	\ i \neq j \implies \mathit{blockspace}(PartitionTable(i).Memory) \cap \mathit{blockspace}(PartitionTable(j).Memory) = \emptyset) \\
	CurrentPartition \in \{ NULL \_ PARTITION \_ ID \} \cup \dom PartitionTable
\end{schema*}

\paragraph{7.The type of Core Kernel State}
\begin{schema*}{CORE \_ KERNEL \_ STATE \_ TYPE}
	MemoryManagement : AREA \_ MEMORY \_ MANAGEMENT \_ TYPE \\
	KernelPageTable : PAGE \_ TABLE \_ TYPE \\
	KernelProcManagement : KERNEL \_ PROCESS \_ MANAGEMENT \_ TYPE \\
	PartitionManagement : PARTITION \_ MANAGEMENT \_ TYPE
	\where
	\dom KernelPageTable \subseteq \mathit{totalpage}(KERNEL \_ VIRTUAL \_ ADDRESS \_ SPACE) \\
	\ran KernelPageTable = \mathit{totalpage}(MemoryManagement.Memory)
\end{schema*}

\subsection{ARINC 653 O/S Layer(Partition Kernel)} \label{section:types&constant_PartitionOS}
\subsubsection{Basic Definition}
\paragraph{1.The definition of injective sequence of a generic type}
\begin{zed}
	\begin{array}{lll}
		iseq[\textit{X}] & == & \{ s : seq[\textit{X}] | s \in \nat \inj \textit{X} \}
	\end{array}
\end{zed}

\paragraph{2.The constant PROCESS \_ NUMBER \_ LIMIT which is the Limit Number of Processes in any Partition of the System}
\begin{axdef}
	PROCESS \_ NUMBER \_ LIMIT : UNSIGNED \_ INTEGER \_ TYPE
\end{axdef}

\paragraph{3.The type of Disk Address}
\begin{zed}
	\begin{array}{lll}
		[DISK \_ ADDRESS]
	\end{array}
\end{zed}

\paragraph{4.The constant NULL \_ DISK \_ ADDRESS which is a Disk Address for denoting empty}
\begin{axdef}
	NULL \_ DISK \_ ADDRESS : DISK \_ ADDRESS
\end{axdef}

\paragraph{5.The type of Path}
\begin{zed}
	\begin{array}{lll}
		PATH \_ TYPE & == & DISK \_ ADDRESS
	\end{array}
\end{zed}

\paragraph{6.The constant MIN \_ STACK \_ SIZE and MAX \_ STACK \_ SIZE}
\begin{axdef}
	MIN \_ STACK \_ SIZE : UNSIGNED \_ INTEGER \_ TYPE \\
	MAX \_ STACK \_ SIZE : UNSIGNED \_ INTEGER \_ TYPE
	\where
	MIN \_ STACK \_ SIZE > 0 \\
	MAX \_ STACK \_ SIZE > MIN \_ STACK \_ SIZE
\end{axdef}

\paragraph{7.The range of Process Stack Size}
\begin{zed}
	\begin{array}{lll}
		STACK \_ SIZE \_ RANGE & == & MIN \_ STACK \_ SIZE \upto MAX \_ STACK \_ SIZE
	\end{array}
\end{zed}

\paragraph{8.The constant MIN \_ PRIORITY \_ VALUE and MAX \_ PRIORITY \_ VALUE}
\begin{axdef}
	MIN \_ PRIORITY \_ VALUE : SIGNED \_ INTEGER \_ TYPE \\
	MAX \_ PRIORITY \_ VALUE : SIGNED \_ INTEGER \_ TYPE
	\where
	MIN \_ PRIORITY \_ VALUE > 0 \\
	MAX \_ PRIORITY \_ VALUE > MIN \_ PRIORITY \_ VALUE
\end{axdef}

\paragraph{9.The range of Process Priority}
\begin{zed}
	\begin{array}{lll}
		PRIORITY \_ RANGE & == & MIN \_ PRIORITY \_ VALUE \upto MAX \_ PRIORITY \_ VALUE
	\end{array}
\end{zed}

\paragraph{10.The constant MIN \_ LOCK \_ LEVEL and MAX \_ LOCK \_ LEVEL}
\begin{axdef}
	MIN \_ LOCK \_ LEVEL : UNSIGNED \_ INTEGER \_ TYPE \\
	MAX \_ LOCK \_ LEVEL : UNSIGNED \_ INTEGER \_ TYPE
	\where
	MIN \_ LOCK \_ LEVEL = 0 \\
	MAX \_ LOCK \_ LEVEL > MIN \_ LOCK \_ LEVEL
\end{axdef}

\paragraph{11.The range of Partition Lock Level}
\begin{zed}
	\begin{array}{lll}
		LOCK \_ LEVEL \_ RANGE & == & MIN \_ LOCK \_ LEVEL \upto MAX \_ LOCK \_ LEVEL
	\end{array}
\end{zed}

\paragraph{12.The type of System Call}
\begin{zed}
	\begin{array}{lll}
		SYSTEM \_ CALL \_ TYPE & ::= & GET \_ OPERATING \_ MODE | GET \_ PARTITION \_STATUS \\
									 && | SET \_ OPERATING \_ MODE \\
									 && | GET \_ PROCESS \_ NAMES | GET \_ PROCESS \_ IDS \\
									 && | GET \_ FREE \_ PROCESS \_ IDS | GET \_ FREE \_ SPACE \\
									 && | GET \_ PARTITION \_ PERIOD | GET \_ PROCESS \_ STATE \\
									 && | GET \_ NEXT \_ PERIODIC \_ START | GET \_ DELAYED \_ PERIODIC \_ START \\
									 && | GET \_ PROCESS \_ KIND | GET \_ PROCESS \_ PERIOD\\
									 && | GET \_ PROCESS \_ TIME \_ CAPACITY | GET \_ PARTITION \_ LOCK \_ LEVEL \\
									 && | GET \_ PROCESS \_ ID | GET \_ PROCESS \_ STATUS | CREATE \_ PROCESS \\
									 && | SET \_ PRIORITY | SUSPEND \_ SELF | SUSPEND | RESUME | STOP \_ SELF \\
									 && | STOP | START | DELAYED \_ START | LOCK \_ PREEMPTION \\
									 && | UNLOCK \_ PREEMPTION | GET \_ MY \_ ID \\
									 && | GET \_ CURRENT \_ TIME | GET \_ NEXT \_ RELEASE \_ POINT | TIMED \_ WAIT \\
									 && | PERIODIC \_ WAIT | GET \_ TIME | REPLENISH
	\end{array}
\end{zed}

\paragraph{13.The System Call(Partition Management),System Call(Process Management) and System Call(Time Management)}
\begin{axdef}
	SystemCall \_ Partition : \power SYSTEM \_ CALL \_ TYPE \\
	SystemCall \_ Process : \power SYSTEM \_ CALL \_ TYPE \\
	SystemCall \_ Time : \power SYSTEM \_ CALL \_ TYPE
	\where
	SystemCall \_ Partition = \{ GET \_ OPERATING \_ MODE,GET \_ PARTITION \_STATUS,SET \_ OPERATING \_ \\
	\t5 \ \ \ \; MODE\} \\
	SystemCall \_ Process = \{ GET \_ PROCESS \_ NAMES,GET \_ PROCESS \_ IDS,GET \_ FREE \_ PROCESS \_ IDS, \\
	\t5 \ ~ GET \_ FREE \_ SPACE,GET \_ PARTITION \_ PERIOD,GET \_ PROCESS \_ STATE, \\
	\t5 \ ~ GET \_ NEXT \_ PERIODIC \_ START,GET \_ DELAYED \_ PERIODIC \_ START,GET \_ \\
	\t5 \ ~ PROCESS \_ KIND,GET \_ PROCESS \_ PERIOD,GET \_ PROCESS \_ TIME \_ CAPACITY, \\
	\t5 \ ~ GET \_ PARTITION \_ LOCK \_ LEVEL,GET \_ PROCESS \_ ID,GET \_ PROCESS \_ \\
	\t5 \ ~ STATUS,CREATE \_ PROCESS,SET \_ PRIORITY,SUSPEND \_ SELF,SUSPEND, \\
	\t5 \ ~ RESUME,STOP \_ SELF,STOP,START,DELAYED \_ START,LOCK \_ PREEMPTION, \\
	\t5 \ ~ UNLOCK \_ PREEMPTION,GET \_ MY \_ ID \} \\
	SystemCall \_ Time = \{ GET \_ CURRENT \_ TIME,GET \_ NEXT \_ RELEASE \_ POINT,TIMED \_ WAIT, \\
	\t4 \ \ \ \ \: PERIODIC \_ WAIT,GET \_ TIME,REPLENISH \}
\end{axdef}

\subsubsection{Process}
\paragraph{1.The type of Process Identifier}
\begin{zed}
	\begin{array}{lll}
		PROCESS \_ ID \_ TYPE & == & UNSIGNED \_ INTEGER \_ TYPE
	\end{array}
\end{zed}

\paragraph{2.The type of Process Name}
\begin{zed}
	\begin{array}{lll}
		PROCESS \_ NAME \_ TYPE & == & STRING
	\end{array}
\end{zed}

\paragraph{3.The type of Process Kind}
\begin{zed}
	\begin{array}{lll}
		PROCESS \_ KIND \_ TYPE & ::= & PERIODIC | APERIODIC | ERROR \_ HANDLER
	\end{array}
\end{zed}

\paragraph{4.The type of Process Stack Size}
\begin{zed}
	\begin{array}{lll}
		STACK \_ SIZE \_ TYPE & == & \{ n : UNSIGNED \_ INTEGER \_ TYPE | n > 0 \land n \mod PAGE \_ SIZE = 0 \}
	\end{array}
\end{zed}

\paragraph{5.The type of Process Priority}
\begin{zed}
	\begin{array}{lll}
		PRIORITY \_ TYPE & == & SIGNED \_ INTEGER \_ TYPE
	\end{array}
\end{zed}

\paragraph{6.The constant DEFAULT \_ PRIORITY which is a default value for uncertain priority}
\begin{axdef}
	DEFAULT \_ PRIORITY : PRIORITY \_ TYPE
	\where
	DEFAULT \_ PRIORITY = 0
\end{axdef}

\paragraph{7.The type of Process Deadline}
\begin{zed}
	\begin{array}{lll}
		DEADLINE \_ TYPE & ::= & SOFT | HARD
	\end{array}
\end{zed}

\paragraph{8.The type of Process Attributes}
\begin{schema*}{PROCESS \_ ATTRIBUTE \_ TYPE}
	Name : PROCESS \_ NAME \_ TYPE \\
	ProcessKind : PROCESS \_ KIND \_ TYPE \\
	EntryPoint : SYSTEM \_ ADDRESS \_ TYPE \\	
	StackSize : STACK \_ SIZE \_ TYPE \\	
	BasePriority : PRIORITY \_ TYPE \\
	Period : SYSTEM \_ TIME \_ TYPE \\
	TimeCapacity : SYSTEM \_ TIME \_ TYPE \\
	Deadline : DEADLINE \_ TYPE
\end{schema*}

\paragraph{9.The type of Process State}
\begin{zed}
	\begin{array}{lll}
		PROCESS \_ STATE \_ TYPE & ::= & DORMANT | READY | RUNNING | WAITING
	\end{array}
\end{zed}

\paragraph{10.The type of Process Status}
\begin{schema*}{PROCESS \_ STATUS \_ TYPE}
	Attribute : PROCESS \_ ATTRIBUTE \_ TYPE \\
	CurrentPriority : PRIORITY \_ TYPE \\
	DeadlineTime : SYSTEM \_ TIME \_ TYPE \\
	ProcessState : PROCESS \_ STATE \_ TYPE
\end{schema*}

\paragraph{11.The type of Process Control Block}
\begin{schema*}{PROCESS \_ CONTROL \_ BLOCK \_ TYPE}
	Name : PROCESS \_ NAME \_ TYPE \\
	ProcessKind : PROCESS \_ KIND \_ TYPE \\
	ExeFilePath : PATH \_ TYPE \\
	SwapFilePath : PATH \_ TYPE \\
	EntryPoint : SYSTEM \_ ADDRESS \_ TYPE \\
	KernelStack : MEMORY \_ BLOCK \_ TYPE \\
	VirtualMemory : PROCESS \_ VIRTUAL \_ MEMORY \_ TYPE \\
	Period : SYSTEM \_ TIME \_ TYPE \\
	TimeCapacity : SYSTEM \_ TIME \_ TYPE \\
	Deadline : DEADLINE \_ TYPE \\
	BasePriority : PRIORITY \_ TYPE \\
	CurrentPriority : PRIORITY \_ TYPE \\
	ReleasePoint : SYSTEM \_ TIME \_ TYPE \\
	DeadlineTime : SYSTEM \_ TIME \_ TYPE \\
	ProcessState : PROCESS \_ STATE \_ TYPE \\
	UserTempContext : seq[CONTEXT \_ TYPE] \\
	KernelTempContext : seq[CONTEXT \_ TYPE]
	\where
	\mathit{blockspace}(KernelStack) \subset \mathit{blockspace}(KERNEL \_ VIRTUAL \_ ADDRESS \_ SPACE) \\
	KernelStack.Size = KERNEL \_ STACK \_ SIZE
\end{schema*}

\paragraph{12.The constants relates to NULL Process}
\begin{axdef}
	NULL \_ PROCESS \_ ID : PROCESS \_ ID \_ TYPE \\
	NULL \_ PROCESS \_ STATUS : PROCESS \_ STATUS \_ TYPE
	\where
	NULL \_ PROCESS \_ ID \notin 0 \upto PROCESS \_ NUMBER \_ LIMIT
\end{axdef}

\paragraph{13.The constants relates to Idle Process}
\begin{axdef}
	IDLE \_ PROCESS \_ ID : PROCESS \_ ID \_ TYPE \\
	IDLE \_ PROCESS \_ REGSTATE : CPU \_ REGISTER \_ TYPE \\
	IDLE \_ PROCESS \_ CONTROL \_ BLOCK : PROCESS \_ CONTROL \_ BLOCK \_ TYPE
	\where
	IDLE \_ PROCESS \_ ID = 0 \\
	IDLE \_ PROCESS \_ CONTROL \_ BLOCK.Name = IdleProcess \\
	IDLE \_ PROCESS \_ CONTROL \_ BLOCK.ProcessKind = APERIODIC \\
	IDLE \_ PROCESS \_ CONTROL \_ BLOCK.ExeFilePath = NULL \_ DISK \_ ADDRESS \\
	IDLE \_ PROCESS \_ CONTROL \_ BLOCK.SwapFilePath = NULL \_ DISK \_ ADDRESS \\
	IDLE \_ PROCESS \_ CONTROL \_ BLOCK.VirtualMemory.VMBlock.Memory = NULL \_ BLOCK \\	
	IDLE \_ PROCESS \_ CONTROL \_ BLOCK.VirtualMemory.VMBlock.Base = \emptyset \\
	IDLE \_ PROCESS \_ CONTROL \_ BLOCK.Period = \negate 1 \\
	IDLE \_ PROCESS \_ CONTROL \_ BLOCK.TimeCapacity = \negate 1 \\
	IDLE \_ PROCESS \_ CONTROL \_ BLOCK.CurrentPriority = 0 \\
	IDLE \_ PROCESS \_ CONTROL \_ BLOCK.ProcessState \in \{ RUNNING,READY\} \\
	IDLE \_ PROCESS \_ CONTROL \_ BLOCK.UserTempContext = \langle \ \rangle
\end{axdef}

\subsubsection{Partition Kernel}
\paragraph{1.The type of Partition Identifier}
\begin{zed}
	\begin{array}{lll}
		PARTITION \_ ID \_ TYPE & == & KERNEL \_ ID \_ TYPE \setminus \{ CORE \_ KERNEL \}
	\end{array}
\end{zed}

\paragraph{2.The constant NULL \_ PARTITION \_ ID}
\begin{axdef}
	NULL \_ PARTITION \_ ID : PARTITION \_ ID \_ TYPE
	\where
	NULL \_ PARTITION \_ ID \notin 1 \upto PARTITION \_ NUMBER \_ LIMIT
\end{axdef}

\paragraph{3.The type of Partition Name}
\begin{zed}
	\begin{array}{lll}
		PARTITION \_ NAME \_ TYPE & == & STRING
	\end{array}
\end{zed}

\paragraph{4.The type of Partition Periodicity}
\begin{schema*}{PARTITION \_ PERIODICITY \_ TYPE}
	Period : SYSTEM \_ TIME \_ TYPE \\
	Duration : SYSTEM \_ TIME \_ TYPE
	\where
	Period \geq Duration \\
	Period \mod \mathit{bintosigned}(ClockTickInterval) = 0 \\
	Duration > 0 \\
	Duration \mod \mathit{bintosigned}(ClockTickInterval) = 0
\end{schema*}

\paragraph{5.The type of Partition Lock Level}
\begin{zed}
	\begin{array}{lll}
		LOCK \_ LEVEL \_ TYPE & == & UNSIGNED \_ INTEGER \_ TYPE
	\end{array}
\end{zed}

\paragraph{6.The type of Partition Operating Mode}
\begin{zed}
	\begin{array}{lll}
		OPERATING \_ MODE \_ TYPE & ::= & IDLE | COLD \_ START | WARM \_ START | NORMAL
	\end{array}
\end{zed}

\paragraph{7.The type of Partition Start Condition}
\begin{zed}
	\begin{array}{lll}
		START \_ CONDITION \_ TYPE & ::= & NORMAL \_ START | PARTITION \_ RESTART \\
										 && | HM \_ MODULE \_ RESTART | HM \_ PARTITION \_ RESTART
	\end{array}
\end{zed}

\paragraph{8.The type of Partition Status}
\begin{schema*}{PARTITION \_ STATUS \_ TYPE}
	PartitionId : PARTITION \_ ID \_ TYPE \\
	Period : SYSTEM \_ TIME \_ TYPE \\
	Duration : SYSTEM \_ TIME \_ TYPE \\
	Locklevel : LOCK \_ LEVEL \_ TYPE \\
	OperatingMode : OPERATING \_ MODE \_ TYPE \\
	StartCondition : START \_ CONDITION \_ TYPE
\end{schema*}

\paragraph{9.The type of Partition Control Block}
\begin{schema*}{PARTITION \_ CONTROL \_ BLOCK \_ TYPE}
	Name : PARTITION \_ NAME \_ TYPE \\
	Memory : PHYSICAL \_ MEMORY \_ BLOCK \_ TYPE \\
	Periodicity : PARTITION \_ PERIODICITY \_ TYPE \\
	LockLevel : LOCK \_ LEVEL \_ TYPE \\
	OperatingMode : OPERATING \_ MODE \_ TYPE \\
	StartCondition : START \_ CONDITION \_ TYPE
\end{schema*}

\paragraph{10.The type of Process Table}
\begin{zed}
	\begin{array}{lll}
		PROCESS \_ TABLE \_ TYPE & == & PROCESS \_ ID \_ TYPE \pinj PROCESS \_ CONTROL \_ BLOCK \_ TYPE
	\end{array}
\end{zed}

\paragraph{11.The type of Time Counter}
\begin{schema*}{TIME \_ COUNTER \_ TYPE}
	ProcessId : PROCESS \_ ID \_ TYPE \\
	Alarm : SYSTEM \_ TIME \_ TYPE
\end{schema*}

\paragraph{12.The type of Partition Kernel State}
\begin{schema*}{PARTITION \_ KERNEL \_ STATE \_ TYPE}
	MemoryManagement : AREA \_ MEMORY \_ MANAGEMENT \_ TYPE \\
	KernelPageTableRef : PAGE \_ TABLE \_ TYPE \\
	Periodicity : PARTITION \_ PERIODICITY \_ TYPE \\
	TotalProcess : UNSIGNED \_ INTEGER \_ TYPE \\
	ProcessTable : PROCESS \_ TABLE \_ TYPE \\
	TimeCounterQueue : iseq[TIME \_ COUNTER \_ TYPE] \\
	ReadyQueue : iseq[PROCESS \_ ID \_ TYPE] \\
	WaitingQueue : iseq[PROCESS \_ ID \_ TYPE] \\
	CurrentProcess : PROCESS \_ ID \_ TYPE
	\where
	(\forall i : PROCESS \_ ID \_ TYPE | i \in \dom ProcessTable \spot i \in 0 \upto TotalProcess) \\
	(\forall i,j : PROCESS \_ ID \_ TYPE | i,j \in \dom ProcessTable \spot \\
	\ i \neq j \implies \mathit{blockspace}(ProcessTable(i).KernelStack) \cap \mathit{blockspace}(ProcessTable(j).KernelStack) = \emptyset) \\
	(\forall i,j : PROCESS \_ ID \_ TYPE | i,j \in \dom ProcessTable \spot \\
	\ i \neq j \implies ProcessTable(i).VirtualMemory.VMBlock.Base \cap ProcessTable(j).VirtualMemory.VMBlock.Base = \emptyset) \\
	(\forall tc : TIME \_ COUNTER \_ TYPE | tc \in \ran TimeCounterQueue \spot tc.ProcessId \in \dom ProcessTable) \\
	(\forall i,j : \nat_1 | i,j \in \dom TimeCounterQueue \spot \\
	\ i < j \implies |TimeCounterQueue(i).Alarm| \leq |TimeCounterQueue(j).Alarm|) \\
	(\forall i : PROCESS \_ ID \_ TYPE | i \in \ran ReadyQueue \spot \\
	\ i \in \dom ProcessTable \land ProcessTable(i).ProcessState \in \{ READY,WAITING \}) \\
	(\forall i : PROCESS \_ ID \_ TYPE | i \in \ran WaitingQueue \spot \\
	\ i \in \dom ProcessTable \land ProcessTable(i).ProcessState = WAITING) \\
	CurrentProcess \in \{ NULL \_ PROCESS \_ ID \} \cup \dom ProcessTable \\
	(\forall i,j : PROCESS \_ ID \_ TYPE | i,j \in 0 \upto (TotalProcess - 1) \cap \dom ProcessTable \spot \\
	\ ProcessTable(i).ProcessState = RUNNING \land ProcessTable(j).ProcessState = RUNNING \implies i = j)
\end{schema*}

\paragraph{13.The type of Partition Kernel Configuration}
\begin{schema*}{PARTITION \_ KERNEL \_ CONFIG \_ TYPE}
	Name : PARTITION \_ NAME \_ TYPE \\
	Memory : PHYSICAL \_ MEMORY \_ BLOCK \_ TYPE \\
	Periodicity : PARTITION \_ PERIODICITY \_ TYPE \\
	TotalProcess : UNSIGNED \_ INTEGER \_ TYPE
	\where
	Memory.Size > 0 \\
	TotalProcess + 1 \leq PROCESS \_ NUMBER \_ LIMIT
\end{schema*}

\subsection{ARINC 653 O/S Layer(Operating System Kernel)} \label{section:types&constant_OS}
\subsubsection{Operating System Kernel}
\paragraph{1.The type of OS Kernel Configuration}
\begin{schema*}{OS \_ KERNEL \_ CONFIG \_ TYPE}
	PhysicalMemory : PHYSICAL \_ MEMORY \_ BLOCK \_ TYPE \\
	KernelMemory : PHYSICAL \_ MEMORY \_ BLOCK \_ TYPE \\
	KernelImage : PHYSICAL \_ MEMORY \_ BLOCK \_ TYPE \\
	Partitions : iseq_1[PARTITION \_ KERNEL \_ CONFIG \_ TYPE] \\
	PartitionOperationSequence : iseq_1[PARTITION \_ TIME \_ WINDOW \_ CONFIG \_ TYPE]
	\where
	PhysicalMemory.Start = 0_{PSU} \\
	PhysicalMemory.Size \leq KERNEL \_ VIRTUAL \_ ADDRESS \_ SPACE.Size \\
	KernelMemory.Start = 0_{PSU} \\
	KernelMemory.Size < PhysicalMemory.Size \\
	\mathit{blockspace}(KernelImage) \subset \mathit{blockspace}(KernelMemory) \\
	\# Partitions \leq PARTITION \_ NUMBER \_ LIMIT \\
	\mathit{blockspace}(KernelMemory) \cup \bigcup \limits _{i \in \dom Partitions} \mathit{blockspace}(Partitions(i).Memory) = \mathit{blockspace}(PhysicalMemory) \\
	(\forall i : \nat_1 | i,j \in \dom Partitions \spot \\
	\ \mathit{blockspace}(Partitions(i).Memory) \cap \mathit{blockspace}(KernelMemory) = \emptyset) \\
	(\forall i,j : \nat_1 | i,j \in \dom Partitions \spot \\
	\ i \neq j \implies Partitions(i).Name \neq Partitions(j).Name \\
	\t1 \ \ \ \ \ \: ~\land \mathit{blockspace}(Partitions(i).Memory) \cap \mathit{blockspace}(Partitions(j).Memory) = \emptyset) \\
	(\forall i,j : \nat_1 | i,j \in \dom Partitions \spot \\
	\ Partitions(i).Periodicity.Period \geq Partitions(j).Periodicity.Period \implies \\
	\ Partitions(i).Periodicity.Period \mod Partitions(j).Periodicity.Period = 0)
\end{schema*}

\subsection{Application/Executive(APEX) Layer} \label{section:types&constant_APEX}
\subsubsection{Basic Definition}
\paragraph{1.The type of Return Code}
\begin{zed}
	\begin{array}{lll}
		RETURN \_ CODE \_ TYPE & ::= & NO \_ ERROR | NO \_ ACTION | NOT \_ AVAILABLE | INVALID \_ PARAM \\
									 && | INVALID \_ CONFIG | INVALID \_ MODE | TIMED \_ OUT
	\end{array}
\end{zed}

\subsection{Core Module} \label{section:types&constant_CoreModule}
\subsubsection{Core Module}
\paragraph{1.The type of Module Identifier}
\begin{zed}
	\begin{array}{lll}
		MODULE \_ ID \_ TYPE & == & UNSIGNED \_ INTEGER \_ TYPE
	\end{array}
\end{zed}

\paragraph{2.The type of Module Name}
\begin{zed}
	\begin{array}{lll}
		MODULE \_ NAME \_ TYPE & == & STRING
	\end{array}
\end{zed}

\paragraph{3.The type of Module Base Information}
\begin{schema*}{MODULE \_ BASE \_ TYPE}
	ModuleId : MODULE \_ ID \_ TYPE \\
	ModuleName : MODULE \_ NAME \_ TYPE
\end{schema*}

\paragraph{4.The type of Module Configuration}
\begin{schema*}{MODULE \_ CONFIG \_ TYPE}
	ModuleBase : MODULE \_ BASE \_ TYPE \\
	InterruptHandlerConfig : INTERRUPT \_ HANDLER \_ CONFIG \_ TYPE \\
	OSKernelConfig : OS \_ KERNEL \_ CONFIG \_ TYPE
\end{schema*}

\section{Z Operations} \label{section:zoperation}
\subsection{Core Hardware Layer} \label{section:zoperation_CoreHardware}
\subsubsection{Generic CPU}
\paragraph{1.The initialization of CPU Registers}
\begin{schema*}{InitCPURegister}
	CPU \_ REGISTER \_ TYPE'
	\where
	(\forall i : \nat | i \in \dom GPRegisters' \spot GPRegisters'(i) = 0_{PSU}) \\
	
	(SSRegisters').SS = 0_{PSU} \\
	(SSRegisters').SL = 0_{PSU} \\
	(SSRegisters').BP = 0_{PSU} \\
	(SSRegisters').SP = 0_{PSU} \\
	
	(DSRegisters').DS = 0_{PSU} \\
	(DSRegisters').DL = 0_{PSU} \\
	(DSRegisters').SI = 0_{PSU} \\
	(DSRegisters').DI = 0_{PSU} \\
	
	(CSRegisters').CS = 0_{PSU} \\
	(CSRegisters').CL = 0_{PSU} \\
	(CSRegisters').IP = 0_{PSU} \\
	
	(PSWRegisters').CF = 0 \\
	(PSWRegisters').PF = 0 \\
	(PSWRegisters').AF = 0 \\
	(PSWRegisters').ZF = 0 \\
	(PSWRegisters').SF = 0 \\
	(PSWRegisters').OF = 0 \\
	
	(PSWRegisters').TF = 0 \\
	(PSWRegisters').DF = 0 \\
	
	(PSWRegisters').IF = 0	
\end{schema*}

\paragraph{2.The saving of part CPU Registers}
\begin{schema*}{SavePartCPURegister}
	cpureg! : CPU \_ REGISTER \_ TYPE
	\where
	(\exists reg : CPU \_ REGISTER \_ TYPE \spot \\
	\ (\exists stackreg : SSREGISTER \_ TYPE;dsreg : DSREGISTER \_ TYPE; \\
	\ \ \ \ ~ csreg : CSREGISTER \_ TYPE;pswreg : PSWREGISTER \_ TYPE \spot \\
	\ \ (\forall i : \nat | i \in \dom (reg.GPRegisters) \spot reg.GPRegisters(i) = 0_{PSU}) \\
	
	\ \ \land reg.SSRegisters.SS = stackreg.SS \\
	\ \ \land reg.SSRegisters.SL = stackreg.SL \\
	\ \ \land reg.SSRegisters.BP = stackreg.BP \\
	\ \ \land reg.SSRegisters.SP = stackreg.SP \\
	
	\ \ \land reg.DSRegisters.DS = dsreg.DS \\
	\ \ \land reg.DSRegisters.DL = dsreg.DL \\
	\ \ \land reg.DSRegisters.SI = dsreg.SI \\
	\ \ \land reg.DSRegisters.DI = dsreg.DI \\
	
	\ \ \land reg.CSRegisters.CS = csreg.CS \\
	\ \ \land reg.CSRegisters.CL = csreg.CL \\
	\ \ \land reg.CSRegisters.IP = csreg.IP \\
	
	\ \ \land reg.PSWRegisters.CF = pswreg.CF \\
	\ \ \land reg.PSWRegisters.PF = pswreg.PF \\
	\ \ \land reg.PSWRegisters.AF = pswreg.AF \\
	\ \ \land reg.PSWRegisters.ZF = pswreg.ZF \\
	\ \ \land reg.PSWRegisters.SF = pswreg.SF \\
	\ \ \land reg.PSWRegisters.OF = pswreg.OF \\
	
	\ \ \land reg.PSWRegisters.TF = pswreg.TF \\
	\ \ \land reg.PSWRegisters.DF = pswreg.DF \\
	
	\ \ \land reg.PSWRegisters.IF = pswreg.IF) \\
	\ \land cpureg! = reg)
\end{schema*}

\paragraph{3.The saving of all CPU Registers}
\begin{schema*}{SaveAllCPURegister}
	cpureg! : CPU \_ REGISTER \_ TYPE
	\where
	(\exists reg : CPU \_ REGISTER \_ TYPE; \\
	\ (\exists gpreg : GPREGISTER \_ TYPE;stackreg : SSREGISTER \_ TYPE;dsreg : DSREGISTER \_ TYPE; \\
	\ \ \ \ ~ csreg : CSREGISTER \_ TYPE;pswreg : PSWREGISTER \_ TYPE \spot \\
	\ \ (\forall i : \nat | i \in \dom (reg.GPRegisters) \spot reg.GPRegisters(i) = gpreg(i)) \\
	
	\ \ \land reg.SSRegisters.SS = stackreg.SS \\
	\ \ \land reg.SSRegisters.SL = stackreg.SL \\
	\ \ \land reg.SSRegisters.BP = stackreg.BP \\
	\ \ \land reg.SSRegisters.SP = stackreg.SP \\
	
	\ \ \land reg.DSRegisters.DS = dsreg.DS \\
	\ \ \land reg.DSRegisters.DL = dsreg.DL \\
	\ \ \land reg.DSRegisters.SI = dsreg.SI \\
	\ \ \land reg.DSRegisters.DI = dsreg.DI \\
	
	\ \ \land reg.CSRegisters.CS = csreg.CS \\
	\ \ \land reg.CSRegisters.CL = csreg.CL \\
	\ \ \land reg.CSRegisters.IP = csreg.IP \\
	
	\ \ \land reg.PSWRegisters.CF = pswreg.CF \\
	\ \ \land reg.PSWRegisters.PF = pswreg.PF \\
	\ \ \land reg.PSWRegisters.AF = pswreg.AF \\
	\ \ \land reg.PSWRegisters.ZF = pswreg.ZF \\
	\ \ \land reg.PSWRegisters.SF = pswreg.SF \\
	\ \ \land reg.PSWRegisters.OF = pswreg.OF \\
	
	\ \ \land reg.PSWRegisters.TF = pswreg.TF \\
	\ \ \land reg.PSWRegisters.DF = pswreg.DF \\
	
	\ \ \land reg.PSWRegisters.IF = pswreg.IF) \\ 
	\ \land cpureg! = reg)
\end{schema*}

\paragraph{4.The restoring of part CPU Registers}
\begin{schema*}{RestorePartCPURegister}
	cpureg? : CPU \_ REGISTER \_ TYPE
	\where
	(\exists stackreg : SSREGISTER \_ TYPE;dsreg : DSREGISTER \_ TYPE; \\
	\ \ \; \; csreg : CSREGISTER \_ TYPE;pswreg : PSWREGISTER \_ TYPE \spot \\
	\ stackreg.SS = (cpureg?).SSRegisters.SS \\
	\ \land stackreg.SL = (cpureg?).SSRegisters.SL \\
	\ \land stackreg.BP = (cpureg?).SSRegisters.BP \\
	\ \land stackreg.SP = (cpureg?).SSRegisters.SP \\
	
	\ \land dsreg.DS = (cpureg?).DSRegisters.DS \\
	\ \land dsreg.DL = (cpureg?).DSRegisters.DL \\
	\ \land dsreg.SI = (cpureg?).DSRegisters.SI \\
	\ \land dsreg.DI = (cpureg?).DSRegisters.DI \\
	
	\ \land csreg.CS = (cpureg?).CSRegisters.CS \\
	\ \land csreg.CL = (cpureg?).CSRegisters.CL \\
	\ \land csreg.IP = (cpureg?).CSRegisters.IP \\
	
	\ \land pswreg.CF = (cpureg?).PSWRegisters.CF \\
	\ \land pswreg.PF = (cpureg?).PSWRegisters.PF \\
	\ \land pswreg.AF = (cpureg?).PSWRegisters.AF \\
	\ \land pswreg.ZF = (cpureg?).PSWRegisters.ZF \\
	\ \land pswreg.SF = (cpureg?).PSWRegisters.SF \\
	\ \land pswreg.OF = (cpureg?).PSWRegisters.OF \\
	
	\ \land pswreg.TF = (cpureg?).PSWRegisters.TF \\
	\ \land pswreg.DF = (cpureg?).PSWRegisters.DF \\
	
	\ \land pswreg.IF = (cpureg?).PSWRegisters.IF)
\end{schema*}

\paragraph{5.The restoring of all CPU Registers}
\begin{schema*}{RestoreAllCPURegister}
	cpureg? : CPU \_ REGISTER \_ TYPE
	\where
	(\exists gpreg : GPREGISTER \_ TYPE;stackreg : SSREGISTER \_ TYPE;dsreg : DSREGISTER \_ TYPE; \\
	\ \ \; \; csreg : CSREGISTER \_ TYPE;pswreg : PSWREGISTER \_ TYPE \spot \\
	\ (\forall i : \nat | i \in \dom gpreg \spot gpreg(i) = (cpureg?).GPRegisters(i)) \\
	
	\ \land stackreg.SS = (cpureg?).SSRegisters.SS \\
	\ \land stackreg.SL = (cpureg?).SSRegisters.SL \\
	\ \land stackreg.BP = (cpureg?).SSRegisters.BP \\
	\ \land stackreg.SP = (cpureg?).SSRegisters.SP \\
	
	\ \land dsreg.DS = (cpureg?).DSRegisters.DS \\
	\ \land dsreg.DL = (cpureg?).DSRegisters.DL \\
	\ \land dsreg.SI = (cpureg?).DSRegisters.SI \\
	\ \land dsreg.DI = (cpureg?).DSRegisters.DI \\
	
	\ \land csreg.CS = (cpureg?).CSRegisters.CS \\
	\ \land csreg.CL = (cpureg?).CSRegisters.CL \\
	\ \land csreg.IP = (cpureg?).CSRegisters.IP \\
	
	\ \land pswreg.CF = (cpureg?).PSWRegisters.CF \\
	\ \land pswreg.PF = (cpureg?).PSWRegisters.PF \\
	\ \land pswreg.AF = (cpureg?).PSWRegisters.AF \\
	\ \land pswreg.ZF = (cpureg?).PSWRegisters.ZF \\
	\ \land pswreg.SF = (cpureg?).PSWRegisters.SF \\
	\ \land pswreg.OF = (cpureg?).PSWRegisters.OF \\
	
	\ \land pswreg.TF = (cpureg?).PSWRegisters.TF \\
	\ \land pswreg.DF = (cpureg?).PSWRegisters.DF \\
	
	\ \land pswreg.IF = (cpureg?).PSWRegisters.IF)
\end{schema*}

\paragraph{6.The setting of Stack Segment Registers}
\begin{schema*}{SetStackSegRegister}
	\Delta CPU \_ REGISTER \_ TYPE \\
	ssreg? : SSREGISTER \_ TYPE
	\where
	(SSRegister').SS = (ssreg?).SS \\
	(SSRegister').SL = (ssreg?).SL \\
	(SSRegister').BP = (ssreg?).BP \\
	(SSRegister').SP = (ssreg?).SP
\end{schema*}

\subsubsection{Memory Management Unit (MMU)}
\paragraph{1.The translation of linear address}
\begin{schema*}{TransLinearAddress}
\Xi MEMORY \_ MANAGEMENT \_ UNIT \_ TYPE \\
la? : SYSTEM \_ ADDRESS \_ TYPE \\
pa! : SYSTEM \_ ADDRESS \_ TYPE
\where
(\exists a : SYSTEM \_ ADDRESS \_ TYPE | a \in \ran PageTableBuffer \spot \\
\ \mathit{getpage}(a) = PageTableBuffer(\mathit{getpage}(la?)) \land \mathit{getoffset}(a) = \mathit{getoffset}(la?) \land pa! = a)
\end{schema*}

\subsubsection{Main Memory}
\paragraph{1.The loading of a page in Main Memory}
\begin{schema*}{LoadPage}
\Delta MAIN \_ MEMORY \_ TYPE \\
page? : PAGE \_ TYPE \\
piece? : PIECE \_ TYPE
\where
(\forall a : SYSTEM \_ ADDRESS \_ TYPE | a \in \dom piece? \spot Store'(page? + \mathit{getoffset}(a)) = (piece?)(a))
\end{schema*}

\paragraph{2.The extraction of a page in Main Memory}
\begin{schema*}{ExtractPage}
\Xi MAIN \_ MEMORY \_ TYPE \\
page? : PAGE \_ TYPE \\
piece! : PIECE \_ TYPE
\where
(\exists p : PIECE \_ TYPE \spot \\
\ (\forall a : SYSTEM \_ ADDRESS \_ TYPE | a \in \dom Store \land a \in \mathit{pagespace}(page?) \spot p(a) = Store(a)) \\
\ \land piece! = p)
\end{schema*}

\subsubsection{Timer}
\paragraph{1.The initialization of Timer}
\begin{schema*}{InitTimer}
	TIMER \_ TYPE' \\
	p? : PSU
	\where
	Value' = 0_{PSU} \\
	Period' = p?
\end{schema*}

\subsubsection{High-Precision Timer}
\paragraph{1.The initialization of High-Precision Timer}
\begin{schema*}{InitHPTimer}
	HPTIMER \_ TYPE'
	\where
	Value' = 0_{PSU} \\
	Alarm' = LIMIT_{PSU}
\end{schema*}

\subsubsection{Interrupt Controller}
\paragraph{1.The initialization of Interrupt Controller}
\begin{schema*}{InitInterruptController}
	INTERRUPT \_ CONTROLLER \_ TYPE' 	
	\where
	(\forall i : \nat | i \in \dom IRRegisters' \spot IRRegister'(i) = 0) \\
	(\forall j : \nat | j \in \dom IMRegisters' \spot IMRegister'(j) = 0) \\
	(\forall k : \nat | k \in \dom ISRegisters' \spot ISRegister'(k) = 0) 
\end{schema*}

\paragraph{2.The information of Requesting interrupts}
\begin{schema*}{GetRequestInterrupts}
	\Xi INTERRUPT \_ CONTROLLER \_ TYPE \\
	ints! : \power INTERRUPT \_ TYPE
	\where
	ints! = \{i : \nat | i \in \dom IRRegister \land IRRegister(i) = 1 \}
\end{schema*}

\paragraph{3.The information of Masked interrupts}
\begin{schema*}{GetMaskedInterrupts}
	\Xi INTERRUPT \_ CONTROLLER \_ TYPE \\
	ints! : \power INTERRUPT \_ TYPE
	\where
	ints! = \{i : \nat | i \in \dom IMRegister \land IMRegister(i) = 1 \}
\end{schema*}

\paragraph{4.The information of In-Service interrupts}
\begin{schema*}{GetInServiceInterrupts}
	\Xi INTERRUPT \_ CONTROLLER \_ TYPE \\
	ints! : \power INTERRUPT \_ TYPE
	\where
	ints! = \{i : \nat | i \in \dom ISRegister \land ISRegister(i) = 1 \}
\end{schema*}

\subsection{Hardware Interface Software Layer} \label{section:zoperation_HardwareInterfaceSoftware}
\subsubsection{Context Switching}
\paragraph{1.The creating of Stack Segment}
\begin{schema*}{CreateStackSegment}
	ss? : PSU \\
	sl? : PSU \\
	bp? : PSU \\
	sp? : PSU \\
	ssreg! : SSREGISTER \_ TYPE
	\where
	(\exists reg : SSREGISTER \_ TYPE \spot \\
	\ reg.SS = ss? \land reg.Sl = sl? \land reg.BP = bp? \land reg.SP = sp? \land ssreg! = reg)
\end{schema*}

\subsubsection{Memory Management}
\paragraph{1.The initialization of Memory Management}
\begin{schema*}{InitMemoryManagement}
	MEMORY \_ MANAGEMENT \_ STATE \_ TYPE' \\
	areamm? : AREA \_ MEMORY \_ MANAGEMENT \_ TYPE
	\where
	AreaMemoryManagementRef' = areamm?
\end{schema*}

\paragraph{2.The information of the max Free Block}
\begin{schema*}{GetMaxFreeBlock}
	\Xi MEMORY \_ MANAGEMENT \_ STATE \_ TYPE \\
	maxblock! : MEMORY \_ BLOCK \_ TYPE
	\where
	(\exists maxblk : MEMORY \_ BLOCK \_ TYPE \spot \\
	\ \mathit{totalpage}(maxblk) \subseteq AreaMemoryManagementRef.Free \\
	\ \land (\forall blk : MEMORY \_ BLOCK \_ TYPE \spot \\
	\ \ \ \ \ \mathit{totalpage}(blk) \subseteq AreaMemoryManagementRef.Free \implies blk.Size \leq maxblk.Size) \\
	\ \land maxblock! = maxblk)
\end{schema*}

\paragraph{3.The allocation of Physical Memory Page}
\begin{schema*}{AllocPhysicalMemoryPage}
	\Delta MEMORY \_ MANAGEMENT \_ STATE \_ TYPE \\
	n? : UNSIGNED \_ INTEGER \_ TYPE \\
	pages! : \power PAGE \_ TYPE
	\where
	(\exists ps : \power PAGE \_ TYPE \spot \\
	\ ps \subseteq AreaMemoryManagementRef.Free \land \# ps = n? \land pages! = ps \\
	\ \land (AreaMemoryManagementRef').Allocated = AreaMemoryManagementRef.Allocated \cup ps \\
	\ \land (AreaMemoryManagementRef').Free = AreaMemoryManagementRef.Free \setminus ps)
\end{schema*}

\paragraph{4.The allocation of Physical Memory Block}
\begin{schema*}{AllocPhysicalMemoryBlock}
	\Delta MEMORY \_ MANAGEMENT \_ STATE \_ TYPE \\
	size? : UNSIGNED \_ INTEGER \_ TYPE \\
	block! : MEMORY \_ BLOCK \_ TYPE
	\where
	(\exists blk : MEMORY \_ BLOCK \_ TYPE \spot \\
	\ \mathit{totalpage}(blk) \subseteq AreaMemoryManagementRef.Free \land blk.Size = size? \land block! = blk \\
	\ \land (AreaMemoryManagementRef').Allocated = AreaMemoryManagementRef.Allocated \cup \mathit{totalpage}(blk) \\
	\ \land (AreaMemoryManagementRef').Free = AreaMemoryManagementRef.Free \setminus \mathit{totalpage}(blk))
\end{schema*}

\paragraph{5.The deallocation of Physical Memory Page}
\begin{schema*}{DeallocPhysicalMemoryPage}
	\Delta MEMORY \_ MANAGEMENT \_ STATE \_ TYPE \\
	pages? : \power PAGE \_ TYPE
	\where
	(AreaMemoryManagementRef').Allocated = AreaMemoryManagementRef.Allocated \setminus pages? \\
	(AreaMemoryManagementRef').Free = AreaMemoryManagementRef.Free \cup pages?
\end{schema*}

\paragraph{6.The creating of Memory Block}
\begin{schema*}{CreateMemoryBlock}
	start? : PAGE \_ TYPE \\
	size? : UNSIGNED \_ INTEGER \_ TYPE \\
	block! : MEMORY \_ BLOCK \_ TYPE
	\where
	(\exists blk : MEMORY \_ BLOCK \_ TYPE \spot blk.Start = start? \land blk.Size = size? \land block! = blk)
\end{schema*}

\paragraph{7.The creating of Process Virtual Memory}
\begin{schema*}{CreateProcessVirtualMemory}
	block? : MEMORY \_ BLOCK \_ TYPE \\
	base? : \power PAGE \_ TYPE \\
	stack? : MEMORY \_ BLOCK \_ TYPE \\
	data? : MEMORY \_ BLOCK \_ TYPE \\
	code? : MEMORY \_ BLOCK \_ TYPE \\
	procvm! : PROCESS \_ VIRTUAL \_ MEMORY \_ TYPE
	\where
	(\exists vm : PROCESS \_ VIRTUAL \_ MEMORY \_ TYPE \spot \\
	\ vm.VMBlock.Memory = block? \land vm.VMBlock.Base = base? \land vm.VMBlock.PageTable = \emptyset \\
	\ Stack = stack? \land Data = data? \land Code = code? \land procvm! = vm)
\end{schema*}

\paragraph{8.The full loading of Process Page Table}
\begin{schema*}{FullLoadProcPageTable}
	procvm? : PROCESS \_ VIRTUAL \_ MEMORY \_ TYPE \\
	entrypoint? : SYSTEM \_ ADDRESS \_ TYPE \\
	pagetable! : PAGE \_ TABLE \_ TYPE
	\where
	(\exists pt : PAGE \_ TABLE \_ TYPE \spot \\
	\ \dom pt \subseteq \mathit{totalpage}((procvm?).VMBlock.Memory) \land \ran pt = (procvm?).VMBlock.Base \\
	\ \land \mathit{getpage}(entrypoint?) \in \dom pt \land pagetable! = pt)
\end{schema*}

\subsubsection{System Clock}
\paragraph{1.The initialization of System Clock}
\begin{schema*}{InitSystemClock}
	SYSTEM \_ CLOCK \_ STATE \_ TYPE' \\
	bt? : SYSTEM \_ TIME \_ TYPE \\
	ti? : SYSTEM \_ TIME \_ TYPE
	\where
	BaseTime' = bt? \\
	TickInterval' = ti? \\
	TickCounter' = 0 \\
	Time' = \negate 2*ti? \\
	InterruptTime' = max(SYSTEM \_ TIME \_ TYPE)
\end{schema*}

\paragraph{2.The update of System Clock}
\begin{schema*}{UpdateSystemClock}
	\Delta SYSTEM \_ CLOCK \_ STATE \_ TYPE
	\where
	TickCounter' = TickCounter + 1 \\
	Time' = Time + TickInterval
\end{schema*}

\subsubsection{Interrupt Handler}
\paragraph{1.The initialization of Interrupt Handler}
\begin{schema*}{InitInterruptHandler}
	INTERRUPT \_ HANDLER \_ STATE \_ TYPE' \\
	intstack? : MEMORY \_ BLOCK \_ TYPE \\
	pagetable? : PAGE \_ TABLE \_ TYPE
	\where
	KernelPageTableRef' = pagetable? \\
	InterruptStack' = intstack? \\
	TempContexts' = \langle \ \rangle
\end{schema*}

\subsection{ARINC 653 O/S Layer} \label{section:zoperation_OS}
\subsubsection{Core Kernel}
\paragraph{1.The initialization of Core Kernel}
\begin{schema*}{InitCoreKernel}
	CORE \_ KERNEL \_ STATE \_ TYPE' \\
	config? : OS \_ KERNEL \_ CONFIG \_ TYPE
	\where
	(MemoryManagement').Memory = (config?).KernelMemory \\
	(MemoryManagement').Allocated = \mathit{totalpage}((config?).KernelImage) \\
	(MemoryManagement').Free = \mathit{totalpage}((MemoryManagement').Memory) \setminus (MemoryManagement').Allocated \\
	
	\dom (KernelPageTable') = \{ p : PAGE \_ TYPE | p \in \mathit{totalpage}((config?).PhysicalMemory) \spot \\
	\t6 ~ p + KERNEL \_ VIRTUAL \_ ADDRESS \_ SPACE.Start \}  \\
	\ran (KernelPageTable') = \mathit{totalpage}((config?).PhysicalMemory) \\
	(\forall p : PAGE \_ TYPE | p \in \dom (KernelPageTable') \spot \\
	\ (KernelPageTable')(p) = p - KERNEL \_ VIRTUAL \_ ADDRESS \_ SPACE.Start) \\
	
	(KernelProcManagement').ProcessTable = \{ IDLE \_ PROCESS \_ ID \mapsto IDLE \_ PROCESS \_ CONTROL \_ BLOCK \} \\
	
	(KernelProcManagement').ProcessTable(IDLE \_ PROCESS \_ ID).KernelTempContext = \langle \mathit{makecontext}(IDLE \_ \\
	PROCESS \_ REGSTATE,KernelPageTable') \rangle \\
	
	(KernelProcManagement').CurrentProcess = NULL \_ PROCESS \_ ID \\
	
	(\forall i : \nat_1 | i \in \dom ((config?).Partitions) \spot \\
	\ (PartitionManagement').PartitionTable(i).Name = (config?).Partitions(i).Name \\
	\ (PartitionManagement').PartitionTable(i).Memory = (config?).Partitions(i).Memory \\
	\ (PartitionManagement').PartitionTable(i).Periodicity = (config?).Partitions(i).Periodicity \\
	\ (PartitionManagement').PartitionTable(i).LockLevel = MIN \_ LOCK \_ LEVEL \\
	\ (PartitionManagement').PartitionTable(i).OperatingMode = IDLE \\
	\ (PartitionManagement').PartitionTable(i).StartCondition = NORMAL \_ START) \\
	
	(\forall i : \nat_1 | i \in \dom ((config?).PartitionOperationSequence) \spot \\
	\ (\exists_1 j : PARTITION \_ ID \_ TYPE | j \in \dom ((PartitionManagement').PartitionTable) \spot \\
	\ \ (config?).PartitionOperationSequence(i).Name = (PartitionManagement').PartitionTable(j).Name \\
	\ \ \land (PartitionManagement').MajorTimeFrame.PartitionQueue(i).PartitionId = j \\
	\ \ \land (PartitionManagement').MajorTimeFrame.PartitionQueue(i).Periodicity = \\
	\ \ \ \ \ (PartitionManagement').PartitionTable(i).Periodicity \\
	\ \ \land (PartitionManagement').MajorTimeFrame.PartitionQueue(i).PeriodicProcStart = \\
	\ \ \ \ \ (config?).PartitionOperationSequence(i).PeriodicProcStart)) \\
	
	(PartitionManagement').MajorTimeFrame.Length = max \\
	\ \ \ \ \ (\{ i : \nat_1 | i \in \dom ((PartitionManagement').MajorTimeFrame.PartitionQueue) \spot \\
	\t1 \ \: (PartitionManagement').MajorTimeFrame.PartitionQueue(i).Periodicity.Period \}) \\
	
	(PartitionManagement').MajorTimeFrame.PartitionQueue(1).Offset = 0 \\
	
	(\forall i : \nat_1 | i \in \dom ((PartitionManagement').MajorTimeFrame.PartitionQueue) \spot \\
	\ i > 1 \implies (\forall j : \nat_1 | j < i \spot \\
	\t2 \ \ (PartitionManagement').MajorTimeFrame.PartitionQueue(i).PartitionId \neq \\
	\t2 \ \ (PartitionManagement').MajorTimeFrame.PartitionQueue(j).PartitionId \implies \\
	\t2 \ \ (PartitionManagement').MajorTimeFrame.PartitionQueue(i).Offset = \\
	\t2 \ \ (PartitionManagement').MajorTimeFrame.PartitionQueue(i-1).Offset + \\
	\t2 \ \ (PartitionManagement').MajorTimeFrame.PartitionQueue(i-1).Periodicity.Duration) \\
	\t1 \ \ \ ~ \lor (\exists k : \nat_1 \spot \\
	\t2 \ \ k = max(\{ j : \nat_1 | j \in \dom ((PartitionManagement').MajorTimeFrame.PartitionQueue) \\
	\t4 \ \ \; \land (PartitionManagement').MajorTimeFrame.PartitionQueue(i).PartitionId = \\
	\t4 \ \ \ \ \ \: (PartitionManagement').MajorTimeFrame.PartitionQueue(j).PartitionId \\
	\t4 \ \ \; \land j < i \})  \\
	\t2 \ \ \land (PartitionManagement').MajorTimeFrame.PartitionQueue(i).Offset = \\
	\t2 \ \ \ \ \ (PartitionManagement').MajorTimeFrame.PartitionQueue(k).Offset + \\
	\t2 \ \ \ \ \ (PartitionManagement').MajorTimeFrame..PartitionQueue(k).Periodicity.Period)) \\
	
	(PartitionManagement').CurrentPartition = NULL \_ PARTITION \_ ID
\end{schema*}

\paragraph{2.The information of Status of current Partition}
\begin{schema*}{GetPartitionStatus}
	\Xi CORE \_ KERNEL \_ STATE \_ TYPE \\
	parts! : PARTITION \_ STATUS \_ TYPE
	\where
	(\exists status : PARTITION \_ STATUS \_ TYPE \spot \\
	\ \land status.PartitionId = PartitionManagement.CurrentPartition \\
	\ \land status.Period = PartitionManagement.PartitionTable(PartitionManagement.CurrentPartition). \\
	\ \ \ ~ Periodicity.Period \\
	\ \land status.Duration = PartitionManagement.PartitionTable(PartitionManagement.CurrentPartition). \\
	\ \ \ ~ Periodicity.Duration \\
	\ \land status.LockLevel = PartitionManagement.PartitionTable(PartitionManagement.CurrentPartition). \\
	\ \ \ ~ LockLevel \\
	\ \land status.OperatingMode = PartitionManagement.PartitionTable(PartitionManagement.CurrentPartition). \\
	\ \ \ ~ OperatingMode \\
	\ \land status.StartCondition = PartitionManagement.PartitionTable(PartitionManagement.CurrentPartition). \\
	\ \ \ ~ StartCondition \\
	\ \land parts! = status)
\end{schema*}

\paragraph{3.The information of next Periodic Start of current Partition}
\begin{schema*}{GetNextPeriodicStart}
	\Xi CORE \_ KERNEL \_ STATE \_ TYPE \\
	ct? : SYSTEM \_ TIME \_ TYPE \\
	nps! : SYSTEM \_ TIME \_ TYPE
	\where
	((\exists j : \nat_1 | j \in \dom (PartitionManagement.MajorTimeFrame.PartitionQueue) \spot \\
	\ \ PartitionManagement.MajorTimeFrame.PartitionQueue(j).PartitionId = PartitionManagement. \\
	\ \ CurrentPartition \\
	\ \ \land PartitionManagement.MajorTimeFrame.PartitionQueue(j).PeriodicProcStart = TRUE \\
	\ \ \land nps! = min(\{ k : \nat_1;n : \nat | k \in \dom (PartitionManagement.MajorTimeFrame.PartitionQueue) \\
	\t3 \ \ \ \; ~ \land PartitionManagement.MajorTimeFrame.PartitionQueue(k).PartitionId = \\
	\t4 \: ~ PartitionManagement.CurrentPartition \\
	\t3 \ \ \ \; ~ \land PartitionManagement.MajorTimeFrame.PartitionQueue(k).PeriodicProcStart = TRUE \\
	\t3 \ \ \ \; ~ \land PartitionManagement.MajorTimeFrame.PartitionQueue(k).Offset + PartitionManagement. \\
	\t4 \: ~ MajorTimeFrame.Length * n \geq ct? \spot \\
	\t3 \ \ \ \; ~ PartitionManagement.MajorTimeFrame.PartitionQueue(k).Offset + PartitionManagement. \\
	\t3 \ \ \ \; ~ MajorTimeFrame.Length * n \})) \\
	\ \lor (\forall j : \nat_1 | j \in \dom (PartitionManagement.MajorTimeFrame.PartitionQueue) \spot \\
	\ \ \ \ ~ PartitionManagement.MajorTimeFrame.PartitionQueue(j).PartitionId = PartitionManagement. \\
	\ \ \ \ ~ CurrentPartition \\
	\ \ \ \ ~ \land PartitionManagement.MajorTimeFrame.PartitionQueue(j).PeriodicProcStart = FALSE \implies \\
	\ \ \ \ ~ nps! = max(SYSTEM \_ TIME \_ TYPE)))
\end{schema*}

\paragraph{4.The information of next Periodic Start of current Partition including the delay time}
\begin{schema*}{GetDelayedPeriodicStart}
	\Xi CORE \_ KERNEL \_ STATE \_ TYPE \\
	ct? : SYSTEM \_ TIME \_ TYPE \\
	dt? : SYSTEM \_ TIME \_ TYPE \\
	dps! : SYSTEM \_ TIME \_ TYPE
	\where
	((\exists j : \nat_1 | j \in \dom (PartitionManagement.MajorTimeFrame.PartitionQueue) \spot \\
	\ \ PartitionManagement.MajorTimeFrame.PartitionQueue(j).PartitionId = PartitionManagement. \\
	\ \ CurrentPartition \\
	\ \ \land PartitionManagement.MajorTimeFrame.PartitionQueue(j).PeriodicProcStart = TRUE \\
	\ \ \land dps! = min(\{ k : \nat_1;n : \nat | k \in \dom (PartitionManagement.MajorTimeFrame.PartitionQueue) \\
	\t3 \ \ \ \; ~ \land PartitionManagement.MajorTimeFrame.PartitionQueue(k).PartitionId = \\
	\t4 \: ~ PartitionManagement.CurrentPartition \\
	\t3 \ \ \ \; ~ \land PartitionManagement.MajorTimeFrame.PartitionQueue(k).PeriodicProcStart = TRUE \\
	\t3 \ \ \ \; ~ \land PartitionManagement.MajorTimeFrame.PartitionQueue(k).Offset + PartitionManagement. \\
	\t4 \: ~ MajorTimeFrame.Length * n \geq (ct? + dt?) \spot \\
	\t3 \ \ \ \; ~ PartitionManagement.MajorTimeFrame.PartitionQueue(k).Offset + PartitionManagement. \\
	\t3 \ \ \ \; ~ MajorTimeFrame.Length * n \})) \\
	\ \lor (\forall j : \nat_1 | j \in \dom (PartitionManagement.MajorTimeFrame.PartitionQueue) \spot \\
	\ \ \ \ ~ PartitionManagement.MajorTimeFrame.PartitionQueue(j).PartitionId = PartitionManagement. \\
	\ \ \ \ ~ CurrentPartition \\
	\ \ \ \ ~ \land PartitionManagement.MajorTimeFrame.PartitionQueue(j).PeriodicProcStart = FALSE \implies \\
	\ \ \ \ ~ dps! = max(SYSTEM \_ TIME \_ TYPE)))
\end{schema*}

\paragraph{5.The information of next Partition}
\begin{schema*}{GetNextPartition}
	\Xi CORE \_ KERNEL \_ STATE \_ TYPE \\
	t? : SYSTEM \_ TIME \_ TYPE \\
	partid! : PARTITION \_ ID \_ TYPE
	\where
	((\exists_1 i : \nat_1 | i \in \dom (PartitionManagement.MajorTimeFrame.PartitionQueue) \spot \\
	\ \ PartitionManagement.MajorTimeFrame.PartitionQueue(i).Offset \leq \\
	\ \ (t? \mod PartitionManagement.MajorTimeFrame.Length) \\
	\ \ \land PartitionManagement.MajorTimeFrame.PartitionQueue(i).Offset + \\
	\ \ \ \ \; PartitionManagement.MajorTimeFrame.PartitionQueue(i).Periodicity.Duration > \\
	\ \ \ \ \; (t? \mod PartitionManagement.MajorTimeFrame.Length) \\
	\ \ \land partid! = PartitionManagement.MajorTimeFrame.PartitionQueue(i).PartitionId) \\
	\ \lor (\forall i : \nat_1 | i \in \dom (PartitionManagement.MajorTimeFrame.PartitionQueue) \spot \\
	\ \ \ \; \: PartitionManagement.MajorTimeFrame.PartitionQueue(i).Offset + \\
	\ \ \ \; \: PartitionManagement.MajorTimeFrame.PartitionQueue(i).Periodicity.Duration \leq \\
	\ \ \ \; \: (t? \mod PartitionManagement.MajorTimeFrame.Length) \\
	\ \ \ \; \: \lor PartitionManagement.MajorTimeFrame.PartitionQueue(i).Offset > \\ 
	\t1 \; \: (t? \mod PartitionManagement.MajorTimeFrame.Length) \implies \\
	\ \ \ \; \: partid! = NULL \_ PARTITION \_ ID))
\end{schema*}

\subsubsection{Partition Kernel}
\paragraph{1.The initialization of Partition Kernel}
\begin{schema*}{InitPartitionKernel}
	PARTITION \_ KERNEL \_ STATE \_ TYPE' \\
	config? : PARTITION \_ KERNEL \_ CONFIG \_ TYPE \\
	pagetable? : PAGE \_ TABLE \_ TYPE
	\where
	(MemoryManagement').Memory = (config?).Memory \\
	(MemoryManagement').Allocated = \emptyset \\
	(MemoryManagement').Free = \mathit{totalpage}((MemoryManagement').Memory) \\
	KernelPageTableRef' = pagetable? \\
	Periodicity' = (config?).Periodicity \\
	TotalProcess' = (config?).TotalProcess \\
	ProcessTable' = \{ IDLE \_ PROCESS \_ ID \mapsto IDLE \_ PROCESS \_ CONTROL \_ BLOCK \} \\
	ProcessTable'(IDLE \_ PROCESS \_ ID).KernelTempContext = \langle \mathit{makecontext}(IDLE \_ PROCESS \_ REGSTATE, \\
	pagetable?) \rangle \\
	TimeCounterQueue' = \langle \ \rangle \\
	ReadyQueue' = \langle \ \rangle \\
	WaitingQueue' = \langle \ \rangle \\
	CurrentProcess' = NULL \_ PROCESS \_ ID
\end{schema*}

\paragraph{2.The information of Names of Processes}
\begin{schema*}{GetProcessNames}
	\Xi PARTITION \_ KERNEL \_ STATE \_ TYPE \\
	procns! : \power PROCESS \_ NAME \_ TYPE
	\where
	procns! = \{ i : PROCESS \_ ID \_ TYPE | i \in \dom ProcessTable \spot ProcessTable(i).Name \}
\end{schema*}

\paragraph{3.The information of Process Id}
\begin{schema*}{GetProcessId}
	\Xi PARTITION \_ KERNEL \_ STATE \_ TYPE \\
	procn? : PROCESS \_ NAME \_ TYPE \\
	procid! : PROCESS \_ ID \_ TYPE
	\where
	(\exists_1 i : PROCESS \_ ID \_ TYPE | i \in \dom ProcessTable \spot ProcessTable(i).Name = procn? \land procid! = i)
\end{schema*}

\paragraph{4.The information of Process Status}
\begin{schema*}{GetProcessStatus}
	\Xi PARTITION \_ KERNEL \_ STATE \_ TYPE \\
	procid? : PROCESS \_ ID \_ TYPE \\
	procs! : PROCESS \_ STATUS \_ TYPE
	\where
	(\exists status : PROCESS \_ STATUS \_ TYPE \spot \\
	\ status.Attribute.Name = ProcessTable(procid?).Name \\
	\ \land status.Attribute.EntryPoint = ProcessTable(procid?).EntryPoint \\
	\ \land status.Attribute.StackSize = ProcessTable(procid?).VirtualMemory.Stack.Size \\
	\ \land status.Attribute.BasePriority = ProcessTable(procid?).BasePriority \\
	\ \land status.Attribute.Period = ProcessTable(procid?).Period \\
	\ \land status.Attribute.TimeCapacity = ProcessTable(procid?).TimeCapacity \\
	\ \land status.Attribute.Deadline = ProcessTable(procid?).Deadline \\
	\ \land status.CurrentPriority = ProcessTable(procid?).CurrentPriority \\
	\ \land status.DeadlineTime = ProcessTable(procid?).DeadlineTime \\
	\ \land status.ProcessState = ProcessTable(procid?).ProcessState \\
	\ \land procs! = status)
\end{schema*}

\paragraph{5.The initialization of Register State of Process Context}
\begin{schema*}{InitProcRegisterState}
	entrypoint? : SYSTEM \_ ADDRESS \_ TYPE \\
	procvm? : PROCESS \_ VIRTUAL \_ MEMORY \_ TYPE \\
	regstate! : CPU \_ REGISTER \_ TYPE
	\where
	(\exists reg : CPU \_ REGISTER \_ TYPE \spot \\
	\ (\forall i : \nat | i \in \dom (reg.GPRegisters) \spot reg.GPRegisters(i) = 0_{PSU}) \\
	
	\ \land reg.SSRegisters.SS = (procvm?).Stack.Start \\
	\ \land reg.SSRegisters.BP = 0_{PSU} \\
	\ \land reg.SSRegisters.SP = 0_{PSU} \\
	
	\ \land reg.DSRegisters.DS = (procvm?).Data.Start \\
	\ \land reg.DSRegisters.SI = 0_{PSU} \\
	\ \land reg.DSRegisters.DI = 0_{PSU} \\
	
	\ \land reg.CSRegisters.CS = (procvm?).Code.Start \\
	\ \land reg.CSRegisters.IP = entrypoint? \\
	
	\ \land reg.PSWRegisters.CF = 0 \\
	\ \land reg.PSWRegisters.PF = 0 \\
	\ \land reg.PSWRegisters.AF = 0 \\
	\ \land reg.PSWRegisters.ZF = 0 \\
	\ \land reg.PSWRegisters.SF = 0 \\
	\ \land reg.PSWRegisters.OF = 0 \\
	
	\ \land reg.PSWRegisters.TF = 0 \\
	\ \land reg.PSWRegisters.DF = 0 \\
	
	\ \land reg.PSWRegisters.IF = 1	\\
	\ \land regstate! = reg)
\end{schema*}

\paragraph{6.The creating of Process Control Block}
\begin{schema*}{CreateProcessControlBlock}
	proca? : PROCESS \_ ATTRIBUTE \_ TYPE \\
	exefilepath? : PATH \_ TYPE \\
	swapfilepath? : PATH \_ TYPE \\
	kstack? : MEMORY \_ BLOCK \_ TYPE \\
	procvm? : PROCESS \_ VIRTUAL \_ MEMORY \_ TYPE \\
	proccb! : PROCESS \_ CONTROL \_ BLOCK \_ TYPE
	\where
	(\exists pcb : PROCESS \_ CONTROL \_ BLOCK \_ TYPE \spot \\
	\ pcb.Name = (proca?).Name \\
	\ \land pcb.ProcessKind = (proca?).ProcessKind \\
	\ \land pcb.ExeFilePath = exefilepath? \\
	\ \land pcb.SwapFilePath = swapfilepath? \\
	\ \land pcb.EntryPoint = (proca?).EntryPoint \\
	\ \land pcb.KernelStack = kstack? \\
	\ \land pcb.VirtualMemory = procvm? \\
	\ \land pcb.Period = (proca?).Period \\
	\ \land pcb.TimeCapacity = (proca?).TimeCapacity \\
	\ \land pcb.Deadline = (proca?).Deadline \\
	\ \land pcb.BasePriority = (proca?).BasePriority \\
	\ \land pcb.CurrentPriority = DEFAULT \_ PRIORITY \\
	\ \land pcb.ReleasePoint = DEFAULT \_ TIME \\
	\ \land pcb.DeadlineTime = DEFAULT \_ TIME \\
	\ \land pcb.ProcessState = DORMANT \\
	\ \land pcb.UserTempContext = \langle \ \rangle \\
	\ \land pcb.KernelTempContext = \langle \ \rangle \\
	\ \land proccb! = pcb)
\end{schema*}

\paragraph{7.The enqueuing an element into Ready Queue(From Head)}
\begin{schema*}{EnqReadyQueue_1}
	\Delta PARTITION \_ KERNEL \_ STATE \_ TYPE \\
	procid? : PROCESS \_ ID \_ TYPE
	\where
	ReadyQueue' = \langle procid? \rangle \cat ReadyQueue
\end{schema*}

\paragraph{8.The enqueuing an element into Ready Queue(From Tail)}
\begin{schema*}{EnqReadyQueue_2}
	\Delta PARTITION \_ KERNEL \_ STATE \_ TYPE \\
	procid? : PROCESS \_ ID \_ TYPE
	\where
	ReadyQueue' = ReadyQueue \cat \langle procid? \rangle
\end{schema*}

\paragraph{9.The dequeuing an element from Ready Queue}
\begin{schema*}{DeqReadyQueue}
	\Delta PARTITION \_ KERNEL \_ STATE \_ TYPE \\
	procid? : PROCESS \_ ID \_ TYPE
	\where
	ReadyQueue' = squash ~ (ReadyQueue \nrres \{ procid? \})
\end{schema*}

\paragraph{10.The enqueuing an element into Waiting Queue}
\begin{schema*}{EnqWaitingQueue}
	\Delta PARTITION \_ KERNEL \_ STATE \_ TYPE \\
	procid? : PROCESS \_ ID \_ TYPE
	\where
	WaitingQueue' = WaitingQueue \cat \langle procid? \rangle
\end{schema*}

\paragraph{11.The dequeuing an element from Waiting Queue}
\begin{schema*}{DeqWaitingQueue}
	\Delta PARTITION \_ KERNEL \_ STATE \_ TYPE \\
	procid? : PROCESS \_ ID \_ TYPE
	\where
	WaitingQueue' = squash ~ (WaitingQueue \nrres \{ procid? \})
\end{schema*}

\paragraph{12.The creating of Time Counter}
\begin{schema*}{CreateTimeCounter}
	procid? : PROCESS \_ ID \_ TYPE \\
	alarm? : SYSTEM \_ TIME \_ TYPE \\
	tc! : TIME \_ COUNTER \_ TYPE
	\where
	(\exists counter : TIME \_ COUNTER \_ TYPE \spot \\
	\ counter.ProcessId = procid? \land counter.Alarm = alarm? \land tc! = counter)
\end{schema*}

\paragraph{13.The information of Timing Process Alarm}
\begin{schema*}{GetTimingProcessAlarm}
	\Xi PARTITION \_ KERNEL \_ STATE \_ TYPE \\
	procid? : PROCESS \_ ID \_ TYPE \\
	alarm! : SYSTEM \_ TIME \_ TYPE
	\where
	(\exists_1 i : \nat_1 | i \in \dom TimeCounterQueue \spot \\
	\ TimeCounterQueue(i).ProcessId = procid? \land alarm! = TimeCounterQueue(i).Alarm)
\end{schema*}

\paragraph{14.The turning over of Time Counter}
\begin{schema*}{TurnOverTimeCounter}
	\Delta PARTITION \_ KERNEL \_ STATE \_ TYPE \\
	procid? : PROCESS \_ ID \_ TYPE
	\where
	(\exists_1 i : \nat_1 | i \in \dom TimeCounterQueue \spot \\
	\ TimeCounterQueue(i).ProcessId = procid? \land TimeCounterQueue'(i).Alarm = \negate TimeCounterQueue(i).Alarm)
\end{schema*}

\paragraph{15.The information of processes in Time Counter Queue}
\begin{schema*}{GetTimingProcess}
	\Xi PARTITION \_ KERNEL \_ STATE \_ TYPE \\
	tprocids! : \power PROCESS \_ ID \_ TYPE
	\where
	tprocids! = \{ i : \nat_1 | i \in \dom TimeCounterQueue \spot TimeCounterQueue(i).ProcessId \}
\end{schema*}

\paragraph{16.The enqueuing an element into Time Counter Queue}
\begin{schema*}{EnqTimeCounterQueue}
	\Delta PARTITION \_ KERNEL \_ STATE \_ TYPE \\
	tc? : TIME \_ COUNTER \_ TYPE
	\where
	((\exists s,t : iseq[TIME \_ COUNTER \_ TYPE] | s \cat t = TimeCounterQueue \land s,t \neq \langle \ \rangle \spot \\
	\ \ |(last ~ s).Alarm| \leq (tc?).Alarm \land |(head ~ t).Alarm| > (tc?).Alarm \land TimeCounterQueue' = s \cat \langle tc? \rangle \cat t) \\
	\ \lor (\forall i : \nat_1 | i \in \dom TimeCounterQueue \spot \\
	\ \ \ \ \: |TimeCounterQueue(i).Alarm| \leq (tc?).Alarm \implies TimeCounterQueue' = TimeCounterQueue \cat \langle tc? \rangle) \\
	\ \lor (\forall i : \nat_1 | i \in \dom TimeCounterQueue \spot \\
	\ \ \ \ \: |TimeCounterQueue(i).Alarm| > (tc?).Alarm \implies TimeCounterQueue' = \langle tc? \rangle \cat TimeCounterQueue))
\end{schema*}

\paragraph{17.The dequeuing an element from Time Counter Queue}
\begin{schema*}{DeqTimeCounterQueue}
	\Delta PARTITION \_ KERNEL \_ STATE \_ TYPE \\
	procid? : PROCESS \_ ID \_ TYPE
	\where
	(\exists_1 tc : TIME \_ COUNTER \_ TYPE | tc \in \ran TimeCounterQueue \land tc.ProcessId = procid? \spot \\
	\ TimeCounterQueue' = squash ~ (TimeCounterQueue \nrres \{ tc \}))
\end{schema*}

\paragraph{18.The information of the Hishest Priority of processes in the Ready Queue}
\begin{schema*}{GetHighestPriority}
	\Xi PARTITION \_ KERNEL \_ STATE \_ TYPE \\
	hp! : PRIORITY \_ TYPE
	\where
	(\exists i : PROCESS \_ ID \_ TYPE | i \in \ran ReadyQueue \spot \\
	\ (\forall j : PROCESS \_ ID \_ TYPE | j \in \ran ReadyQueue \spot \\
	\ \ j \neq i \implies |ProcessTable(j).CurrentPriority| \leq |ProcessTable(i).CurrentPriority|) \\
	\ \land (hp! = |ProcessTable(i).CurrentPriority|))
\end{schema*}

\paragraph{19.The update of Time Counter Queue}
\begin{schema*}{UpdateTimeCounterQueue}
	\Delta PARTITION \_ KERNEL \_ STATE \_ TYPE \\
	hpt? : SYSTEM \_ TIME \_ TYPE \\
	procids! : iseq[PROCESS \_ ID \_ TYPE]
	\where
	((\forall i : \nat_1 | i \in \dom TimeCounterQueue \spot \\
	\ \ |TimeCounterQueue(i).Alarm| > hpt? \implies procids! = \emptyset \land TimeCounterQueue' = TimeCounterQueue) \\
	\ \lor (\exists k : \nat_1 \spot \\
	\ \ \ \ \: ~ k = max(\{ i : \nat_1 | i \in \dom TimeCounterQueue \land |TimeCounterQueue(i).Alarm| \leq hpt? \}) \\
	\ \ \ \ \: ~ \land procids! = squash~((1 \upto k \dres TimeCounterQueue) \rres \{ i : \nat_1 | i \in \dom TimeCounterQueue \\
	\t5 \ \ ~ \land TimeCounterQueue(i).Alarm > 0 \}) \\
	\ \ \ \ \: ~ \land TimeCounterQueue' = squash~(1 \upto k \ndres TimeCounterQueue)))
\end{schema*}

\paragraph{20.The Checking of Release Points}
\begin{schema*}{CheckReleasePionts}
	\Xi PARTITION \_ KERNEL \_ STATE \_ TYPE \\ \\
	t? : SYSTEM \_ TIME \_ TYPE \\
	procids! : iseq[PROCESS \_ ID \_ TYPE]
	\where
	((\forall i : PROCESS \_ ID \_ TYPE | i \in \dom ProcessTable \spot \\
	\ \ ProcessTable(i).ReleasePoint \neq t? \implies procids! = \langle \ \rangle) \\
	\ \lor (\exists seq : iseq[PROCESS \_ ID \_ TYPE] | \ran seq \subseteq \dom ProcessTable \spot \\
	\ \ \ \ \: ~ (\forall i : PROCESS \_ ID \_ TYPE | i \in \ran seq \spot ProcessTable(i).ReleasePoint = t?) \\
	\ \ \ \ \: ~ \land (\forall i : PROCESS \_ ID \_ TYPE | i \in (\dom ProcessTable \setminus \ran seq) \spot ProcessTable(i).ReleasePoint \neq t?) \\
	\ \ \ \ \: ~ \land procids! = seq))
\end{schema*}

\newpage

\section{Circus Channels} \label{section:circuschannel}
\subsection{Channels for signals which indicates the life stage of a ARINC 653 Module}
\begin{zed}
\begin{array}{ll}
		\circchannel module \_ init,module \_ end
	\end{array}
\end{zed}

\subsection{Channels for signals which indicates the life stage of the Real-Time Clock}
\begin{zed}
	\begin{array}{ll}
		\circchannel rtclock \_ power \_ on,rtclock \_ power \_ off
	\end{array}
\end{zed}

\subsection{Channel for signals which indicates the Clock Pause}
\begin{zed}
	\begin{array}{ll}
		\circchannel clock \_ pause
	\end{array} 
\end{zed}

\subsection{Channel for signals which indicates the Hardware Interrupts}
\begin{zed}
	\begin{array}{ll}
		\circchannel hardware \_ interrupts : INTERRUPT \_ TYPE
	\end{array}
\end{zed}

\subsection{Channel for signals which indicates the Exceptions}
\begin{zed}
	\begin{array}{ll}
		\circchannel exception : EXCEPTION \_ TYPE
	\end{array}
\end{zed}

\subsection{Channels for communication within Interrupt Controller}
\begin{zed}
	\begin{array}{ll}
		\circchannel irregister \_ report \_ new \_ interrupt : INTERRUPT \_ TYPE \\
		\circchannel isregister \_ demand \_ new \_ interrupt \\
		\circchannel isregister \_ service \_ new \_ interrupt : INTERRUPT \_ TYPE \\
		\circchannel priority \_ resolver \_ report \_ new \_ interrupt : INTERRUPT \_ TYPE
	\end{array}
\end{zed}

\subsection{Channels for communication between different parts of Core Hardware Layer}
\begin{zed}
	\begin{array}{ll}
		\circchannel interrupt : INTERRUPT \_ TYPE \\
		\circchannel interrupt \_ acknowledged \\
		\circchannel interrupt \_ handled : INTERRUPT \_ TYPE \\
		
		\circchannel cpu \_ request \_ logical \_ address \_ translation : PAGE \_ TYPE \cross PSU \cross SYSTEM \_ ADDRESS \_ TYPE \\
		\circchannel logical \_ address \_ translation \_ fail \\
		\circchannel cpu \_ get \_ logical \_ address \_ translation : SYSTEM \_ ADDRESS \_ TYPE \\
		
		\circchannel cpu \_ receive \_ data \_ from \_ main \_ memory : SYSTEM \_ ADDRESS \_ TYPE \\
		\circchannel cpu \_ received \_ data \_ from \_ main \_ memory : PSU \\
		
		\circchannel cpu \_ send \_ data \_ to \_ main \_ memory : SYSTEM \_ ADDRESS \_ TYPE \cross PSU \\
		\circchannel cpu \_ sent \_ data \_ to \_ main \_ memory \\
	\end{array}
\end{zed}

\subsection{Channels for communication betwween Core Hardware Layer and Hardware Interface Software Layer}
\begin{zed}
	\begin{array}{ll}
		\circchannel context \_ switching \_ save \_ part \_ cpu \_ registers \\
		\circchannel context \_ switching \_ saved \_ part \_ cpu \_ registers : CPU \_ REGISTER \_ TYPE \\
		
		\circchannel context \_ switching \_ save \_ all \_ cpu \_ registers \\
		\circchannel context \_ switching \_ saved \_ all \_ cpu \_ registers : CPU \_ REGISTER \_ TYPE \\
		
		\circchannel context \_ switching \_ restore \_ part \_ cpu \_ registers : CPU \_ REGISTER \_ TYPE \\
		\circchannel context \_ switching \_ restored \_ part \_ cpu \_ registers \\
		
		\circchannel context \_ switching \_ restore \_ all \_ cpu \_ registers : CPU \_ REGISTER \_ TYPE \\
		\circchannel context \_ switching \_ restored \_ all \_ cpu \_ registers \\
\end{array}
\end{zed}

\subsection{Channels for communication within Interrupt Handler}
\begin{zed}
	\begin{array}{ll}
		\circchannel interrupt \_ handler \_ invoke \_ service \_ routine \_ clocktick \\
		\circchannel interrupt \_ service \_ routine \_ end : INTERRUPT \_ TYPE \\
	\end{array}
\end{zed}

\subsection{Channels for communication between different parts of Hardware Interface Software Layer}
\begin{zed}
	\begin{array}{ll}
		\circchannel interrupt \_ handler \_ save \_ context \\
		\circchannel interrupt \_ handler \_ saved \_ context : CPU \_ REGISTER \_ TYPE \cross PAGE \_ TABLE \_ TYPE \\
		
		\circchannel interrupt \_ handler \_ restore \_ context : CPU \_ REGISTER \_ TYPE \cross PAGE \_ TABLE \_ TYPE \\
		\circchannel interrupt \_ handler \_ restored \_ context \\
		
		\circchannel interrupt \_ handler \_ switch \_ stack : SSREGISTER \_ TYPE \cross PAGE \_ TABLE \_ TYPE \\
		\circchannel interrupt \_ handler \_ switched \_ stack \\
		
		\circchannel isr \_ clocktick \_ update \_ system \_ time \\
		\circchannel isr \_ clocktick \_ updated \_ system \_ time \\
		
		\circchannel isr \_ clocktick \_ request \_ high \_ precision \_ time \\
		\circchannel isr \_ clocktick \_ get \_ high \_ precision \_ time : SYSTEM \_ TIME \_ TYPE \\
		
		\circchannel interrupt \_ handler \_ mask \_ interrupt : INTERRUPT \_ TYPE \\
		\circchannel interrupt \_ handler \_ masked \_ interrupt \\
		
		\circchannel interrupt \_ handler \_ unmask \_ interrupt : INTERRUPT \_ TYPE \\
		\circchannel interrupt \_ handler \_ unmasked \_ interrupt \\
		
		\circchannel interrupt \_ handler \_ enable \_ interrupt \\
		\circchannel interrupt \_ handler \_ enabled \_ interrupt
		
		\circchannel interrupt \_ handler \_ disable \_ interrupt \\
		\circchannel interrupt \_ handler \_ disabled \_ interrupt \\
	\end{array}
\end{zed}

\subsection{Channels for communication betwween Hardware Interface Software Layer and ARINC 653 O/S Layer}
\begin{zed}
	\begin{array}{ll}
		\circchannel kernel \_ save \_ context : KERNEL \_ ID \_ TYPE \cross OPTION \_ TYPE \\
		\circchannel kernel \_ saved \_ context : KERNEL \_ ID \_ TYPE \cross CPU \_ REGISTER \_ TYPE \cross PAGE \_ TABLE \_ TYPE \\
		
		\circchannel kernel \_ restore \_ context : KERNEL \_ ID \_ TYPE \cross OPTION \_ TYPE \cross CPU \_ REGISTER \_ TYPE \cross \\
		\t1 \ \ \ \ \ \; PAGE \_ TABLE \_ TYPE \\
		\circchannel kernel \_ restored \_ context : KERNEL \_ ID \_ TYPE \\
		
		\circchannel kernel \_ switch \_ stack : KERNEL \_ ID \_ TYPE \cross SSREGISTER \_ TYPE \cross PAGE \_ TABLE \_ TYPE \\
		\circchannel kernel \_ switched \_ stack : KERNEL \_ ID \_ TYPE \\
		
		\circchannel kernel \_ init \_ memory \_ management : AREA \_ MEMORY \_ MANAGEMENT \_ TYPE \\
\end{array}
	\\
	
	\znewpage
	
	\begin{array}{ll}
		\circchannel kernel \_ create \_ virtual \_ memory \_ block : KERNEL \_ ID \_ TYPE \cross PAGE \_ TYPE \cross UNSIGNED \_ INTEGER \_ \\
		\t1 \ \ \ \ \ \; TYPE \\
		\circchannel kernel \_ created \_ virtual \_ memory \_ block : KERNEL \_ ID \_ TYPE \cross MEMORY \_ BLOCK \_ TYPE \\
		
		\circchannel kernel \_ create \_ process \_ virtual \_ memory : KERNEL \_ ID \_ TYPE \cross \power PAGE \_ TYPE \cross UNSIGNED \_ \\
		\t1 \ \ \ \ \ \; INTEGER \_ TYPE \cross UNSIGNED \_ INTEGER \_ TYPE \cross UNSIGNED \_ INTEGER \_ TYPE \\
		\circchannel kernel \_ created \_ process \_ virtual \_ memory : KERNEL \_ ID \_ TYPE \cross PROCESS \_ VIRTUAL \_ MEMORY \_ \\
		\t1 \ \ \ \ \ \; TYPE \\
		
		\circchannel kernel \_ full \_ load \_ process \_ pages : KERNEL \_ ID \_ TYPE \cross SYSTEM \_ ADDRESS \_ TYPE \cross \\
\end{array}
\end{zed}

\subsection{Channels for communication of Error Handler betwween its User Mode and Kernel Mode}
\begin{zed}
	\begin{array}{ll}
		\circchannel error \_ process \_ id : PROCESS \_ ID \_ TYPE \\
	\end{array}
\end{zed}

\subsection{Channels for communication within each Partition Kernel}
\begin{zed}
	\begin{array}{ll}
		\circchannel system \_ call \_ handler \_ invoke \_ service \_ routine \_ get \_ operating \_ mode : PARTITION \_ ID \_ TYPE \cross PROCESS \_ \\
		\t1 \ \ \ \ \ \; ID \_ TYPE \\
		\circchannel system \_ call \_ handler \_ invoke \_ service \_ routine \_ get \_ partition \_ status : PARTITION \_ ID \_ TYPE \cross PROCESS \_ \\
		\t1 \ \ \ \ \ \; ID \_ TYPE \\
		\circchannel system \_ call \_ handler \_ invoke \_ service \_ routine \_ set \_ operating \_ mode : PARTITION \_ ID \_ TYPE \cross PROCESS \_ \\
		\t1 \ \ \ \ \ \; ID \_ TYPE \\
		\circchannel system \_ call \_ handler \_ invoke \_ service \_ routine \_ get \_ process \_ names : PARTITION \_ ID \_ TYPE \cross PROCESS \_ \\
		\t1 \ \ \ \ \ \; ID \_ TYPE \\
\end{array}
	\\
	
	\znewpage
	
	\begin{array}{ll}
		\circchannel system \_ call \_ handler \_ invoke \_ service \_ routine \_ get \_ current \_ time : PARTITION \_ ID \_ TYPE \cross PROCESS \_ \\
		\t1 \ \ \ \ \ \; ID \_ TYPE \\
		\circchannel system \_ call \_ handler \_ invoke \_ service \_ routine \_ get \_ partition \_ time : PARTITION \_ ID \_ TYPE \cross PROCESS \_ \\
		\t1 \ \ \ \ \ \; ID \_ TYPE \\
		\circchannel system \_ call \_ handler \_ invoke \_ service \_ routine \_ get \_ next \_ release \_ point : PARTITION \_ ID \_ TYPE \cross \\
		\t1 \ \ \ \ \ \; PROCESS \_ ID \_ TYPE \\
		\circchannel system \_ call \_ handler \_ invoke \_ service \_ routine \_ timed \_ wait : PARTITION \_ ID \_ TYPE \cross PROCESS \_ ID \_ \\
		\t1 \ \ \ \ \ \; TYPE \\
\end{array}
\end{zed}

\subsection{Channels for communication between different parts of ARINC 653 O/S Layer}
\begin{zed}
	\begin{array}{ll}
		\circchannel core \_ kernel \_ init \_ partition \_ kernel : PARTITION \_ ID \_ TYPE \cross PAGE \_ TABLE \_ TYPE \\
		\circchannel core \_ kernel \_ inited \_ partition \_ kernel : PARTITION \_ ID \_ TYPE \\
		
		\circchannel core \_ kernel \_ service \_ get \_ operating \_ mode : PARTITION \_ ID \_ TYPE \\
		\circchannel core \_ kernel \_ service \_ get \_ operating \_ mode \_ output : PARTITION \_ ID \_ TYPE \cross OPERATING \_ MODE \_ TYPE \\
		\circchannel core \_ kernel \_ service \_ get \_ operating \_ mode \_ return : PARTITION \_ ID \_ TYPE \\
		
		\circchannel core \_ kernel \_ service \_ get \_ partition \_ status : PARTITION \_ ID \_ TYPE \\
		\circchannel core \_ kernel \_ service \_ get \_ partition \_ status \_ output : PARTITION \_ ID \_ TYPE \cross PARTITION \_ STATUS \_ TYPE \\
		\circchannel core \_ kernel \_ service \_ get \_ partition \_ status \_ return : PARTITION \_ ID \_ TYPE \\
\end{array}
	\\
	
	\znewpage
	
	\begin{array}{ll}
		\circchannel core \_ kernel \_ service \_ decrease \_ lock \_ level : PARTITION \_ ID \_ TYPE \\
		\circchannel core \_ kernel \_ service \_ decrease \_ lock \_ level \_ return : PARTITION \_ ID \_ TYPE \\
		
		\circchannel partition \_ kernel \_ continue \_ execution : PARTITION \_ ID \_ TYPE \\
		\circchannel partition \_ kernel \_ suspend \_ execution : PARTITION \_ ID \_ TYPE \\
		\circchannel partition \_ kernel \_ suspended \_ execution : PARTITION \_ ID \_ TYPE \\
		\circchannel partition \_ kernel \_ idle \_ execution : PARTITION \_ ID \_ TYPE \\
		\circchannel partition \_ kernel \_ prepare \_ execution : PARTITION \_ ID \_ TYPE \\
		\circchannel partition \_ kernel \_ update \_ system \_ time : PARTITION \_ ID \_ TYPE \cross SYSTEM \_ TIME \_ TYPE \\
		\circchannel partition \_ kernel \_ start \_ execution : PARTITION \_ ID \_ TYPE \\
	\end{array}
\end{zed}

\subsection{Channels for communication betwween ARINC 653 O/S Layer and APEX Layer}
\begin{zed}
	\begin{array}{ll}
		\circchannel system \_ call : PARTITION \_ ID \_ TYPE \cross PROCESS \_ ID \_ TYPE \cross SYSTEM \_ CALL \_ TYPE \\
		
		\circchannel system \_ call \_ service \_ routine \_ get \_ operating \_ mode \_ output : PARTITION \_ ID \_ TYPE \cross PROCESS \_ ID \_ \\
		\t1 \ \ \ \ \ \; TYPE \cross OPERATING \_ MODE \_ TYPE \\
		
		\circchannel system \_ call \_ service \_ routine \_ get \_ partition \_ status \_ output : PARTITION \_ ID \_ TYPE \cross PROCESS \_ ID \_ \\
		\t1 \ \ \ \ \ \; TYPE \cross PARTITION \_ STATUS \_ TYPE \\
		
		\circchannel system \_ call \_ service \_ routine \_ set \_ operating \_ mode \_ input : PARTITION \_ ID \_ TYPE \cross PROCESS \_ ID \_ \\
		\t1 \ \ \ \ \ \; TYPE \cross OPERATING \_ MODE \_ TYPE \\
		
		\circchannel system \_ call \_ service \_ routine \_ get \_ process \_ names \_ output : PARTITION \_ ID \_ TYPE \cross PROCESS \_ ID \_ \\
\end{array}
	\\
	
	\znewpage
	
	\begin{array}{ll}
		\circchannel system \_ call \_ service \_ routine \_ get \_ partition \_ lock \_ level \_ output : PARTITION \_ ID \_ TYPE \cross PROCESS \_ \\
		\t1 \ \ \ \ \ \; ID \_ TYPE \cross LOCK \_ LEVEL \_ TYPE \\
		
		\circchannel system \_ call \_ service \_ routine \_ get \_ process \_ id \_ input : PARTITION \_ ID \_ TYPE \cross PROCESS \_ ID \_ TYPE \cross \\
		\t1 \ \ \ \ \ \; PROCESS \_ NAME \_ TYPE \\
		\circchannel system \_ call \_ service \_ routine \_ get \_ process \_ id \_ output : PARTITION \_ ID \_ TYPE \cross PROCESS \_ ID \_ TYPE \cross \\
		\t1 \ \ \ \ \ \; PROCESS \_ ID \_ TYPE \\
		
		\circchannel system \_ call \_ service \_ routine \_ get \_ process \_ status \_ input : PARTITION \_ ID \_ TYPE \cross PROCESS \_ ID \_ \\
		\t1 \ \ \ \ \ \; TYPE \cross PROCESS \_ ID \_ TYPE \\
\end{array}
	\\
	
	\znewpage
	
	\begin{array}{ll}
		\circchannel system \_ call \_ service \_ routine \_ replenish \_ input : PARTITION \_ ID \_ TYPE \cross PROCESS \_ ID \_ TYPE \cross \\
		\t1 \ \ \ \ \ \; SYSTEM \_ TIME \_ TYPE \cross SYSTEM \_ TIME \_ TYPE \\
		\circchannel system \_ call \_ return : PARTITION \_ ID \_ TYPE \cross PROCESS \_ ID \_ TYPE \\
	\end{array}
\end{zed}

\subsection{Channels for communication betwween APEX Layer and File System}
\begin{zed}
	\begin{array}{ll}
		\circchannel apex \_ function \_ request \_ executable \_ file : PARTITION \_ ID \_ TYPE \cross PROCESS \_ ID \_ TYPE \cross \\
		\t1 \ \ \ \ \ \; PATH \_ TYPE \\
		\circchannel apex \_ function \_ get \_ executable \_ file : PARTITION \_ ID \_ TYPE \cross PROCESS \_ ID \_ TYPE \cross \\
		\t1 \ \ \ \ \ \; EXECUTABLE \_ FILE \_ TYPE \\
	\end{array}
\end{zed}

\subsection{Channels for Return Code}
\begin{zed}
	\begin{array}{ll}
		\circchannel apex \_ function \_ return \_ code : PARTITION \_ ID \_ TYPE \cross PROCESS \_ ID \_ TYPE \cross RETURN \_ CODE \_ \\
		\t1 \ \ \ \ \ \; TYPE
	\end{array}
\end{zed}

\subsection{Channels for communication betwween APEX Layer and Application of each Partition}
\begin{zed}
	\begin{array}{ll}
		\circchannel apex \_ function \_ get \_ partition \_ status : PARTITION \_ ID \_ TYPE \cross PROCESS \_ ID \_ TYPE \\
		\circchannel apex \_ function \_ get \_ partition \_ status \_ output : PARTITION \_ ID \_ TYPE \cross PROCESS \_ ID \_ TYPE \cross \\
		\t1 \ \ \ \ \ \; PARTITION \_ STATUS \_ TYPE \\
		\circchannel apex \_ function \_ get \_ partition \_ status \_ return : PARTITION \_ ID \_ TYPE \cross PROCESS \_ ID \_ TYPE \\
		
		\circchannel apex \_ function \_ set \_ operating \_ mode : PARTITION \_ ID \_ TYPE \cross PROCESS \_ ID \_ TYPE \\
		\circchannel apex \_ function \_ set \_ operating \_ mode \_ input : PARTITION \_ ID \_ TYPE \cross PROCESS \_ ID \_ TYPE \cross \\
		\t1 \ \ \ \ \ \; OPERATING \_ MODE \_ TYPE \\
		\circchannel apex \_ function \_ set \_ operating \_ mode \_ return : PARTITION \_ ID \_ TYPE \cross PROCESS \_ ID \_ TYPE \\
\end{array}
	\\
	
	\znewpage
	
	\begin{array}{ll}
		\circchannel apex \_ function \_ suspend \_ self : PARTITION \_ ID \_ TYPE \cross PROCESS \_ ID \_ TYPE \\
		\circchannel apex \_ function \_ suspend \_ self \_ input : PARTITION \_ ID \_ TYPE \cross PROCESS \_ ID \_ TYPE \cross \\
		\t1 \ \ \ \ \ \; SYSTEM \_ TIME \_ TYPE \\
		\circchannel apex \_ function \_ suspend \_ self \_ return : PARTITION \_ ID \_ TYPE \cross PROCESS \_ ID \_ TYPE \\
		
		\circchannel apex \_ function \_ suspend : PARTITION \_ ID \_ TYPE \cross PROCESS \_ ID \_ TYPE \\
		\circchannel apex \_ function \_ suspend \_ input : PARTITION \_ ID \_ TYPE \cross PROCESS \_ ID \_ TYPE \cross \\
		\t1 \ \ \ \ \ \; PROCESS \_ ID \_ TYPE \\
		\circchannel apex \_ function \_ suspend \_ return : PARTITION \_ ID \_ TYPE \cross PROCESS \_ ID \_ TYPE \\
\end{array}
\end{zed}

\section{Circus Processes} \label{section:circusprocess}
\subsection{Core Hardware Layer} \label{section:circusprocess_CoreHardware}
\subsubsection{Generic CPU} \label{circusprocess_GenericCPU}
\begin{zed}
	\circprocess GenericCPU \circdef \\
	\circbegin \\
	\circstate \ CPU \_ REGISTER \_ TYPE \\
	\\
	\begin{array}{lll}
		Initialize & \circdef & InitCPURegisters \\
		\\
	\end{array}
	\\
	\begin{array}{lll}
		SavePart & \circdef & \circvar cpureg : CPU \_ REGISTER \_ TYPE \circspot \\
							&& context \_ switching \_ save \_ part \_ cpu \_ registers \circthen SavePartCPURegister[cpureg/cpureg!] \circseq \\
							&& context \_ switching \_ saved \_ part \_ cpu \_ registers \ !cpureg \circthen \Skip \\
		\\
	\end{array}
	\\
	\begin{array}{lll}
		SaveAll & \circdef & \circvar cpureg : CPU \_ REGISTER \_ TYPE \circspot \\
						   && context \_ switching \_ save \_ all \_ cpu \_ registers \circthen SaveAllCPURegister[cpureg/cpureg!] \circseq \\
						   && context \_ switching \_ saved \_ all \_ cpu \_ registers \ !cpureg \circthen \Skip \\
		\\
	\end{array}
	\\
	\begin{array}{lll}
		RestorePart & \circdef & context \_ switching \_ restore \_ part \_ cpu \_ registers \ ?cpureg \circthen \\
							   && RestorePartCPURegister[cpureg/cpureg?] \circseq \\
							   && context \_ switching \_ restored \_ part \_ cpu \_ registers \circthen \Skip \\
		\\
	\end{array}
	\\
	\begin{array}{lll}
		RestoreAll & \circdef & context \_ switching \_ restore \_ all \_ cpu \_ registers \ ?cpureg \circthen \\
							  && RestoreAllCPURegister[cpureg/cpureg?] \circseq \\
							  && context \_ switching \_ restored \_ all \_ cpu \_ registers \circthen \Skip \\
		\\
	\end{array}
	\\
	\begin{array}{lll}
		SetStackSegment & \circdef & context \_ switching \_ set \_ stack \_ segment \_ registers \ ?ssreg \circthen SetStackSegRegister[ssreg/ssreg?] \circseq \\
								   && context \_ switching \_ have \_ set \_ stack \_ segment \_ registers \circthen \Skip \\
		\\
	\end{array}
	\\
	\begin{array}{lll}
		SetIFOn & \circdef & interrupt \_ controlling \_ enable \_ interrupt \circthen PSWRegisters.IF := 1 \circseq \\
						   && interrupt \_ controlling \_ enabled \_ interrupt \circthen \Skip \\
		\\
	\end{array}
	\\
	\begin{array}{lll}
		SetIFOff & \circdef & interrupt \_ controlling \_ disable \_ interrupt \circthen PSWRegisters.IF := 0 \circseq \\
							&& interrupt \_ controlling \_ disabled \_ interrupt \circthen \Skip \\
		\\
	\end{array}
	\\
	\begin{array}{lll}
		CPURegisterService & \circdef & SavePart \extchoice SaveAll \extchoice RestorePart \extchoice RestoreAll \extchoice SetIFOn \extchoice SetIFOff \\
		\\
	\end{array}
	\\
	\begin{array}{lll}
		InterruptResponse \_ Pre & \circdef & \lcircguard PSWRegisters.IF=1 \rcircguard \circguard interrupt \ ?i \circthen interrupt \_ acknowledged \circthen \\
											&& interrupt \_ handler \_ start \_ interrupt \_ handling \ !i \circthen \Skip \\
											&& \extchoice \lcircguard PSWRegisters.IF=0 \rcircguard \circguard \Skip \\
		\\
	\end{array}
	\\
	\begin{array}{lll}
		InterruptResponse \_ Pro & \circdef & interrupt \_ handler \_ finish \_ interrupt \_ handling \ ?i \circthen \\
											&& interrupt \_ handled \ !i \circthen \Skip \\
		\\
	\end{array}
	\\
	\begin{array}{lll}
		InterruptResponse & \circdef & InterruptResponse \_ Pre \extchoice InterruptResponse \_ Pro \\
		\\
	\end{array}
	\\
	\begin{array}{lll}
		ExceptionResponse \_ Pre & \circdef & exception \ ?e \circthen exception \_ handler \_ start \_ exception \_ handling \ !e \circthen \Skip \\
		\\
	\end{array}
	\\
	\begin{array}{lll}
		ExceptionResponse \_ Pro & \circdef & exception \_ handler \_ finish \_ exception \_ handling \circthen \Skip \\
		\\
	\end{array}
	\\
	\begin{array}{lll}
		ExceptionResponse & \circdef & ExceptionResponse \_ Pre \extchoice ExceptionResponse \_ Pro \\
		\\
	\end{array}
	\\
	
	\znewpage
	
	\begin{array}{lll}
		CPUService & \circdef & CPURegisterService \extchoice InterruptResponse \extchoice ExceptionResponse \\
		\\
	\end{array}
	\\
	\begin{array}{lll}
		ExecCPUService & \circdef & \circmu X \circspot CPUService \circseq X \\
		\\
	\end{array}
	\\
	\begin{array}{lll}
		\circspot module \_ init & \circthen & (Initialize \circseq ExecCPUService) \Delta module \_ end \circthen \Skip \\
		\\
	\end{array}
	\\	
	\circend
	\\
\end{zed}

\subsubsection{Memory Management Unit (MMU)} \label{circusprocess_MMU}
\begin{zed}
	\circprocess MemoryManagementUnit \circdef \\
	\circbegin \\
	\circstate \ MEMORY \_ MANAGEMENT \_ UNIT \_ TYPE \\
	\\
	\begin{array}{lll}
		Initialize & \circdef & PageTableBuffer := \emptyset \\
		\\
	\end{array}
	\\
	\begin{array}{lll}
		SetPageTableBuffer & \circdef & context \_ switching \_ set \_ page \_ table \ ?pt \circthen PageTableBuffer := pt \circseq \\
									  && context \_ switching \_ have \_ set \_ page \_ table \circthen \Skip \\
		\\
	\end{array}%
	
	\\
	\circend
	\\
\end{zed}

\subsubsection{Main Memory} \label{circusprocess_MainMemory}
\begin{zed}
	\circprocess MainMemory \circdef \\
	\circbegin \\
	\circstate \ MAIN \_ MEMORY \_ TYPE \\
	\\
	\begin{array}{lll}
		Initialize & \circdef & Store := \emptyset \\
		\\
	\end{array}
	\\
	\begin{array}{lll}
		Read & \circdef & cpu \_ receive \_ data \_ from \_ main \_ memory \ ?a \circthen \\
						&& cpu \_ received \_ data \_ from \_ main \_ memory \ !Store(a) \circthen \Skip \\
		\\
	\end{array}
	\\
	\begin{array}{lll}
		Write & \circdef & cpu \_ send \_ data \_ to \_ main \_ memory \ ?a \ ?data \circthen Store(a) := data \\
						 && cpu \_ sent \_ data \_ to \_ main \_ memory \circthen \Skip \\
		\\
	\end{array}
	\\
	\begin{array}{lll}
		Load & \circdef & memory \_ management \_ load \_ page \_ into \_ main \_ memory \ ?page \ ?piece \circthen \\
						&& LoadPage[page/page?,piece/piece?] \circseq \\
						&& memory \_ management \_ loaded \_ page \_ into \_ main \_ memory \circthen \Skip \\
		\\
	\end{array}
	\\
	\begin{array}{lll}
		Extract & \circdef & \circvar piece : PIECE \_ TYPE \circspot \\
						   && memory \_ management \_ extract \_ page \_ from \_ main \_ memory \ ?page \circthen \\
						   && ExtractPage[page/page?,piece/piece!] \circseq \\
						   && memory \_ management \_ extracted \_ page \_ from \_ main \_ memory \ !piece \circthen \Skip \\
		\\
	\end{array}
	\\
	\begin{array}{lll}
		MainMemoryService & \circdef & Read \extchoice Write \extchoice Load \extchoice Extract \\
		\\
	\end{array}
	\\
	\begin{array}{lll}
		ExecMainMemoryService & \circdef & \circmu X \circspot MainMemoryService \circseq X \\
		\\
	\end{array}
	\\
	\begin{array}{lll}
		\circspot module \_ init \circthen (Initialize \circseq ExecMainMemoryService) \Delta module \_ end \circthen \Skip \\
		\\
	\end{array}	
	\\
	\circend
	\\
\end{zed}

\subsubsection{Real-Time Clock} \label{circusprocess_RTClock}
\begin{zed}
	\circprocess RTClock \circdef \\
	\circbegin \\
	\circstate \ RTCLOCK \_ TYPE \\
	\\
	\begin{array}{lll}
		Initialize & \circdef & Value := 0_{PSU} \\
		\\
	\end{array}
	\\
	\begin{array}{lll}
		RunClockPause & \circdef & \circmu X \circspot (\textbf{wait}(1) \circseq Value := Value + 1_{PSU} \circseq clock \_ pause) \circseq X \\
		\\
	\end{array}
	\\
	\begin{array}{lll}
		RTClockService & \circdef & system \_ clock \_ request \_ current \_ time \circthen \\
								  && system \_ clock \_ get \_ current \_ time \ !Value \circthen \Skip \\
		\\
	\end{array}
	\\
	\begin{array}{lll}
		ExecRTClockService & \circdef & \circmu X \circspot RTClockService \circseq X \\
		\\
	\end{array}
	\\
	\begin{array}{lll}
		\circspot rtclock \_ power \_ on \circthen (Initialize \circseq RunClockPause \interleave ExecRTClockService) \Delta rtclock \_ power \_ off \circthen \Skip \\
		\\
	\end{array}
	\\		
	\circend
	\\
\end{zed}

\subsubsection{Timer} \label{circusprocess_Timer}
\begin{zed}
	\circprocess Timer \circdef \\
	\circbegin \\
	\circstate \ TIMER \_ TYPE \\
	\\
	\begin{array}{lll}
		Initialize & \circdef & InitTimer[ClockTickInterval/p?] \\
		\\
	\end{array}%
	
	\\
	\circend
	\\
\end{zed}

\subsubsection{High-Precision Timer} \label{circusprocess_HPTimer}
\begin{zed}
	\circprocess HPTimer \circdef \\
	\circbegin \\
	\circstate \ HPTIMER \_ TYPE \\
	\\
	\begin{array}{lll}
		Initialize & \circdef & InitHPTimer \\
		\\
	\end{array}%
	
	\\
	\circend
	\\
\end{zed}

\subsubsection{ChannelSets for communication within Interrupt Controller}
\begin{zed}
	\begin{array}{lll}
		\circchannelset IRR \_ PR & == & \lchanset irregister \_ report \_ new \_ interrupt \rchanset \\
		\circchannelset PR \_ ISR & == & \lchanset priority \_ resolver \_ report \_ new \_ interrupt \rchanset \\
		\circchannelset ISR \_ IRR & == & \lchanset isregister \_ service \_ new \_ interrupt,isregister \_ demand \_ new \_ interrupt \rchanset \\
		\\
	\end{array}
	\\
	\begin{array}{lll}
		\circchannelset ICChannels & == & IRR \_ PR \cup PR \_ ISR \cup ISR \_ IRR \\
		\\
	\end{array}
	\\
\end{zed}

\subsubsection{Interrupt Controller} \label{circusprocess_InterruptController}
\begin{zed}
	\circprocess InterruptController \circdef \\
	\circbegin \\
	\circstate \ INTERRUPT \_ HANDLER \_ STATE \_ TYPE \\
	\\
	\begin{array}{lll}
		Initialize & \circdef & InitInterruptController \\
		\\
	\end{array}%
	\\
	\circend
	\\
\end{zed}

\subsubsection{ChannelSets for communication between different parts of Core Hardware Layer}
\begin{zed}
	\begin{array}{lll}
		\circchannelset CPU \_ IC & == & \lchanset interrupt,interrupt \_ acknowledged,interrupt \_ handled \rchanset \\
		\circchannelset CPU \_ MMU & == & \lchanset cpu \_ request \_ logical \_ address \_ translation, \\
										&& \ \: ~ logical \_ address \_ translation \_ fail,exception, \\
										&& \ \: ~ cpu \_ get \_ logical \_ address \_ translation \rchanset \\
		\circchannelset CPU \_ MEMORY & == & \lchanset cpu \_ receive \_ data \_ from \_ main \_ memory, \\
										   && \ \: ~ cpu \_ received \_ data \_ from \_ main \_ memory, \\
										   && \ \: ~ cpu \_ send \_ data \_ to \_ main \_ memory, \\
										   && \ \: ~ cpu \_ sent \_ data \_ to \_ main \_ memory \rchanset \\
		\circchannelset RTC \_ TIMER & == & \lchanset clock \_ pause \rchanset \\
		\circchannelset RTC \_ HPTIMER & == & \lchanset clock \_ pause \rchanset \\
		\circchannelset TIMER \_ IC & == & \lchanset hardware \_ interrupts \rchanset \\
		\circchannelset HPTIMER \_ IC & == & \lchanset hardware \_ interrupts \rchanset \\
		\\
	\end{array}
	\\
	\begin{array}{lll}
		\circchannelset CHLayerChannels & == & CPU \_ IC \cup CPU \_ MMU \cup CPU \_ MEMORY \\
											 && \cup RTC \_ TIMER \cup RTC \_ HPTIMER \\
											 && \cup TIMER \_ IC \cup HPTIMER \_ IC \\
		\\
	\end{array}
	\\
\end{zed}

\subsubsection{Core Hardware Layer Model} \label{circusprocess_CoreHardwareLayer}
\begin{zed}
	\circprocess Core \_ Hardware \_ Layer \circdef \\
	\\
	\begin{array}{lll}
		CPU \lpar CHLayerChannels \rpar MemoryManagementUnit \lpar CHLayerChannels \rpar MainMemory \\
		\lpar CHLayerChannels \rpar RTClock \lpar CHLayerChannels \rpar Timer \lpar CHLayerChannels \rpar HPTimer \\
		\lpar CHLayerChannels \rpar InterruptController \\
		\\
	\end{array}
\end{zed}

\subsection{Hardware Interface Software Layer} \label{section:circusprocess_HardwareInterfaceSoftware}
\subsubsection{Context Switching} \label{circusprocess_ContextSwitching}
\begin{zed}
	\circprocess ContextSwitching \circdef \\
	\circbegin \\
	\\
	\begin{array}{lll}
		SwitchOut \_ IntHandler & \circdef & interrupt \_ handler \_ save \_ context \circthen \\
										   && context \_ switching \_ save \_ all \_ cpu \_ registers \circthen \\
										   && context \_ switching \_ saved \_ all \_ cpu \_ registers \ ?cpureg \circthen \\
										   && context \_ switching \_ request \_ page \_ table \circthen \\
										   && context \_ switching \_ get \_ page \_ table \ ?pt \circthen \\
										   && interrupt \_ handler \_ saved \_ context \ !cpureg \ !pt \circthen \Skip \\
		\\
	\end{array}
	\\
	\begin{array}{lll}
		SwitchIn \_ IntHandler & \circdef & interrupt \_ handler \_ restore \_ context \ ?cpureg \ ?pt \circthen \\
										  && context \_ switching \_ restore \_ all \_ cpu \_ registers \ !cpureg  \circthen \\
										  && context \_ switching \_ restored \_ all \_ cpu \_ registers \circthen \\
										  && context \_ switching \_ set \_ page \_ table \ !pt \circthen \\
										  && context \_ switching \_ have \_ set \_ page \_ table \circthen \\
										  && interrupt \_ handler \_ restored \_ context \circthen \Skip \\
		\\
	\end{array}
	\\
	\begin{array}{lll}
		SwitchStack \_ IntHandler & \circdef & interrupt \_ handler \_ switch \_ stack \ ?ssreg \ ?pt \circthen \\
											 && context \_ switching \_ set \_ stack \_ segment \_ registers \ !ssreg \circthen \\
											 && context \_ switching \_ have \_ set \_ stack \_ segment \_ registers \circthen \\
											 && context \_ switching \_ set \_ page \_ table \ !pt \circthen \\
											 && context \_ switching \_ have \_ set \_ page \_ table \circthen \\
											 && interrupt \_ handler \_ switched \_ stack \circthen \Skip \\
		\\
	\end{array}
	\\
	\begin{array}{lll}
		ContextSwitching \_ IntHandler & \circdef & SwitchOut \_ IntHandler \extchoice SwitchIn \_ IntHandler \extchoice SwitchStack \_ IntHandler \\
		\\
	\end{array}%
	\\
	\circend
	\\
\end{zed}

\subsubsection{Memory Management} \label{circusprocess_MemoryManagement}
\begin{zed}
	\circprocess MemoryManagement \circdef \\
	\circbegin \\
	\circstate \ MEMORY \_ MANAGEMENT \_ STATE \_ TYPE \\
	\\
	\begin{array}{lll}
		Initialize & \circdef & kernel \_ init \_ memory \_ management \ ?areamm \circthen InitMemoryManagement[areamm/areamm?] \circseq \\
							  && kernel \_ inited \_ memory \_ management \circthen \Skip \\
		\\
	\end{array}
	\\
	\begin{array}{lll}
		SetAreaMemoryManagementRef & \circdef & kernel \_ set \_ area \_ memory \_ management \_ reference.kernelid \ ?areamm \circthen \\
											  && AreaMemoryManagementRef := areamm \circseq \\
											  && kernel \_ have \_ set \_ area \_ memory \_ management \_ reference.kernelid \circthen \Skip \\
		\\
	\end{array}
	\\
	\begin{array}{lll}
		GetAreaMemoryManagementRef & \circdef & kernel \_ request \_ area \_ memory \_ management \_ reference.kernelid \circthen \\
											  && kernel \_ get \_ area \_ memory \_ management \_ reference.kernelid \\
											  && !AreaMemoryManagementRef \circthen \Skip \\
		\\
	\end{array}
	\\
	\begin{array}{lll}
		GetAreaFreePages & \circdef & kernel \_ request \_ area \_ free \_ pages.kernelid \circthen  \\
									&& kernel \_ get \_ area \_ free \_ pages.kernelid \ !\#(AreaMemoryManagementRef.Free) \circthen \Skip \\
		\\
	\end{array}
	\\
	\begin{array}{lll}
		GetAreaFreeBlock & \circdef & \circvar maxblock : MEMORY \_ BLOCK \_ TYPE \circspot \\
									&& kernel \_ request \_ area \_ free \_ block.kernelid \circthen GetMaxFreeBlock[maxblock/maxblock!] \circseq \\
									&& kernel \_ get \_ area \_ free \_ block.kernelid \ !(maxblock.Size) \circthen \Skip \\
		\\
	\end{array}
	\\
	\begin{array}{lll}
		AllocAreaFreePage & \circdef & \circvar pages : \power PAGE \_ TYPE \circspot \\
									 && kernel \_ allocate \_ area \_ free \_ pages.kernelid \ ?n \circthen \\
									 && AllocPhysicalMemoryPage[n/n?,pages/pages!] \circseq \\
									 && kernel \_ allocated \_ area \_ free \_ pages.kernelid \ !pages \circthen \Skip \\
		\\
	\end{array}
	\\
	\begin{array}{lll}
		AllocAreaFreeBlock & \circdef & \circvar block : MEMORY \_ BLOCK \_ TYPE \circspot \\
									  && kernel \_ allocate \_ area \_ free \_ block.kernelid \ ?size \circthen \\
									  && AllocPhysicalMemoryBlock[size/size?,block/block!] \circseq \\
									  && kernel \_ allocated \_ area \_ free \_ block.kernelid \ !block \circthen \Skip \\
		\\
	\end{array}
	\\
	\begin{array}{lll}
		AllocAreaFreeSpace & \circdef & AllocAreaFreePage \extchoice AllocAreaFreeBlock \\
		\\
	\end{array}
	\\
	\begin{array}{lll}
		DeallocAreaUsedPage & \circdef & kernel \_ deallocate \_ area \_ used \_ pages.kernelid \ ?pages \circthen \\
									   && DeallocPhysicalMemoryPage[pages/pages?] \circseq \\
									   && kernel \_ deallocated \_ area \_ used \_ pages.kernelid \circthen \Skip \\
		\\
	\end{array}
	\\
	\begin{array}{lll}
		DeallocAreaUsedBlock & \circdef & kernel \_ deallocate \_ area \_ used \_ block.kernelid \ ?block \circthen \\
										&& DeallocPhysicalMemoryPage[\mathit{totalpage}(block)/pages?] \circseq \\
										&& kernel \_ deallocated \_ area \_ used \_ block.kernelid \circthen \Skip \\
		\\
	\end{array}
	\\
	\begin{array}{lll}
		DeallocAreaUsedSpace & \circdef & DeallocAreaUsedPage \extchoice DeallocAreaUsedBlock \\
		\\
	\end{array}
	\\
	\begin{array}{lll}
		CreateVirtualMemoryBlock & \circdef & \circvar block : MEMORY \_ BLOCK \_ TYPE \circspot \\
											&& kernel \_ create \_ virtual \_ memory \_ block.kernelid \ ?start \ ?size \circthen \\
											&& CreateMemoryBlock[start/start?,size/size?,block/block!] \circseq \\
											&& kernel \_ created \_ virtual \_ memory \_ block.kernelid \ !block \circthen \Skip \\
		\\
	\end{array}%
	\\
	\circend
	\\
\end{zed}

\subsubsection{System Clock} \label{circusprocess_SystemClock}
\begin{zed}
	\circprocess SystemClock \circdef \\
	\circbegin \\
	\circstate SYSTEM \_ CLOCK \_ STATE \_ TYPE \\
	\\
	\begin{array}{lll}
		Initialize & \circdef & kernel \_ init \_ system \_ clock \circthen \\
							  && system \_ clock \_ request \_ current \_ time \circthen system \_ clock \_ get \_ current \_ time \ ?bt \circthen \\
							  && InitSystemClock[\mathit{bintosigned}(bt)/bt?,\mathit{bintosigned}(ClockTickInterval)/ti?] \circseq \\
							  && kernel \_ inited \_ system \_ clock \circthen \Skip \\
		\\
	\end{array}
	\\
	\begin{array}{lll}
		GetTime & \circdef & kernel \_ request \_ system \_ time.kernelid \circthen \\
						   && kernel \_ get \_ system \_ time.kernelid \ !Time \circthen \Skip \\
		\\
	\end{array}
	\\
	\begin{array}{lll}
		GetHPTime & \circdef & kernel \_ request \_ high \_ precision \_ time.kernelid \circthen \\
							 && system \_ clock \_ request \_ high \_ precision \_ time \circthen \\
							 && system \_ clock \_ get \_ high \_ precision \_ time \ ?hpt \circthen \\
							 && kernel \_ get \_ high \_ precision \_ time.kernelid \ !\mathit{bintosigned}(hpt) \circthen \Skip \\
		\\
	\end{array}
	\\
	\begin{array}{lll}
		ResetHPTimer & \circdef & kernel \_ reset \_ high \_ precision \_ timer.kernelid \circthen \\
								&& system \_ clock \_ reset \_ high \_ precision \_ timer \circthen \\
								&& system \_ clock \_ have \_ reset \_ high \_ precision \_ timer \circthen \\
								&& kernel \_ have \_ reset \_ high \_ precision \_ timer.kernelid \circthen \Skip \\
		\\
	\end{array}
	\\
	\begin{array}{lll}
		SetHPAlarm & \circdef & kernel \_ set \_ high \_ precision \_ timer \_ alarm.kernelid \ ?a \circthen \\
							  && system \_ clock \_ set \_ high \_ precision \_ timer \_ alarm \ !\mathit{signedtobin}(a) \circthen \\
							  && system \_ clock \_ have \_ set \_ high \_ precision \_ timer \_ alarm \circthen \\
							  && kernel \_ have \_ set \_ high \_ precision \_ timer \_ alarm.kernelid \circthen \Skip \\
		\\
	\end{array}
	\\
	\begin{array}{lll}
		EndHPTimer & \circdef & kernel \_ end \_ high \_ precision \_ timer.kernelid \circthen \\
							  && system \_ clock \_ end \_ high \_ precision \_ timer \circthen \\
							  && system \_ clock \_ ended \_ high \_ precision \_ timer \circthen \\
							  && kernel \_ ended \_ high \_ precision \_ timer.kernelid \circthen \Skip \\
		\\
	\end{array}
	\\
	\begin{array}{lll}
		UpdateTime & \circdef & isr \_ clocktick \_ update \_ system \_ time \circthen UpdateSystemClock \circseq \\
							  && isr \_ clocktick \_ updated \_ system \_ time \circthen \Skip \\
		\\
	\end{array}
	\\
	\begin{array}{lll}
		UpdateInterruptTime & \circdef & isr \_ clocktick \_ request \_ high \_ precision \_ time \circthen \\
									   && system \_ clock \_ request \_ high \_ precision \_ time \circthen \\
									   && system \_ clock \_ get \_ high \_ precision \_ time \ ?hpt \circthen InterruptTime := \mathit{bintosigned}(hpt) \circseq \\
									   && isr \_ clocktick \_ get \_ high \_ precision \_ time \ !hpt \circthen \Skip \\
		\\
	\end{array}
	\\
	\begin{array}{lll}
		GetInterruptTime & \circdef & kernel \_ request \_ interrupt \_ time.kernelid \circthen \\
									&& kernel \_ get \_ interrupt \_ time.kernelid \ !InterruptTime \circthen \Skip \\
		\\
	\end{array}
	\\
	\begin{array}{lll}
		SystemClockService & \circdef & GetTime \\
									  && \extchoice GetHPTime \\
									  && \extchoice ResetHPTimer \\
									  && \extchoice SetHPAlarm \\
									  && \extchoice EndHPTimer \\
									  && \extchoice UpdateTime \\
									  && \extchoice UpdateInterruptTime \\
									  && \extchoice GetInterruptTime \\
		\\
	\end{array}
	\\
	\begin{array}{lll}
		ExecSystemClockService & \circdef & \circmu X \circspot SystemClockService \circseq X \\
		\\
	\end{array}
	\\
	
	\znewpage
	
	\begin{array}{lll}
		\spot module \_ init \circthen (Initialize \circseq ExecSystemClockService) \Delta module \_ end \circthen \Skip \\
		\\
	\end{array}
	\\	
	\circend
\end{zed}

\subsubsection{Interrupt Controlling} \label{circusprocess_InterruptControlling}
\begin{zed}
	\circprocess InterruptControlling \circdef \circspot \\
	\circbegin \\
	\\
	\begin{array}{lll}
		Mask \_ IntHandler & \circdef & interrupt \_ handler \_ mask \_ interrupt \ ?i \circthen \\
									  && interrupt \_ controlling \_ mask \_ interrupt \ !i \circthen \\
									  && interrupt \_ controlling \_ masked \_ interrupt \circthen \\
									  && interrupt \_ handler \_ masked \_ interrupt \circthen \Skip \\
		\\
	\end{array}
	\\
	\begin{array}{lll}
		Unmask \_ IntHandler & \circdef & interrupt \_ handler \_ unmask \_ interrupt \ ?i \circthen \\
										&& interrupt \_ controlling \_ unmask \_ interrupt \ !i \circthen \\
										&& interrupt \_ controlling \_ unmasked \_ interrupt \circthen \\
										&& interrupt \_ handler \_ unmasked \_ interrupt \circthen \Skip \\
		\\
	\end{array}
	\\
	\begin{array}{lll}
		InterruptMaskService \_ IntHandler & \circdef & Mask \_ IntHandler \extchoice Unmask \_ IntHandler \\
		\\
	\end{array}
	\\
	\begin{array}{lll}
		Mask \_ Kernel & \circdef & kernel \_ mask \_ interrupt.kernelid \ ?i \circthen interrupt \_ controlling \_ mask \_ interrupt \ !i \circthen \\
								  && interrupt \_ controlling \_ masked \_ interrupt \circthen kernel \_ masked \_ interrupt.kernelid \circthen \Skip \\
		\\
	\end{array}
	\\
	\begin{array}{lll}
		Unmask \_ Kernel & \circdef & kernel \_ unmask \_ interrupt.kernelid \ ?i \circthen interrupt \_ controlling \_ unmask \_ interrupt \ !i \circthen \\
						  			&& interrupt \_ controlling \_ unmasked \_ interrupt \circthen kernel \_ unmasked \_ interrupt.kernelid \circthen \Skip \\
		\\
	\end{array}
	\\
	\begin{array}{lll}
		InterruptMaskService \_ Kernel & \circdef & Mask \_ Kernel \extchoice Unmask \_ Kernel \\
		\\
	\end{array}
	\\
	\begin{array}{lll}
		InterruptMaskService & \circdef & InterruptMaskService \_ IntHandler \extchoice InterruptMaskService \_ Kernel \\
		\\
	\end{array}
	\\
	\begin{array}{lll}
		ExecInterruptMaskService & \circdef & \circmu X \circspot InterruptMaskService \circseq X \\
		\\
	\end{array}
	\\
	\begin{array}{lll}
		Enable \_ IntHandler & \circdef & interrupt \_ handler \_ enable \_ interrupt \circthen \\
										&& interrupt \_ controlling \_ enable \_ interrupt \circthen \\
										&& interrupt \_ controlling \_ enabled \_ interrupt \circthen \\
										&& interrupt \_ handler \_ enabled \_ interrupt \circthen \Skip \\
		\\
	\end{array}
	\\
	\begin{array}{lll}
		Disable \_ IntHandler & \circdef & interrupt \_ handler \_ disable \_ interrupt \circthen \\
										 && interrupt \_ controlling \_ disable \_ interrupt \circthen \\
										 && interrupt \_ controlling \_ disabled \_ interrupt \circthen \\
										 && interrupt \_ handler \_ disabled \_ interrupt \circthen \Skip \\
		\\
	\end{array}
	\\
	\begin{array}{lll}
		InterruptEnableService \_ IntHandler & \circdef & Enable \_ IntHandler \extchoice Disable \_ IntHandler \\
		\\
	\end{array}
	\\
	\begin{array}{lll}
		Enable \_ Kernel & \circdef & kernel \_ enable \_ interrupt.kernelid \circthen interrupt \_ controlling \_ enable \_ interrupt \circthen \\
						  			&& interrupt \_ controlling \_ enabled \_ interrupt \circthen kernel \_ enabled \_ interrupt.kernelid \circthen \Skip \\
		\\
	\end{array}
	\\
	\begin{array}{lll}
		Disable \_ Kernel & \circdef & kernel \_ disable \_ interrupt.kernelid \circthen interrupt \_ controlling \_ disable \_ interrupt \circthen \\
						   			 && interrupt \_ controlling \_ disabled \_ interrupt \circthen kernel \_ disabled \_ interrupt.kernelid \circthen \Skip \\
		\\
	\end{array}
	\\
	\begin{array}{lll}
		InterruptEnableService \_ Kernel & \circdef & Enable \_ Kernel \extchoice Disable \_ Kernel \\
		\\
	\end{array}
	\\
	\begin{array}{lll}
		InterruptEnableService & \circdef & InterruptEnableService \_ IntHandler \extchoice InterruptEnableService \_ Kernel \\
		\\
	\end{array}
	\\
	\begin{array}{lll}
		ExecInterruptEnableService & \circdef & \circmu X \circspot InterruptEnableService \circseq X \\
		\\
	\end{array}
	\\
	\begin{array}{lll}
		ExecInterruptControlling & \circdef ExecInterruptMaskService \interleave ExecInterruptEnableService \\
		\\
	\end{array}
	\\
	\begin{array}{lll}
		\spot module \_ init \circthen ExecInterruptControlling \Delta module \_ end \circthen \Skip \\
		\\
	\end{array}
	\\	
	\circend
\end{zed}

\subsubsection{ChannelSets for communication within Interrupt Handler}
\begin{zed}
	\begin{array}{lll}
		\circchannelset SelectISR & == & \lchanset interrupt \_ handler \_ invoke \_ service \_ routine \_ clocktick \rchanset \\
		\circchannelset ISRReturn & == & \lchanset interrupt \_ service \_ routine \_ end \rchanset \\
		\\
	\end{array}
	\\
\end{zed}

\subsubsection{Interrupt Handler} \label{circusprocess_InterruptHandler}
\begin{zed}
	\circprocess InterruptHandler \circdef IntHandlerConfig : INTERRUPT \_ HANDLER \_ CONFIG \_ TYPE \circspot \\
	\circbegin \\
	\circstate \ INTERRUPT \_ HANDLER \_ STATE \_ TYPE \\
	\\
	\begin{array}{lll}
		Initialize & \circdef & kernel \_ init \_ interrupt \_ handler \ ?pagetable \circthen \\
							  && interrupt \_ handler \_ request \_ interrupt \_ stack \ !(IntHandlerConfig.InterruptStackSize) \circthen \\
							  && interrupt \_ handler \_ get \_ interrupt \_ stack \ ?intstack \circthen \\
							  && InitInterruptHandler[intstack/intstack?,pagetable/pagetable?] \circseq \\
							  && kernel \_ inited \_ interrupt \_ handler \circthen \Skip \\
		\\
	\end{array}%
	\\
	\circend
	\\
\end{zed}

\subsubsection{ChannelSets for communication between different parts of Hardware Interface Software Layer}
\begin{zed}
	\begin{array}{lll}
		\circchannelset IH \_ CS & == & \lchanset interrupt \_ handler \_ save \_ context,interrupt \_ handler \_ saved \_ context, \\
									  && \ \: ~ interrupt \_ handler \_ restore \_ context,interrupt \_ handler \_ restored \_ context, \\
									  && \ \: ~ interrupt \_ handler \_ switch \_ stack,interrupt \_ handler \_ switched \_ stack \rchanset \\
		\circchannelset IH \_ CLOCK & == & \lchanset isr \_ clocktick \_ update \_ system \_ time,isr \_ clocktick \_ updated \_ system \_ time, \\
										 && \ \: ~ isr \_ clocktick \_ request \_ high \_ precision \_ time,isr \_ clocktick \_ get \_ high \_ precision \_ time \rchanset \\
		\circchannelset IH \_ IC & == & \lchanset interrupt \_ handler \_ mask \_ interrupt,interrupt \_ handler \_ masked \_ interrupt, \\
									  && \ \: ~ interrupt \_ handler \_ unmask \_ interrupt,interrupt \_ handler \_ unmasked \_ interrupt, \\
									  && \ \: ~ interrupt \_ handler \_ enable \_ interrupt,interrupt \_ handler \_ enabled \_ interrupt, \\
									  && \ \: ~ interrupt \_ handler \_ disable \_ interrupt,interrupt \_ handler \_ disabled \_ interrupt \rchanset \\
		\\
	\end{array}
	\\
	\begin{array}{lll}
		\circchannelset HISLayerChannels & == & IH \_ CS \cup IH \_ CLOCK \cup IH \_ IC \\
		\\
	\end{array}
	\\
\end{zed}

\subsubsection{Hardware Interface Software Layer Model} \label{circusprocess_HardwareInterfaceSoftwareLayer}
\begin{zed}
	\circprocess Hardware \_ Interface \_ Software \_ Layer \circdef IntHandlerConfig : INTERRUPT \_ HANDLER \_ CONFIG \_ TYPE \circspot \\
	\\
	\begin{array}{lll}
		(ContextSwitching \lpar HISLayerChannels \rpar SystemClock \lpar HISLayerChannels \rpar InterruptControlling \\
		\lpar HISLayerChannels \rpar InterruptHandler(IntHandlerConfig)) \interleave MemoryManagement \\
		\\
	\end{array}
	\\
\end{zed}

\subsection{ARINC 653 O/S Layer} \label{section:circusprocess_OS}
\subsubsection{Core Kernel} \label{circusprocess_CoreKernel}
\begin{zed}
	\circprocess CoreKernel \circdef OSKernelConfig : OS \_ KERNEL \_ CONFIG \_ TYPE \circspot \\
	\circbegin \\
	\circstate CORE \_ KERNEL \_ STATE \_ TYPE \\
	\\
	\begin{array}{lll}
		Initialize \_ CoreKernel & \circdef & InitCoreKernel[OSKernelConfig/config?] \\
		\\
	\end{array}
	\\
	\begin{array}{lll}
		Initialize \_ PartitionKernels & \circdef & \Interleave partid : \dom (PartitionManagement.PartitionTable) \circspot \\
												  && core \_ kernel \_ init \_ partition \_ kernel.partid \ !KernelPageTable \circthen \\
												  && core \_ kernel \_ inited \_ partition \_ kernel.partid \circthen \Skip \\
		\\
	\end{array}
	\\
	\begin{array}{lll}
		Initialize \_ MemoryManagement & \circdef & kernel \_ init \_ memory \_ management \ !MemoryManagement \circthen \\
												  && kernel \_ inited \_ memory \_ management \circthen \Skip \\
		\\
	\end{array}
	\\
	\begin{array}{lll}
		Initialize \_ InterruptHandler & \circdef & kernel \_ init \_ interrupt \_ handler \ !KernelPageTable \circthen \\
												  && interrupt \_ handler \_ request \_ interrupt \_ stack \ ?size \circthen \\
												  && kernel \_ allocate \_ area \_ free \_ block.CORE \_ KERNEL \ !size \circthen \\
												  && kernel \_ allocated \_ area \_ free \_ block.CORE \_ KERNEL \ ?block \circthen \\
												  && kernel \_ create \_ virtual \_ memory \_ block.CORE \_ KERNEL \\
												  && !(KernelPageTable \inv (block.Start)) \ !size \circthen \\
												  && kernel \_ created \_ virtual \_ memory \_ block.CORE \_ KERNEL \ ?intstack \circthen \\
												  && interrupt \_ handler \_ get \_ interrupt \_ stack \ !intstack \circthen \\
												  && kernel \_ inited \_ interrupt \_ handler \circthen \Skip \\
		\\
	\end{array}
	\\
	\begin{array}{lll}
		Initialize \_ SystemClock & \circdef & kernel \_ init \_ system \_ clock \circthen kernel \_ inited \_ system \_ clock \circthen \Skip \\
		\\
	\end{array}
	\\
	\begin{array}{lll}
		Initialize & \circdef & Initialize \_ CoreKernel \circseq Initialize \_ PartitionKernels \circseq Initialize \_ MemoryManagement \circseq \\
							  && Initialize \_ InterruptHandler \circseq Initialize \_ SystemClock \\
		\\
	\end{array}
	\\
	\begin{array}{lll}
		StartPartitionKernel & \circdef & \Interleave partid : \dom (PartitionManagement.PartitionTable) \circspot \\
								  		&& PartitionManagement.PartitionTable(partid).OperatingMode = COLD \_ START \circseq \\
								   		&& PartitionManagement.PartitionTable(partid).LockLevel = MAX \_ LOCK \_ LEVEL \\
		\\
	\end{array}
	\\
	\begin{array}{lll}
		StartCoreKernel & \circdef & KernelProcManagement.CurrentProcess := IDLE \_ PROCESS \_ ID \circseq \\
								   && kernel \_ restore \_ context.CORE \_ KERNEL \ !ALL \ !(first ~ (head ~ KernelProcManagement. \\
								   && ProcessTable(KernelProcManagement.CurrentProcess).KernelTempContext)) \\
								   && !(second ~ (head ~ KernelProcManagement.ProcessTable(KernelProcManagement. \\
								   && CurrentProcess).KernelTempContext)) \circthen \\
								   && kernel \_ restored \_ context.CORE \_ KERNEL \circthen \\
								   && KernelProcManagement.ProcessTable(KernelProcManagement.CurrentProcess). \\
								   && KernelTempContext := \\
								   && tail ~ KernelProcManagement.ProcessTable(KernelProcManagement.CurrentProcess). \\
								   && KernelTempContext \circseq \\
								   && kernel \_ enable \_ interrupt.CORE \_ KERNEL \circthen \\
								   && kernel \_ enabled \_ interrupt.CORE \_ KERNEL \circthen \\
								   && KernelProcManagement.ProcessTable(KernelProcManagement.CurrentProcess). \\
								   && ProcessState := RUNNING \\
		\\
	\end{array}
	\\
	\begin{array}{lll}
		Start & \circdef & StartPartitionKernel \circseq StartCoreKernel \\
		\\
	\end{array}
	\\
	
	\znewpage
	
	\begin{array}{lll}
		Lock & \circdef & kernel \_ disable \_ interrupt.CORE \_ KERNEL \circthen kernel \_ disabled \_ interrupt.CORE \_ KERNEL \circthen \Skip \\
		\\
	\end{array}
	\\
	\begin{array}{lll}
		Unlock & \circdef & kernel \_ enable \_ interrupt.CORE \_ KERNEL \circthen kernel \_ enabled \_ interrupt.CORE \_ KERNEL \circthen \Skip \\
		\\
	\end{array}
	\\
	\begin{array}{lll}
		Core \_ GetOperatingMode & \circdef & \circvar om : OPERATING \_ MODE \_ TYPE \circspot \\
											&& core \_ kernel \_ service \_ get \_ operating \_ mode.partid \circthen \\
											&& om := PartitionManagement.PartitionTable(partid).OperatingMode \circseq \\
											&& core \_ kernel \_ service \_ get \_ operating \_ mode \_ output.partid \ !om \circthen \\
											&& core \_ kernel \_ service \_ get \_ operating \_ mode \_ return.partid \circthen \Skip \\
		\\
	\end{array}
	\\
	\begin{array}{lll}
		Core \_ GetPartitionStatus & \circdef & \circvar parts : PARTITION \_ STATUS \_ TYPE \circspot \\
											  && core \_ kernel \_ service \_ get \_ partition \_ status.partid \circthen \\
											  && GetPartitionStatus[parts/parts!] \circseq \\
											  && core \_ kernel \_ service \_ get \_ partition \_ status \_ output.partid \ !parts \circthen \\
											  && core \_ kernel \_ service \_ get \_ partition \_ status \_ return.partid \circthen \Skip \\
		\\
	\end{array}%
	\\
	\circend
\end{zed}

\subsubsection{ChannelSets for communication within Partition Kernel}
\begin{zed}
	\begin{array}{lll}
		\circchannelset SelectSysCall & == & \lchanset system \_ call \_ handler \_ invoke \_ service \_ routine \_ get \_ operating \_ mode, \\
										   && \ \: ~ system \_ call \_ handler \_ invoke \_ service \_ routine \_ get \_ partition \_ status, \\
										   && \ \: ~ system \_ call \_ handler \_ invoke \_ service \_ routine \_ set \_ operating \_ mode, \\
										   && \ \: ~ system \_ call \_ handler \_ invoke \_ service \_ routine \_ get \_ process \_ names, \\
										   && \ \: ~ system \_ call \_ handler \_ invoke \_ service \_ routine \_ get \_ process \_ ids, \\
										   && \ \: ~ system \_ call \_ handler \_ invoke \_ service \_ routine \_ get \_ free \_ process \_ ids, \\
										   && \ \: ~ system \_ call \_ handler \_ invoke \_ service \_ routine \_ get \_ free \_ space, \\
										   && \ \: ~ system \_ call \_ handler \_ invoke \_ service \_ routine \_ get \_ partition \_ period, \\
\end{array}
	\\
\end{zed}

\subsubsection{Partition Kernel} \label{circusprocess_PartitionKernel}
\begin{zed}
	\circprocess PartitionKernel \circdef PartitionId : PARTITION \_ ID \_ TYPE; \\
	\t5 \ \ \ \ \ \ ~ PartitionConfig : PARTITION \_ KERNEL \_ CONFIG \_ TYPE \circspot \\
	\circbegin \\
	\circstate PARTITION \_ KERNEL \_ STATE \_ TYPE \\
	\\
	%
									  				   \right) \\
									  		 \circelse & (procid \notin \ran WaitingQueue \land procid \in tprocids) \circthen \\
									  				   & ProcessTable(procid).ProcessState := DORMANT \circseq \\
									  				   & core \_ kernel \_ service \_ get \_ operating \_ mode.PartitionId \circthen \\
									  				   & core \_ kernel \_ service \_ get \_ operating \_ mode \_ return.PartitionId \circthen \\
									  				   & core \_ kernel \_ service \_ get \_ operating \_ mode \_ output.PartitionId \\
									  				   & ?om \circthen RemoveTimeCounter(procid,om) \\
									  		 \circelse & (procid \in \ran WaitingQueue \land procid \notin tprocids) \circthen \\
									  				   & DeqWaitingQueue[procid/procid?] \circseq \\
									  				   & ProcessTable(procid).ProcessState := DORMANT \\
									  		 \circelse & (procid \in \ran WaitingQueue \land procid \in tprocids) \circthen \\
									  				   & DeqWaitingQueue[procid/procid?] \circseq \\
									  				   & ProcessTable(procid).ProcessState := DORMANT \circseq \\
												  	   & core \_ kernel \_ service \_ get \_ operating \_ mode.PartitionId \circthen \\
												  	   & core \_ kernel \_ service \_ get \_ operating \_ mode \_ return.PartitionId \circthen \\
												  	   & core \_ kernel \_ service \_ get \_ operating \_ mode \_ output.PartitionId \\
												  	   & ?om \circthen RemoveTimeCounter(procid,om) \\
									  		 \circfi
									  	 \end{array}
									  	 \right) \\
		\\
	\end{array}
	\znewpage
	

	\\
	\circend
\end{zed}

\subsubsection{ChannelSets for communication between different parts of ARINC 653 O/S Layer}
\begin{zed}
	\begin{array}{lll}
		\circchannelset CORE \_ KERNEL \_ SERVICE & == & \lchanset core \_ kernel \_ init \_ partition \_ kernel, \\
													   && \ \: ~ core \_ kernel \_ inited \_ partition \_ kernel, \\
		
													   && \ \: ~ core \_ kernel \_ service \_ get \_ operating \_ mode, \\
													   && \ \: ~ core \_ kernel \_ service \_ get \_ operating \_ mode \_ output, \\
													   && \ \: ~ core \_ kernel \_ service \_ get \_ operating \_ mode \_ return, \\
		
													   && \ \: ~ core \_ kernel \_ service \_ get \_ partition \_ status, \\
													   && \ \: ~ core \_ kernel \_ service \_ get \_ partition \_ status \_ output, \\
													   && \ \: ~ core \_ kernel \_ service \_ get \_ partition \_ status \_ return, \\
\end{array}
	\\
	\begin{array}{lll}
		\circchannelset OSLayerChannels & == & CORE \_ KERNEL \_ SERVICE \cup CORE \_ KERNEL \_ CONTROL \\
		\\
	\end{array}
	\\
\end{zed}

\subsubsection{ARINC 653 O/S Layer Model} \label{circusprocess_ARINC653OSLayer}
\begin{zed}
	\circprocess ARINC653 \_ OS \_ Layer \circdef OSKernelConfig : OS \_ KERNEL \_ CONFIG \_ TYPE \circspot \\
	\\
	\begin{array}{lll}
		CoreKernel(OSKernelConfig) \lpar OSLayerChannels \rpar (\Interleave partid : \dom (OSKernelConfig.Partitions) \circspot \\
		PartitionKernel(partid,OSKernelConfig.Partitions(partid))) \\
		\\
	\end{array}
\end{zed}

\subsection{Application/Executive(APEX) Layer} \label{section:circusprocess_APEX}
\subsubsection{Application/Executive(APEX) Interface} \label{circusprocess_APEXInterface}
\begin{zed}
	\circprocess APEX \_ Interface \circdef partid : PARTITION \_ ID \_ TYPE;callerid : PROCESS \_ ID \_ TYPE \circspot \\
	\circbegin \\
	\\
	%
							   		  				  		  	\right) \\
							   		  				  \circelse & (to > 0 \land (pt + to) \notin SYSTEM \_ TIME \_ TYPE) \circthen \\
							   		  				  			& apex \_ function \_ return \_ code.partid.callerid \ !INVALID \_ PARAM \circthen \\
							   		  				  			& apex \_ function \_ suspend \_ self \_ return.partid.callerid \circthen \Skip \\
							   		  				  \circelse & (to > 0 \land (pt + to) \in SYSTEM \_ TIME \_ TYPE) \circthen \\
							   		  				  			& system \_ call \_ service \_ routine \_ suspend \_ self \_ input.partid.callerid \ !pt \ !to \\
							   		  				  			& \circthen \\
							   		  				  			& system \_ call.partid.callerid \ !SUSPEND \_ SELF \circthen \\
							   		  				  			& system \_ call \_ return.partid.callerid \circthen \\
							   		  				  			& system \_ call \_ service \_ routine \_ suspend \_ self \_ output.partid.callerid \ ?result \\
							   		  				  			& \circthen \\
							   		  				  			& \left(

	\\
	\circend
\end{zed}

\subsubsection{Application/Executive(APEX) Layer Model} \label{circusprocess_APEXLayer}
\begin{zed}
	\circprocess APEX \_ Layer \circdef OSKernelConfig : OS \_ KERNEL \_ CONFIG \_ TYPE \circspot \\
	\begin{array}{lll}
		\Interleave partid : \dom (OSKernelConfig.Partitions) \circspot \\
		\t1 \Interleave procid : 0 \upto (OSKernelConfig.Partitions(partid).TotalProcess) \circspot APEX(partid,procid) \\
		\\
	\end{array}
	\\
\end{zed}

\subsection{Core Module} \label{section:circusprocess_CoreModule}
\subsubsection{ChannelSets for communication between different layers of Core Module}
\begin{zed}
	\begin{array}{lll}
		\circchannelset CH \_ HIS & == & \lchanset context \_ switching \_ save \_ part \_ cpu \_ registers, \\
									   && \ \: ~ context \_ switching \_ saved \_ part \_ cpu \_ registers, \\
		
									   && \ \: ~ context \_ switching \_ save \_ all \_ cpu \_ registers,context \_ switching \_ saved \_ all \_ cpu \_ registers, \\
		
									   && \ \: ~ context \_ switching \_ restore \_ part \_ cpu \_ registers, \\
									   && \ \: ~ context \_ switching \_ restored \_ part \_ cpu \_ registers, \\
		
									   && \ \: ~ context \_ switching \_ restore \_ all \_ cpu \_ registers, \\
									   && \ \: ~ context \_ switching \_ restored \_ all \_ cpu \_ registers, \\
		
									   && \ \: ~ context \_ switching \_ set \_ stack \_ segment \_ registers, \\
\end{array}
	\\
	
	\znewpage
	
	\begin{array}{lll}
		\circchannelset HIS \_ OS & == & \lchanset kernel \_ save \_ context,kernel \_ saved \_ context, \\
		
									   && \ \: ~ kernel \_ restore \_ context,kernel \_ restored \_ context, \\
		
									   && \ \: ~ kernel \_ switch \_ stack,kernel \_ switched \_ stack, \\
		
									   && \ \: ~ kernel \_ init \_ memory \_ management,kernel \_ inited \_ memory \_ management, \\
		
									   && \ \: ~ kernel \_ set \_ area \_ memory \_ management \_ reference, \\
									   && \ \: ~ kernel \_ have \_ set \_ area \_ memory \_ management \_ reference, \\
		
									   && \ \: ~ kernel \_ request \_ area \_ memory \_ management \_ reference, \\
									   && \ \: ~ kernel \_ get \_ area \_ memory \_ management \_ reference, \\
\end{array}
	\\
	
	\znewpage
	
	\begin{array}{lll}
		\circchannelset OS \_ APEX & == & \lchanset system \_ call, \\
		
										&& \ \: ~ system \_ call \_ service \_ routine \_ get \_ operating \_ mode \_ output , \\
		
										&& \ \: ~ system \_ call \_ service \_ routine \_ get \_ partition \_ status \_ output, \\
		
										&& \ \: ~ system \_ call \_ service \_ routine \_ set \_ operating \_ mode \_ input, \\
		
										&& \ \: ~ system \_ call \_ service \_ routine \_ get \_ process \_ names \_ output, \\
		
										&& \ \: ~ system \_ call \_ service \_ routine \_ get \_ process \_ ids \_ output, \\
		
										&& \ \: ~ system \_ call \_ service \_ routine \_ get \_ free \_ process \_ ids \_ output, \\
		
										&& \ \: ~ system \_ call \_ service \_ routine \_ get \_ free \_ space \_ output, \\
\end{array}
	\\
\end{zed}

\subsubsection{Core Module Model} \label{circusprocess_CoreModule}
\begin{zed}
	\circprocess CoreModule \circdef ModuleConfig : MODULE \_ CONFIG \_ TYPE \circspot \\
	\\
	\begin{array}{lll}
		Core \_ Hardware \_ Layer \lpar CH \_ HIS \rpar Hardware \_ Interface \_ Software \_ Layer(ModuleConfig.InterruptHandlerConfig) \\
		\lpar HIS \_ OS \rpar ARINC653 \_ OS \_ Layer(ModuleConfig.OSKernelConfig) \lpar OS \_ APEX \rpar APEX \_ Layer(ModuleConfig. \\
		OSKernelConfig) \\
		\\
	\end{array}
	\\
\end{zed}

\end{document}